\documentclass[11pt,a4paper]{article}
\pdfoutput=1
\usepackage[T1]{fontenc}
\usepackage[english]{babel}
\usepackage[utf8]{inputenc}

\usepackage[textwidth=16cm, textheight=22cm]{geometry}

\usepackage{tocloft} 
\usepackage[colorlinks,linkcolor=black,citecolor=blue,urlcolor=blue,linktocpage,pagebackref]{hyperref}

\usepackage{tikz}
\usetikzlibrary{decorations.pathmorphing}
\usetikzlibrary{decorations.markings}
\usetikzlibrary{calc}

\usepackage{amsmath}
\usepackage{amsfonts}
\usepackage{amssymb}
\usepackage{mathtools}
\usepackage{latexsym}
\usepackage{mathrsfs}
\usepackage{braket}

\usepackage{enumitem}
\usepackage{booktabs}
\usepackage{graphicx}
\usepackage{subcaption}
\usepackage{cite}
\usepackage[normalem]{ulem}

\numberwithin{equation}{section}

\def\nn{\nonumber}

\let\originalleft\left
\let\originalright\right
\renewcommand{\left}{\mathopen{}\mathclose\bgroup\originalleft}
\renewcommand{\right}{\aftergroup\egroup\originalright}

\newcommand{\ped}[1]{\textormath{\textsubscript{#1}}{_{\mathrm{#1}}}}
\newcommand{\ap}[1]{\textormath{\textsuperscript{#1}}{^{\mathrm{#1}}}}

\newcommand{\hypgeo}[2]{%
  \operatorname{%
    {\vphantom{\mathnormal{F}}}_{#1}%
    \kern-\scriptspace
    \mathnormal{F}_{#2}%
  }%
}

\newcommand{\stirlingI}[2]{\genfrac{[}{]}{0pt}{}{#1}{#2}}
\newcommand{\stirlingII}[2]{\genfrac{\{}{\}}{0pt}{}{#1}{#2}}
\newcommand{\stirlingIpoly}[3]{S_{1}\left(\genfrac{}{}{0pt}{}{#1}{#2} \mathop{;} #3\right)}
\newcommand{\stirlingIIpoly}[3]{S_{2}\left(\genfrac{}{}{0pt}{}{#1}{#2} \mathop{;} #3\right)}

\newcommand{\id}{\text{\usefont{U}{bbold}{m}{n}1}}

\newcommand{\im}{\operatorname{Im}}
\newcommand{\re}{\operatorname{Re}}

\usepackage{xifthen}
\newcommand{\dif}[1][]{\mathord{\ifthenelse{\isempty{#1}}{\mathrm{d}}{\mathrm{d}^{#1}}}}

\newcommand{\hatF}{\ensuremath{\widehat{F}}}

\usepackage{pifont}
\newcommand{\cmark}{\ding{51}}
\newcommand{\xmark}{\ding{55}}

\usepackage{xcolor}
\definecolor{stokes}{rgb}{0.40082222609352647, 0.5220066643438841, 0.85}
\definecolor{branch}{rgb}{0.915, 0.3325, 0.2125}
\definecolor{path}{rgb}{0.880722, 0.611041, 0.142051}
\definecolor{region}{rgb}{0.9728288904374106, 0.621644452187053, 0.07336199581899142}

\definecolor{green}{rgb}{0.3,0.7,0.3}

\begin{document}

\setcounter{page}{0}
\thispagestyle{empty}

\parskip 3pt

\font\mini=cmr10 at 2pt

\hypersetup{pageanchor=false}
\begin{titlepage}
    \vspace{1cm}
    \begin{center}
	\vspace*{-.6cm}
	\begin{center}
		\vspace*{1.1cm}
		{\centering \Large\textbf{On the $1/c$ expansion in $2d$ CFTs with degenerate operators}}
	\end{center}
	\vspace{0.8cm}
	{\bf Agnese Bissi$^{a,b,c}$, Nicola Dondi$^{a,c}$,\\[0.5em] Alessandro Piazza$^{d,c}$, Tomas Reis$^{d,c}$, and Marco Serone$^{d,c}$}
	\vspace{1.cm}
 
 ${}^a\!\!$
	{\em  Abdus Salam International Centre for Theoretical Physics, \\
				Strada Costiera 11, 34151, Trieste, Italy}

	\vspace{.3cm}
   ${}^b\!\!$			
	{\em  Department of Physics and Astronomy, Uppsala University,\\
Box 516, SE-751 20 Uppsala, Sweden}
\vspace{.3cm}

${}^c\!\!$			
	{\em INFN, Sezione di Trieste, Via Valerio 2, I-34127 Trieste, Italy}
	\vspace{.3cm}
	
${}^d\!\!$
	{\em SISSA, Via Bonomea 265, I-34136 Trieste, Italy}

	        \vspace{.8cm}

        {\small{\texttt{abissi@ictp.it, ndondi@ictp.it,  \\ a.piazza@sissa.it, treis@sissa.it, serone@sissa.it}}}
    \end{center}
    \vskip 15pt
    
    \begin{abstract}
        We analytically determine the large central charge asymptotic expansion of Virasoro conformal blocks entering in four-point functions with external degenerate operators on a sphere in $2d$ CFTs, and study its resurgence properties as a function of the conformal cross-ratio $z$. 
        We focus on the cases of four heavy $(2,1)$ degenerate operators, and two $(2,1)$ heavy degenerate ones plus two arbitrary light operators.
        The $1/c$ asymptotic series is Borel summable for generic values of $z$, but it jumps when a Stokes line is crossed.
        Starting from the $1/c$ series of the identity block, we show how a resurgent analysis allows us to completely determine the other Virasoro block and in fact to reconstruct the full correlator. 
        We also show that forbidden singularities, known to exist in correlators with two heavy and two light operators,  appear with four heavy operators as well.
        In both cases, they are turning points emanating Stokes lines, artefacts of the asymptotic expansion, and we show how they are non perturbatively resolved.
        More general correlators and implications for gravitational theories in $\text{AdS}_{3}$ are briefly discussed.
        Our results are based on new asymptotic expansions for large parameters $(a,b,c)$ of certain hypergeometric functions $\hypgeo{2}{1}(a,b,c;z)$ which can be useful in general.  
    \end{abstract}
\end{titlepage}
\hypersetup{pageanchor=true}

\tableofcontents

\section{Introduction}

Two-dimensional conformal field theories (CFTs) are an incredibly rich class of theories. 
Due to the power of conformal symmetry in $2d$,
entire classes of strongly coupled models can be analytically solved~\cite{Belavin:1984vu}.
Within the AdS/CFT correspondence~\cite{Maldacena:1997re,Witten:1998qj,Gubser:1998bc}, $2d$ CFTs with large central charge $c$ are expected to be related to a semi-classical gravitational theory on $\text{AdS}_3$ with ${c = 3\ell/2G_N}$, 
$G_N$ and $\ell$ being Newton's constant and the AdS$_3$ length of the $3d$ theory, respectively~\cite{Brown:1986nw}. 
The $1/c$ expansion in a $2d$ CFT is then mapped to a gravitational semi-classical expansion around $3d$ asymptotic AdS space-times, and as such it is expected to be factorially divergent.

Virasoro primary states with conformal dimension $\Delta=h+\bar h$, $h$ and $\bar h$ being the holomorphic and anti-holomorphic conformal weights, are associated to specific fields in $\text{AdS}_3$.
The $\Delta = 0$ identity state corresponds to pure $\text{AdS}_3$. 
$2d$ states with fixed $\Delta$ as $c\to\infty$ can be seen as local perturbations of $\text{AdS}_3$, and are typically denoted ``light'' states.
States with $\Delta\propto c$ are denoted ``heavy'': they can no longer be considered as probes and back-react on the metric.
For $\Delta < c/12$, they correspond to matter fields with mass $m^2 = \Delta (\Delta-2)/\ell^2$ and give rise to $\text{AdS}_3$ backgrounds with a deficit/excess angle $\Delta \phi = 2\pi \sqrt{1-12 \Delta/c}$.
States with $\Delta>c/12$ are instead associated with Banados-Teitelboim-Zanelli (BTZ) black holes~\cite{Banados:1992wn}.
CFTs with a finite number of states with $\Delta < c/12$ and large $c$ are particularly interesting because they can admit a dual description in terms of a semi-classical theory of gravity (holographic CFTs)~\cite{Heemskerk:2009pn}.

Correlation functions of local operators in CFTs can be expanded using the operator product expansion (OPE) in terms of products of holomorphic and anti-holomorphic Virasoro conformal blocks $\mathcal{F}_{h}$ and $\overline{\mathcal{F}}_{\bar{h}}$. 
In contrast to the $\text{SL}(2,\mathbb{R})$ global blocks, the analytic form of the Virasoro blocks is unknown.\footnote{Although the form of the blocks is unknown, the crossing kernel relating Virasoro blocks in different channels is explicitly known~\cite{Ponsot:1999uf,Ponsot:2000mt}.}
A convergent power series in the conformal cross-ratio $z$ for Virasoro blocks was provided long ago by Zamolodchikov~\cite{Zamolodchikov:1984eqp, Zamolodchikov:1987avt}. 
Starting from this expression, obtaining a $1/c$ expansion is in principle straightforward, as already pointed out in~\cite{Fitzpatrick:2015zha,Perlmutter:2015iya}. 
For four-point correlators on the sphere, the most studied case, we schematically have
\begin{equation}\label{eq:Int0}
    \mathcal{F}_h(h_i,c; z) \sim \sum_{k=0}^\infty \frac{f_k(h,h_i; z)}{c^k}\,, 
\end{equation}
where $h_i$ $(i=1,2,3,4$) and $h$ denote the conformal weights of the external and the exchanged operators, respectively.
The coefficients $f_k$ are functions of $h$, $h_i$ and $z$, whose explicit form can be recursively computed but whose closed form is generally hard to determine.
In writing~\eqref{eq:Int0} we have assumed that $h$ and $h_i$ do not scale with $c$, in which case $f_0$ coincides with the $\text{SL}(2,\mathbb{R})$ global block.
We do not have an equality sign in~\eqref{eq:Int0} because the $1/c$ expansion is generally asymptotic.

Over the years, great progress has been achieved in finding approximate expressions for the Virasoro blocks. 
This was mostly motivated by the fact that, in holographic CFTs, such expressions can shed light and give a new perspective on the properties of $3d$ gravitational theories.
Special attention has been paid to the Virasoro identity block exchange, $h=0$ in~\eqref{eq:Int0}, and in the cases in which 
two or all four external operators have $h_i\propto c$ and are heavy.
When all operator dimensions scale with $c$, Virasoro blocks are expected to exponentiate~\cite{ZamolodchikovExp} (see~\cite{Besken:2019jyw} for a proof). 
A form of exponentiation is also expected for arbitrary exchanged operators, in particular for the identity exchange.
For four identical external heavy operators with $h_{i}=\eta c$, a useful parametrization of the $1/c$ expansion of the identity block $\mathcal{F}_{1,1} = \mathcal{F}_{h=0}$ is given by
\begin{equation}\label{eq:Int1}
    \mathcal{F}_{1,1}(h_i,c; z) \sim e^{-c S(\eta, z)} A\ped{4H}(\eta,z) \left(1+\sum_{k=1}^\infty \frac{g_k(\eta, z)}{c^k}\right)\,, \qquad \qquad (\text{4H})\,.
\end{equation}
The leading order terms $S$ and $A\ped{4H}$ define a semi-classical approximation to the block and their form is in general unknown, like all other sub-leading terms. 
For specific external heavy operators (twist fields), $\mathcal{F}_{1,1}$ has been shown to govern the Rényi entropy of two intervals in holographic CFTs and the expression of $S$ for $z\to 0$ has been determined~\cite{Hartman:2013mia}. 
For $\mathbb{Z}_2$ twist fields, with $h_i=c/32$, the semi-classical block $S$ can be determined exactly~\cite{Maloney:2016kee}.

The setup which has been more extensively studied is the one of a four-point correlator between two light (L) and two heavy (H) generic states.
In this case, the identity block can be written as 
\begin{equation}\label{eq:Int2}
    \mathcal{F}_{1,1}(h_i,c; z) \sim z^{- 2 h\ped{H}} A_\text{2H2L}(\eta, h\ped{L}, z) \left(1+\sum_{k=1}^\infty \frac{h_k(\eta,h\ped{L},z)}{c^k}\right)\,, \qquad \qquad (\text{2H2L})\,,
\end{equation}
where for simplicity we have taken identical light and heavy fields, with conformal weights $h\ped{H} = \eta c$ and $h\ped{L}$, respectively. 
The semi-classical block in this case is identified with the leading order term $A_\text{2H2L}$, whose form has been worked out in~\cite{Fitzpatrick:2014vua,Fitzpatrick:2015zha}.
From the gravity side, $A_\text{2H2L}$ has been shown to correspond to a Witten diagram on geodesics over a conical geometry, the latter created by the heavy operators~\cite{Hijano:2015qja}. Its form can also be determined by taking appropriate limits of the crossing kernels~\cite{Collier:2018exn}. 

The identity block exchange in 2H2L correlators, in the appropriate channel, is roughly expected to correspond on the $\text{AdS}_{3}$ side to the all-order perturbative series of graviton exchange between the probe light state interacting with the source (point-like for $\Delta\ped{H}<c/12$ or a BTZ black hole for $\Delta\ped{H}>c/12$).\footnote{
    In $3d$ there are no propagating gravitons, but we do have gravitational interactions due to off-shell gravitons. In this sense, the situation is analogous to gauge interactions between charged particles in $2d$ gauge theories.}
When $\Delta\ped{H}>c/12$, assuming the eigenstate thermalization hypothesis, the four-point function to leading order corresponds to the thermal two-point function of the light states in the presence of a BTZ black hole, with mass, spin and temperature governed by the conformal weights of the heavy state~\cite{Fitzpatrick:2014vua}.
In Minkowski signature this correlator, when approximated with the exchange of the semi-classical identity block, decays exponentially to zero at late-times, a signal of information loss~\cite{Maldacena:2001kr}. 
This is expected since in this approximation the exponential decay is given by quasi-normal modes associated with the relaxation to equilibrium of the excited BTZ black hole~\cite{Birmingham:2001pj}. 
The identity exchange in an out-of-time-ordered 2H2L correlator has also been studied, in connection with chaotic properties of $2d$ CFTs~\cite{Roberts:2014ifa,Fitzpatrick:2016thx}. 

The very same semi-classical $A\ped{2H2L}$ in~\eqref{eq:Int2}, in Euclidean signature, has been shown to be singular away from OPE singularities of the correlators, which are the only ones
expected to occur. 
In analogy to the phenomenon of information loss, such forbidden singularities should be an artefact of the semi-classical expansion and are expected to be somehow smoothed out at the ``quantum'' level. 
Understanding how such singularities cancel motivated the study of the first $1/c$ corrections to~\eqref{eq:Int2}, computed in~\cite{Beccaria:2015shq,Fitzpatrick:2015dlt,Fitzpatrick:2015foa}.
A universal behaviour of the identity block close to the forbidden singularities has been proposed in~\cite{Fitzpatrick:2016ive} (see also~\cite{Fitzpatrick:2016mjq} for a related analysis).
When one or more of the external operators are degenerate, a more in-depth analysis is possible~\cite{Chen:2016cms}. 
By performing an analysis of Stokes phenomena of the identity block for external degenerate operators, it has been argued in~\cite{Fitzpatrick:2016ive} that non-perturbative effects of order $\exp(-c)$ are responsible for the resolution of both forbidden singularities and information loss in the Euclidean and Lorentzian regimes, respectively. 
See also~\cite{Chen:2017yze} for a numerical study and~\cite{Faulkner:2017hll} for a proposal of how forbidden singularities can be resolved. Although it was already suspected that an explanation of what forbidden singularities are and how they are resolved could be obtained by a systematic analysis of the $1/c$ asymptotic series~\eqref{eq:Int2} (see e.g.~\cite{Fitzpatrick:2016ive,Fitzpatrick:2016mjq}), such an analysis has not been performed yet.

The theory of resurgence~\cite{ecalle} is the appropriate tool to investigate this problem as it provides, among other things, a convenient way to discover non-perturbative effects.
In the context of $2d$ QFTs, these techniques have mostly been applied on integrable models with asymptotically free weak coupling limits~\cite{Dunne:2012ae,Cherman:2014ofa,Bajnok:2022xgx,Abbott:2020qnl,Schepers:2023dqk,Schepers:2020ehn,Marino:2019eym,Marino:2021dzn,DiPietro:2021yxb}.
On the other hand, in the context of CFT specifically, the investigation has mostly been in four and three dimensions, namely for $\mathcal{N}=4$ SYM and the large charge expansion of sigma models, see e.g.~\cite{Paul:2023rka,Aniceto:2015rua,Dondi:2021buw,Antipin:2022dsm,Dorigoni:2024dhy}. 
For resurgence of the large $c$ expansion in two dimensional CFT, there were some early investigations in~\cite{Fitzpatrick:2016ive} and recent study in~\cite{Benjamin:2023uib,Benjamin:2024cvv}.

The aim of this paper is to perform a resurgent analysis of the $1/c$ expansion for four-point correlators of the type 4H and 2H2L.
Since a rigorous study requires a detailed knowledge of the terms $g_k$ and $h_k$ appearing in~\eqref{eq:Int1} and~\eqref{eq:Int2}, which is hard to achieve in general,
in this paper we focus our attention on correlators involving degenerate operators of the kind $\phi_{2,1}$ (notation as in~\cite{DiFrancesco:1997nk}).\footnote{See~\cite{Benjamin:2023uib} for a first analysis in this direction.} 
As well known, such correlators satisfy a second-order differential equation and the Virasoro blocks can be determined in terms of hypergeometric functions. 
We focus on the cases of four heavy $\phi_{2,1}$ degenerate operators (4H) and two heavy $\phi_{2,1}$ degenerate plus two arbitrary light operators (2H2L).
For the above specific correlators, starting from the all-order $1/c$ series associated with the identity block (related with the all-order graviton corrections around the $3d$ background formed by the heavy states), we answer positively the following questions:
\begin{itemize}
    \item Can we see that the identity block cannot be the full answer and other non-perturbative terms should be present? 
    \item Do forbidden singularities appear also for 4H correlators?
    \item  How are these globally resolved for both 4H and 2H2L?
    \item  More ambitiously, can we reconstruct the full correlators? 
\end{itemize}
Since the resulting holographic theories are necessarily non-unitary, we do not discuss information loss and consider correlators only in the Euclidean regime.

\subsection{Summary of the results}

We summarize here the main results of the paper and refer the reader to the appropriate sections for full explanations and derivations.

Four-point correlators of $\phi_{2,1}$ degenerate fields involve two Virasoro blocks, $\mathcal{F}_{1,1}$ associated to the exchange of the identity operator $\phi_{1,1} = \id$ and $\mathcal{F}_{3,1}$ associated to the exchange of the degenerate $\phi_{3,1}$ field.
Both are exactly determined in terms of hypergeometric functions $\hypgeo{2}{1}(\alpha,\beta,\gamma;z)$, where $\alpha,\beta,\gamma$ scale with $c$. 
We show how such hypergeometric functions can be rewritten in terms of other $\hypgeo{2}{1}$-s where only one out of the three parameters scales with $c$, see~\eqref{eq:2F1-expansion-4H-identity} and~\eqref{eq:2F1-expansion-2H2L-identity-lambda}. 
We then determine the whole asymptotic series of the latter hypergeometric function, reported in~\eqref{eq:2F1-expansion-large-c-no-shift}.
To the best of our knowledge, these results present new asymptotic expansions for large parameters of $\hypgeo{2}{1}$ which can be useful more broadly in mathematics and physics. 
For the reader's convenience, we have collected these results in appendix~\ref{app:summary2F1}.

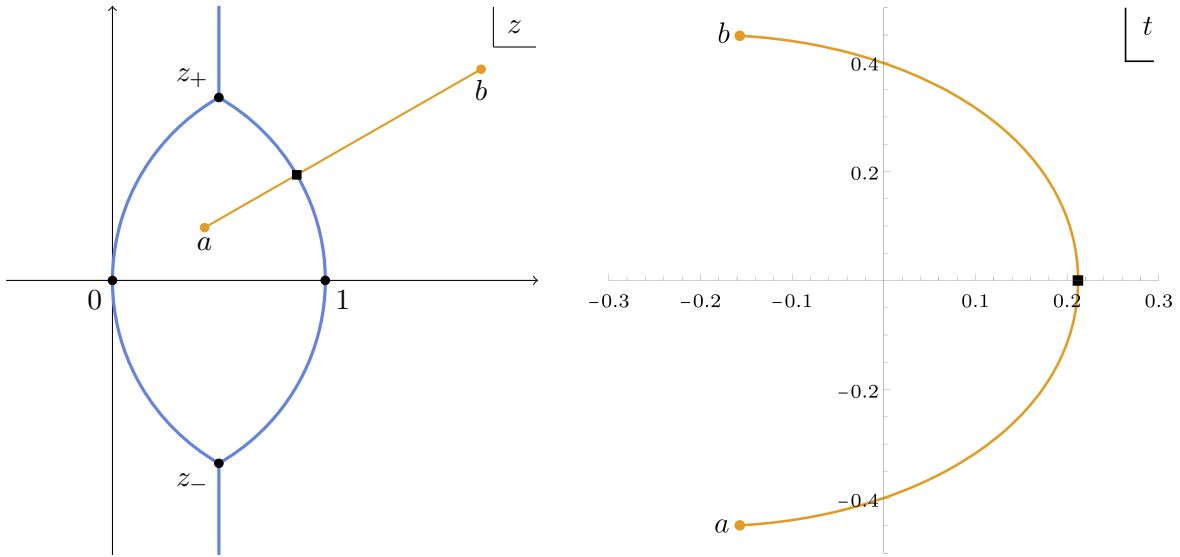
\begin{figure}[t!]
    \centering
    \hspace{0.5em}
    \begin{subfigure}[c]{0.48\textwidth}
        \centering
        \begin{tikzpicture}[scale=2.8]
            \coordinate (zp) at (0.5,{sqrt(3)/2});
            \coordinate (zm) at (0.5,{-sqrt(3)/2});
            \coordinate (A) at ({sqrt(3)/4},{1/4});
            \coordinate (B) at ({sqrt(3)},1);
            \coordinate (C) at ({sqrt(3)/2-0.02},{1/2-0.02});

            \draw[->] (-0.5,0) -- (2,0);
            \draw[->] (0,-1.3) -- (0,1.3);

            \node[draw=none, inner sep=5pt, append after command={
                \pgfextra{\draw (\tikzlastnode.south west) -- (\tikzlastnode.south east);}
                \pgfextra{\draw (\tikzlastnode.south west) -- (\tikzlastnode.north west);}
            }] at (1.89, 1.2) {$z$};

            \draw[stokes, very thick] (zm) arc[start angle=-60, end angle=60, radius=1];
            \draw[stokes, very thick] (zp) arc[start angle=120, end angle=240, radius=1];
            \draw[stokes, very thick] (0.5,{sqrt(3)/2}) -- (0.5,1.3);
            \draw[stokes, very thick] (0.5,{-sqrt(3)/2}) -- (0.5,-1.3);
            \filldraw[black] (zp) circle (0.02) node[above left] {$z_+$};
            \filldraw[black] (zm) circle (0.02) node[below left] {$z_-$};
            \filldraw[black] (1,0) circle (0.02) node[below right] {$1$};
            \filldraw[black] (0,0) circle (0.02) node[below left] {$0$};

            \draw[path, thick] (A) -- (B);
            \filldraw[black] ({sqrt(3)/2-0.02},{1/2-0.02}) rectangle +(0.04,0.04);
            \filldraw[path] (A) circle (0.02) node[below, text=black] {$a$};
            \filldraw[path] (B) circle (0.02) node[below, text=black] {$b$};
        \end{tikzpicture}
    \end{subfigure}
    \hfill
    \begin{subfigure}[c]{0.48\textwidth}
        \includegraphics[width=\textwidth]{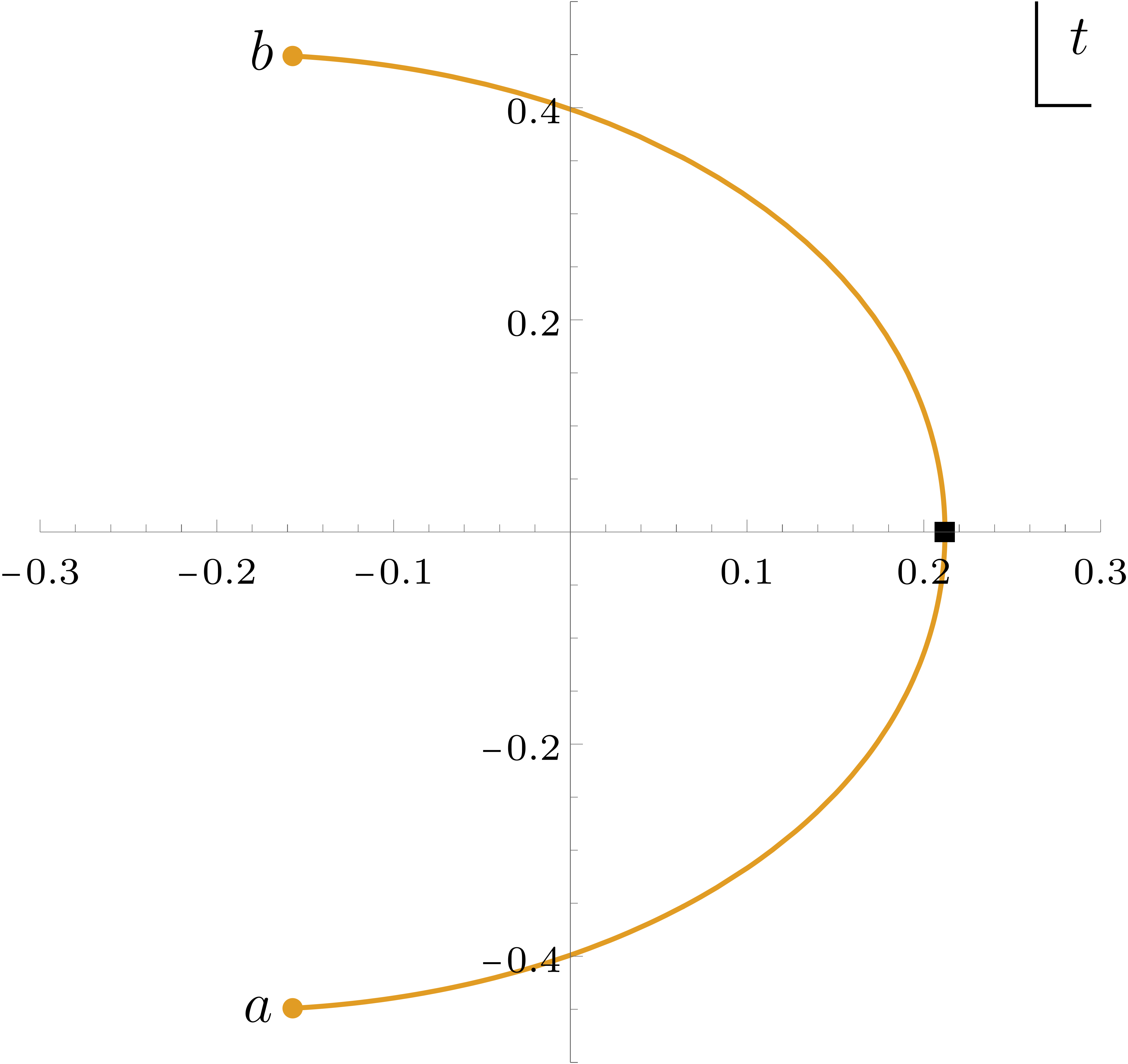}
    \end{subfigure}
    \hspace{0.5em}
    \caption{(left) Stokes lines in the $z$-plane for the  4H correlator (blue lines). The points $z_\pm= \exp(\pm i\pi/3)$ are forbidden singularities. The orange line is $z = \rho \exp(i \pi/6)$ with $1/2\leq \rho\leq 2$. (right) The position of the leading Borel singularity as $z$ varies along the straight orange line in the left panel. When the orange line intersects the blue Stokes line (squared bullet), the Borel singularity is on the positive real axis in the $t$-plane. See figure~\ref{fig:4H-stokes-and-branch} for a more detailed version and explanation of this figure.\label{fig:Fig12Draft}}
\end{figure}

Using the above results, we determine to all orders the terms $g_k$ and $h_k$ appearing in~\eqref{eq:Int1} and~\eqref{eq:Int2}.
The coefficients admit a simple expression using appropriate maps, $z\to r(z)$ (4H) or $z\to v(z)$ (2H2L), see~\eqref{eq:rmapSec} and~\eqref{eq:vmapSec} for their explicit form. 
In both cases, the forbidden singularities are non-analytic points of the above maps. 
At those points, all terms in~\eqref{eq:Int1} and~\eqref{eq:Int2} diverge.

\begin{figure}[t!]
\centering
\begin{tikzpicture}[scale=2.4]
    \coordinate (one) at (1,0);
    \coordinate (two) at (2,0);
    \draw[->] (-1,0) -- (3,0);
    \draw[->] (0,-1.5) -- (0,1.5);

    \node[draw=none, inner sep=5pt, append after command={
        \pgfextra{\draw (\tikzlastnode.south west) -- (\tikzlastnode.south east);}
        \pgfextra{\draw (\tikzlastnode.south west) -- (\tikzlastnode.north west);}
    }] at (2.86, 1.37) {$z$};

    \draw[stokes, very thick] (two) arc[start angle=0, end angle=360, radius=1];
    \draw[stokes, very thick] (one) -- (2.98,0);
    \filldraw[black] (0,0) circle (0.02) node[below left] {\( 0 \)};
    \filldraw[black] (one) circle (0.02) node[below left] {\( 1 \)};
    \filldraw[black] (two) circle (0.02) node[below right] {\( 2 \)};
    \end{tikzpicture}
    \caption{Stokes lines in the $z$-plane for the 2H2L correlator (blue lines). The point $z=2$ is a forbidden singularity.
    See figure~\ref{fig:2H2L-stokes} for a more detailed version and explanation of this figure.\label{fig:Fig2Draft}}
\end{figure}
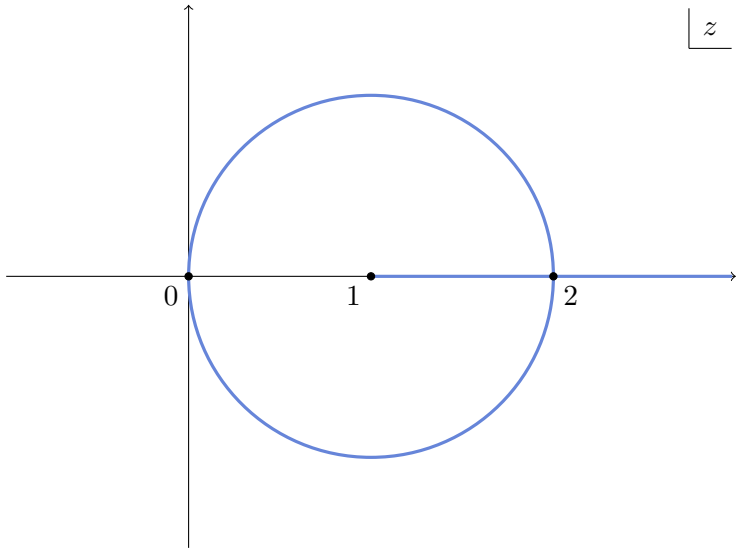
Having an analytic expression for the entire asymptotic series, we determine its associated Borel function, large order behaviour (see~\eqref{eq:LBO4H}  and~\eqref{eq:LBO2H2L})
and its Borel summability, which crucially depends on $z$. 
The $1/c$ asymptotic series is Borel summable for generic values of $z$, when all singularities of the Borel function associated with the two asymptotic series~\eqref{eq:Int1} and~\eqref{eq:Int2} are away from the real positive axis in the Borel $t$-plane.
For special values of $z$, however, some singularities approach the real axis and give rise to a Stokes phenomenon, see~\eqref{eq:stokes-hhhh} (4H) and~\eqref{eq:hhll-pure-stokes} (2H2L). 
This occurs in a codimension-one locus in the complex plane forming a web of Stokes lines, see figure~\ref{fig:Fig12Draft} (4H) and figure~\ref{fig:Fig2Draft} (2H2L) for an illustration.
Forbidden singularities are nothing else than points emanating Stokes lines.
Remarkably, they are not exclusive to 2H2L correlators but feature also in 4H ones.
Interestingly enough, they only appear in the sub-leading term $A_{\text{4H}}(z)$ in \eqref{eq:Int1}, the leading term $S(z)$ being smooth at forbidden singularities.\footnote{This is consistent with $S(z)$ computed for the identity block with $h_i=c/32$ fields in \cite{Maloney:2016kee}.}
We conjecture that, in analogy to 2H2L correlators, forbidden singularities in $A_{\text{4H}}(z)$ persist for more general 4H correlators.

In exact WKB terminology, we get two simple turning points (three Stokes lines converging) for 4H, $z_+$ and $z_-$ in figure~\ref{fig:Fig12Draft},
and one double turning point (four Stokes lines converging) for the 2H2L correlators, $z=2$ in figure~\ref{fig:Fig2Draft}.
Forbidden singularities in both cases correspond to an artefact of the asymptotic expansion. 
We show how Stokes line discontinuities combine with the semi-classical branch cut to turn a forbidden singularity into a perfectly regular point, with trivial monodromy, see~\eqref{eq:trivial-monodromy-1} (4H) and~\eqref{eq:trivial-monodromy-2} (2H2L).

The Airy function provides a simple example of this phenomenon. 
The exact function is analytic at $z=0$, but its asymptotic expression, being an expansion in powers of $z^{-3/2}$, is ill-defined there.
The point $z=0$ is a simple turning point and is the analogue of a forbidden singularity.
The combined monodromies due to Stokes lines jumps and branch cuts appearing in the asymptotic representation of the function nicely cancel,
reflecting the actual analytic nature of the point $z=0$.

The $\mathcal{F}_{3,1}$ block is completely determined in terms of the Stokes discontinuity of the identity block. 
Putting together the identity block and its Stokes discontinuity, and combining holomorphic and anti-holomorphic components, as in~\eqref{eq:four-point-ansatz}, the full four-point function is reconstructed by demanding trivial monodromy around $z=0$ and $z=1$.

For both correlators 4H and 2H2L, two scenarios are possible:
\begin{enumerate}[label=\roman*)]
    \item\label{item:c-positive} $c > 0$, $h_{2,1} < 0$.
    \item\label{item:c-negative} $c < 0$, $h_{2,1} > 0$.
\end{enumerate} 
Starting from the identity block, the two cases are quite distinct. 
In~\ref{item:c-positive} there exist regions in $z$ where the identity block expansion Borel resums to the exact result. 
It should be noted that in this case, the identity block is sub-leading in the OPE limit compared to the $\phi_{3,1}$ exchange, since $h_{3,1}<0$. 
In case~\ref{item:c-negative} the identity block dominates in the OPE limit $(h_{3,1}>0)$, but the asymptotic series of the identity block now never reconstructs the exact block, because the latter includes non-perturbative $e^{-c}$ terms which would be missed. 
The Borel resummation of the series gives rise in this case to the unique combination of the identity and $\mathcal{F}_{3,1}$ block with suitable analytic properties in the $1/c$ plane, as expected from general arguments. 
We summarize the key properties of the two cases in table~\ref{tab:1}. 
The accuracy of the semi-classical approximation of the identity block in the $z$-plane matches the expectations coming from an analysis of Stokes and anti-Stokes lines, see figures~\ref{fig:4H-exact-vs-semiclassic} and~\ref{fig:2H2L-exact-vs-semiclassic} for both $c>0$ and $c<0$ for 4H and 2H2L.

Interestingly enough, the Borel summability to the exact result of the $1/c$ expansion for positive central charge extends down to $c<1$.
We can then interpret the correlators of unitary minimal models with $c<1$ as the $1/c$ resummations in the ``deep quantum gravity'' regime
of an $\text{AdS}_{3}$ theory in the presence of exotic negative energy sources!\footnote{
    This is different from the connections between certain minimal models and $3d$ gravity pointed out in~\cite{Castro:2011zq,Maloney:2016kee}.
} 

\begin{table}[t!]
    \centering
    \renewcommand{\arraystretch}{1.2}
    \begin{tabular}{c|cc}
        & $c>0$ & $c<0$ \\ \midrule 
        $s(\widetilde{{\cal{F}}}_{1,1}) = {\cal F}_{1,1}$ & \cmark & \xmark \\
        Stokes jumps & \cmark & \cmark \\
        Identity leading as $z\to 0$ & \xmark & \cmark \\
        Full correlator from $\widetilde{{\cal F}}_{1,1}$ & \cmark & \cmark \\ 
    \end{tabular}
\caption{Summary of the key properties of the asymptotic expansion $\widetilde{{\cal{F}}}_{1,1}$ of the identity block for both $c>0$ and $c<0$. The results apply for both the 4H and the 2H2L correlators.
The relation $s(\widetilde{{\cal{F}}}_{1,1}) = {\cal F}_{1,1}$ is understood to apply only in a sector of the $z$-complex plane delimited by Stokes lines.
See the main text for further explanations.\label{tab:1}}
\end{table}

\subsection{Structure of the paper}
We start in section~\ref{sec:4pFwDO} by recalling a few basic facts about the 4H and the 2H2L correlators we will be discussing.
Based on the results reported in the appendices, we consider in section~\ref{sec:4H} the 4H correlator with positive large central charge. We show the appearance of forbidden singularities in the semi-classical expansion of the identity block, 
its large order behaviour, how forbidden singularities are characterized in terms of Stokes jumps and resolved, and how the full correlator can be reconstructed 
from the knowledge of the expansion of the identity block alone. We then discuss how the above analysis is modified when the central charge is large but negative. 
The same study is performed for the 2H2L correlator in section~\ref{sec:2H2L}, where at the end we comment on how from large $c$ we can interpolate
our results until reaching the unitary minimal models with $c<1$.
In the final section~\ref{sec:outlook} we briefly discuss possible further directions of investigations both from a gravitational/Chern-Simons and a CFT point of view and we also discuss extensions of our findings to unitary theories.
Three appendices complete the paper. In appendix~\ref{app:hypgeom} we show how Stokes lines of the large parameter expansion of $\hypgeo{2}{1}$ can be determined,
as well as various analytic closed formulas for several asymptotic expansions.
The resurgence properties of such series, including their large order behaviour and Stokes discontinuities, are discussed next in appendix~\ref{app:resurgence}.
We collect in a final appendix~\ref{app:summary2F1} the asymptotic expansions for large parameters $(a,b,c)$ of 
$\hypgeo{2}{1}(a,b,c;z)$ derived in this work which appear, to the best of our knowledge, to be new.
A reader interested uniquely to these mathematical results can directly refer to appendix~\ref{app:summary2F1}, which is independent of the rest of the paper.

\subsection{Notation}

We distinguish an actual function $f$ from its formal asymptotic power series in a small parameter $\epsilon$ by a tilde:
\begin{equation}
    f(\epsilon) \sim \widetilde{f}(\epsilon) = \sum_{n=0}^\infty f_n \epsilon^n \,.
\end{equation}
A hat denotes the associated Borel function defined as:
\begin{equation}\label{eq:borel-function-def}
    \widehat{f}(t) = \sum_{n=1}^\infty \frac{f_{n}}{(n-1)!} t^{n-1} \,.
\end{equation}
We define the Laplace transform in the direction $\theta$ as
\begin{equation}\label{eq:laplace-transform}
    s_\theta(\widetilde f) = f_0 + \int_0^{e^{i \theta} \infty}\dif{t} \, e^{-\frac{t}{\epsilon}} \widehat{f}(t) \, , \qquad \re\left( \frac{e^{i\theta}}{\epsilon} \right) >0 \, .
\end{equation}
We also define $s_{\theta \pm} = s_{\theta \pm \delta}$ with $0< \delta \ll 1$. We denote the Borel resummation of $\widetilde{f}$ as
\begin{equation}
\begin{aligned}
    f^B &= s_0(\widetilde{f})\, , \qquad \epsilon > 0 \, , \\
     f^B &= s_\pi(\widetilde{f})\, , \qquad \epsilon < 0 \, .
\end{aligned}
\end{equation}
See appendix~\ref{app:resurgence} and references within for a quick review of asymptotic analysis and resurgence.

\section{Four-point functions with degenerate operators}
\label{sec:4pFwDO}

In this section, we review basic facts about the Virasoro conformal blocks and the form of four-point functions in the presence of degenerate fields and set up our notations. 
We use conventions as in~\cite{DiFrancesco:1997nk}.
For simplicity, we consider non-chiral CFTs with $c_L=c_R=c$.

Degenerate operators \( \phi_{r,s} \) are Virasoro primaries having a null descendant at level \( rs \). 
By the Kac formula, these operators have holomorphic conformal weight
\begin{equation}\label{eq:degenerate_ops}
    h_{r,s} = - \frac{b^{2}}{4}(r^{2}-1) - \frac{1}{4b^{2}}(s^{2} - 1) - \frac{1}{2}(rs-1) \, ,
\end{equation}
where $r,s$ are positive integers and \( b \) is a convenient parametrization for the central charge:
\begin{equation}
    c = 13 + 6 \left(b^{2} + \frac{1}{b^{2}}\right) \, .
\end{equation}
The operators $\phi_{r,s}$ can be part of a unitary CFT only if $0<c <1$, but they exist for any value of $c$ if the unitarity constraint is relaxed.
We focus on the simplest case of four-point functions on the sphere involving the degenerate field \( \phi_{2,1} \). 
The fusion rule of the field with itself contains just two Virasoro primaries
\begin{equation}\label{eq:fusion}
    \phi_{2,1} \times \phi_{2,1} \sim \phi_{1,1}  + \phi_{3,1}\,,
\end{equation}
\( \phi_{1,1} = \id \) being the identity operator. 
The operators involved have holomorphic weights
\begin{equation}\label{eq:hDegConfWeights}
    h_{2,1} = -\frac{3}{4}b^{2} - \frac{1}{2}\,, \qquad h_{1,1} = 0 \,, \qquad h_{3,1} = -2 b^{2} - 1 \,.
\end{equation}
We take the large $c$ limit by sending \( |b^{2}| \to +\infty \). 
For simplicity, we consider the case in which $b^2$ is real, though it is not difficult to extend our analysis for complex values.\footnote{
    As we will see, Stokes discontinuities will anyhow force us to give an imaginary component to $b^2$.
} 
In both cases the heavy field $H$ is identified with the degenerate field as
\begin{equation}
    H(z,\bar z) = \phi_{2,1}(z) \bar\phi_{2,1}(\bar z)\,,
\end{equation}
with a conformal weight proportional to $c$.
For $b^{2} \to +\infty$, $c>0$ and $\Delta\ped{H} < 0$. For $b^{2} \to -\infty$, $c<0$ while $\Delta\ped{H} > 0$.
We will separately consider both cases in the next sections. 

The fusion rule~\eqref{eq:fusion} implies that in an OPE expansion, the four-point function is given by the sum of the product of two Virasoro blocks. 
Moreover, since $H$ has a null descendant at level 2, the four-point function satisfies a second-order equation of hypergeometric type (both in $z$ and $\bar{z}$). 
This allows for an exact determination of the two blocks, which are nothing but a basis of solutions of such differential equation.
Demanding trivial monodromy at $z=0$ and $z=1$ finally allows us to reconstruct the entire correlator.

The two correlators we will analyse are 
\begin{align}
    {\cal A}\ped{4H} & = \braket{H(\infty) H(1)  H(z,\bar z)  H(0)}\,, \label{eq:4H-correlator} \\
    {\cal A}\ped{2H2L}& = \braket{L(\infty) L(1)  H(z,\bar z)  H(0)}\,. \label{eq:2H2L-correlator}
\end{align}
In the 2H2L correlator, the field $L$ is a generic light scalar operator with scaling dimension $\Delta_\text{L} = 2h_\text{L}$. 
It is convenient to parametrize its conformal weight as
\begin{equation}\label{eq:gammaDef}
    h\ped{L}(b^2) = \gamma + \frac{\gamma(1-\gamma)}{b^2} \, ,
\end{equation}
with $\gamma$ constant. 
The explicit form of these correlators is given by 
\begin{align}
    {\cal A}\ped{4H} 
    & = \left|\mathcal{F}_{1,1}\right|^2  + \lambda\ped{4H} \left|\mathcal{F}_{3,1}\right|^2\,, \label{eq:4H-block-decomposition} \\
    {\cal A}\ped{2H2L}
    & = \left|\mathcal{F}_{1,1}\right|^2  + \lambda_{\text{2H2L}} \left|\mathcal{F}_{3,1}\right|^2 \,. \label{eq:2H2L-block-decomposition}
\end{align}
Here $\mathcal{F}_{1,1}$ and $\mathcal{F}_{3,1}$ are the Virasoro conformal blocks, associated with the exchange of the operators in the fusion rule~\eqref{eq:fusion}. 
$\lambda$ are the product of OPE coefficients entering the $s$-channel decomposition of the corresponding four-point function. 
The two-point functions are unit normalized and in the $s$-channel OPE limit, $z\to 0$, we normalize $\mathcal{F}_{h} \approx z^{-2h_{2,1} + h}$. 
The explicit form of Virasoro blocks is given in terms of $\hypgeo{2}{1}$ hypergeometric functions (see e.g.~\cite{Ribault:2014hia}).
Their expression, together with the OPE coefficients squared, is 
\begin{align}
    (\text{4H}) & \qquad\qquad
    \left\{
    \begin{aligned}
        \mathcal{F}_{1,1}  &= z^{\frac{3}{2} b^{2}+1}(1-z)^{-\frac{1}{2} b^{2}} \hypgeo{2}{1}(-b^{2},1+b^{2},2+2b^{2}; z) \,, \\
        \mathcal{F}_{3,1}  &= z^{-\frac{1}{2}b^{2}} (1-z)^{\frac{3}{2} b^{2}+1}\hypgeo{2}{1}(-b^{2},1+b^{2},-2b^{2}; z) \,, \\[0.5em]
        \lambda\ped{4H}  &= \frac{\Gamma(-b^2)^2 \, \Gamma(2 b^2+1)^2 \, \Gamma(2 b^2+2)^2}{\pi ^2 \left({\csc(\pi  b^2)}^2-{\csc(2 \pi  b^2)}^2\right) \Gamma(3 b^2+2)^2} \,,
    \end{aligned} 
    \right.
    \label{eq:4H-blocks} \\[1.5em]
    (\text{2H2L}) & \qquad\qquad
    \left\{
    \begin{aligned}
        \mathcal{F}_{1,1}  &= z^{\frac{3}{2} b^{2}+1} (1-z)^{\gamma} \hypgeo{2}{1}(2\gamma,1+b^2,2+2b^2; z) \,,  \\
        \mathcal{F}_{3,1}  & = z^{-\frac{1}{2}b^{2}} (1-z)^{\gamma} \hypgeo{2}{1}(-b^2,2\gamma-1-2 b^2,-2b^2;z) \,, \\[0.5em]
        \lambda\ped{2H2L} &= -\frac{\Gamma(-b^2)^2 \, \Gamma(2 + 2 b^2)^2 \, \Gamma(1 - 2 \gamma) \Gamma(-1 - 2 b^2 + 2 \gamma)}{\Gamma(-2 b^2)^2 \, \Gamma(1 + b^2)^2 \, \Gamma(2 + 2 b^2 - 2 \gamma) \Gamma(2 \gamma)}  \,, 
    \end{aligned}
    \right.
    \label{eq:2H2L-blocks} 
\end{align}
where $\gamma$ is defined in~\eqref{eq:gammaDef}.
We show in appendix~\ref{app:hypgeom} how to get an analytic closed form expression for the asymptotic expansion of the above blocks for $|b^2|\to \infty$ using
some appropriate rewriting of the $\hypgeo{2}{1}$ functions. 
More precisely, we will perform an expansion in $|\epsilon| \ll 1$, where
\begin{equation}\label{eq:epsilonDef}
    \frac{1}{\epsilon} = b^2+\frac{3}{2} \,.
\end{equation}
The offset by $3/2$ is useful to simplify various expressions that will follow. 
Note that the operator placement for~\eqref{eq:2H2L-correlator} is not the conventional one to obtain the thermal correlator in the large $c$ limit, but rather its crossing symmetric.

The large central charge limit can also be taken by sending $b^2\to 0$. 
In this case, the role of the heavy and light fields is exchanged, and ${\cal A}\ped{4H}$ turns into a four-point function of light operators. 
In ${\cal A}\ped{2H2L}$, unitarity is now broken only by the light probe and can be considered a milder breaking. 
We do not discuss this case, because the asymptotic expansion of hypergeometric functions with small parameters is more complicated and we currently lack a closed-form expression.

\section{4H correlator with \texorpdfstring{$\phi_{2,1}$}{phi21}}
\label{sec:4H}

The correlator of four $\phi_{2,1}$ fields is known, and given by~\eqref{eq:4H-block-decomposition} and~\eqref{eq:4H-blocks}. 
As anticipated in the introduction, we consider this correlator starting from the $1/c$ asymptotic expansion of the identity block and see what we can learn. 

The major and non-trivial step of the analysis is the determination of the whole large $c$ expansion, which can be found in appendix~\ref{app:hypgeom}.
From the asymptotic series of the identity block, we can ``discover'' the $\phi_{3,1}$ exchange when the series becomes non-Borel summable. 
This happens at specific values of $z$ which draw Stokes lines in the complex $z$-plane. 
These asymptotic series also have forbidden singularities which are not compatible with the full correlator.
We show that the non-perturbative effects at the Stokes lines resolve these singularities and finally we show how to fix the whole four-point function.
We discuss first the case $\epsilon>0$.

We start by expanding the conformal block for identity exchange shown in~\eqref{eq:4H-blocks} in powers of $\epsilon$.
We find that
\begin{equation}\label{eq:pc1}
    \widetilde{\mathcal{F}}_{1,1}(z;\epsilon)=  e^{-\frac{1}{\epsilon}S_1(z)} A_1(z) \widetilde F_1(z; \epsilon)\,,
\end{equation}
where
\begin{gather}
    S_1(z) = S(z) +\frac{1}{2} \log (1-z) -\frac{3}{2} \log z \,, \qquad \qquad  
    A_1(z) = A(z) z^{-\frac{5}{4}}(1-z)^{\frac{3}{4}}  \,, \label{eq:pc2b} \\  
    \widetilde F_1 =f_0+ \sum_{n=1}^\infty f_n(r) \epsilon^n\,, \qquad \qquad  
    f_0=1\,, \qquad 
    f_{n}(r) = \sum_{k=1}^n (-1)^{n+k} \stirlingII{n-1}{k-1} \frac{\big(\frac{1}{6}\big)_k \big(\frac{5}{6}\big)_k }{k!} r^k \,, \label{eq:pc2d}
\end{gather}
with $S$ and $A$ given by
\begin{equation}\label{eq:S0A0}
    S(z) = \frac{1}{2}\log\left(\frac{27}{16} \frac{z^{2}}{(1-z)^{2}}\frac{1-r(z)}{r(z)}\right)\,, \qquad \qquad 
    A(z) = \frac{e^{S(z)}}{(1-z+z^{2})^{\frac{1}{4}}} \,,
\end{equation}
and the map $r(z)$ given by
\begin{equation}\label{eq:rmapSec}
    r(z) = \frac{1}{2} + \frac{(z+1)(2z-1) (2-z)}{4(1-z+z^2)^{\frac{3}{2}}}.
\end{equation}
In~\eqref{eq:pc2d} $\stirlingII{n}{k}$ are the Stirling numbers of the second kind and $(a)_k$ denotes the Pochhammer symbol.
The derivation of~\eqref{eq:pc1} is quite non-trivial. 
We refer the reader to appendix~\ref{app:2F1-expansion-4H-reduction} for a detailed derivation of how the map $r(z)$ is found, and to appendix~\ref{app:2F1-expansion-c-large} for the form of the coefficients $f_n$.
It is quite remarkable that the coefficients $f_n$ have a simple form when expressed in terms of the variable $r(z)$ in~\eqref{eq:rmapSec}.

The map $r(z)$ satisfies the following properties:
\begin{equation}
    r(0) = 0, \qquad r(1) = 1, \qquad r(1-z) = 1-r(z).
\end{equation}
It is easy to show that only for $z=0,1$ we have $r=0,1$, respectively.
Furthermore, for any real value of $z$, $r(z)$ is contained in $(0,1)$.
Around each of these points $r(z)$ behaves as
\begin{equation}\label{eq:r-at-origin}
    r(z) = \frac{27}{16}z^2 + {\cal O}(z^3)\,, \qquad 1-r(z) = \frac{27}{16}(1-z)^2 + {\cal O}((1-z)^3)\,.
\end{equation}
Meanwhile, $r(z)$ is singular at
\begin{equation}
    z_{\pm} = e^{\pm \frac{i\pi}{3}}.
\end{equation}
While $S(z)$ is not singular at $z_{\pm}$, the leading factor $A(z)$ is indeed singular, as every perturbative correction $f_n$ is. 
In fact, the higher the order in perturbation theory, the higher the degree of singularity.
Both the full four-point Euclidean correlator and the Virasoro blocks can be singular at $z=0,1, \infty$, when operators collide, but nowhere else.
This makes $z_{\pm}$ examples of ``forbidden singularities'', in the terminology of~\cite{Fitzpatrick:2016ive}, with the novelty that they are identified for correlators with four heavy operators. Notice that they are elusive in an analysis around $z=0$, as the one presented in~\cite{Benjamin:2023uib}.

\subsection{Borel summation and discontinuity}
\label{subsec:BoSumDisc4H}

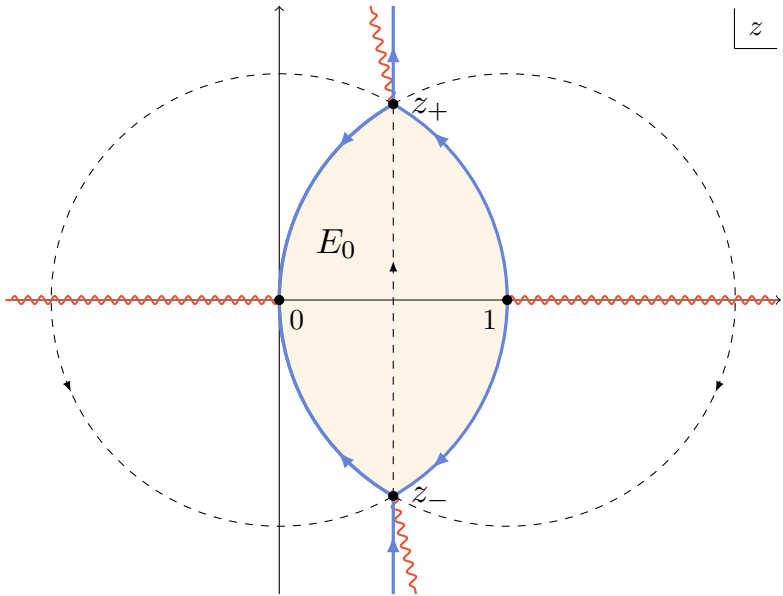
\begin{figure}[t!]
\centering
\begin{tikzpicture}[scale=3]
    \coordinate (zp) at (0.5,{sqrt(3)/2});
    \coordinate (zm) at (0.5,{-sqrt(3)/2});
    \coordinate (zero) at (0,0);
    \coordinate (one) at (1,0);
    \coordinate (A) at ({sqrt(3)/4},{1/4});
    \coordinate (B) at ({sqrt(3)},1);
    \coordinate (C) at ({sqrt(3)/2-0.02},{1/2-0.02});

    \draw[->] (-1.2,0) -- (2.2,0);
    \draw[->] (0,-1.3) -- (0,1.3);

    \node[draw=none, inner sep=5pt, append after command={
        \pgfextra{\draw (\tikzlastnode.south west) -- (\tikzlastnode.south east);}
        \pgfextra{\draw (\tikzlastnode.south west) -- (\tikzlastnode.north west);}
    }] at (2.09, 1.2) {$z$};

    \fill[region, opacity=0.1] (zp) arc[start angle=120, end angle=240, radius=1] -- (zm) arc[start angle=-60, end angle=60, radius=1] -- cycle;
    \node[draw=none, scale=1.2] at (1/4,1/4) {$E_{0}$};
    
    \draw[branch, thick, decorate, decoration={snake, segment length=5, amplitude=1.5}] (one) -- (2.18,0);
    \draw[branch, thick, decorate, decoration={snake, segment length=5, amplitude=1.5}] (zero) -- (-1.2,0);
    \draw[branch, thick, decorate, decoration={snake, segment length=5, amplitude=1.5}] (zp) -- ({1/2-0.1},1.3);
    \draw[branch, thick, decorate, decoration={snake, segment length=5, amplitude=1.5}] (zm) -- ({1/2+0.1},-1.3);

    \draw[stokes, very thick] (one) arc[start angle=0, end angle=-60, radius=1] [postaction={decorate, decoration={
            markings, mark=at position 0.8 with {\arrow{latex}}
        }}];
    \draw[stokes, very thick] (one) arc[start angle=0, end angle=60, radius=1] [postaction={decorate, decoration={
            markings, mark=at position 0.8 with {\arrow{latex}}
        }}];
    \draw[stokes, very thick] (zp) arc[start angle=120, end angle=180, radius=1] [postaction={decorate, decoration={
            markings, mark=at position 0.3 with {\arrow{latex}}
        }}];
    \draw[stokes, very thick] (zm) arc[start angle=240, end angle=180, radius=1] [postaction={decorate, decoration={
            markings, mark=at position 0.3 with {\arrow{latex}}
        }}];
    \draw[stokes, very thick] (zp) arc[start angle=120, end angle=240, radius=1];
    \draw[stokes, very thick] (0.5,{sqrt(3)/2}) -- (0.5,1.3) [postaction={decorate, decoration={
            markings, mark=at position 0.6 with {\arrow{latex}}
        }}];
    \draw[stokes, very thick] (0.5,-1.3) -- (0.5,{-sqrt(3)/2}) [postaction={decorate, decoration={
            markings, mark=at position 0.6 with {\arrow{latex}}
        }}];
    \filldraw[black] (zp) circle (0.02) node[right, xshift=2pt, yshift=-2pt, scale=1.2] {$z_+$};
    \filldraw[black] (zm) circle (0.02) node[right, xshift=2pt, yshift=-1pt, scale=1.2] {$z_-$};
    \filldraw[black] (1,0) circle (0.02) node[below left] {$1$};
    \filldraw[black] (0,0) circle (0.02) node[below right] {$0$};
    
    \draw[dashed] (zm) -- (zp) [postaction={decorate, decoration={
            markings, mark=at position 0.6 with {\arrow{latex}}
        }}];
    \draw[dashed] (zp) arc[start angle=120, end angle=-120, radius=1] [postaction={decorate, decoration={
            markings, mark=at position 0.6 with {\arrow{latex}}
        }}];
    \draw[dashed] (zp) arc[start angle=60, end angle=300, radius=1] [postaction={decorate, decoration={
            markings, mark=at position 0.6 with {\arrow{latex}}
        }}];
\end{tikzpicture}
\caption{Branch cuts (red wiggly lines), Stokes lines (blue) and anti-Stokes lines (dashed) in the $z$-plane for the 4H asymptotic series $(\widetilde{\mathcal{F}}_{1,1},\widetilde{\mathcal{G}})$.
Blue lines ingoing and outgoing from $z_\pm$ indicate the Stokes lines where $\widetilde{\mathcal{F}}_{1,1}$ and $\widetilde{\mathcal{G}}$ jump, respectively.
A refined version of the left panel of figure~\ref{fig:Fig12Draft}.
See section~\ref{subsec:StMono4H} for details.
\label{fig:4H-stokes-and-branch}}
\end{figure}

As derived in appendix~\ref{app:resurgence}, the Borel transform of the $\widetilde{F}_1$ series is given by
\begin{equation}\label{eq:Borel-transform-id-4h}
    \hatF_1(r;t) = \frac{5 r}{36} \hypgeo{2}{1}\left(\frac{7}{6},\frac{11}{6},2;r(1-e^{-t })\right) \, .
\end{equation}
For $\epsilon>0$, when $0 < z < 1$ and thus $0 < r < 1$, we have that $F_1^B = F_1$, as shown in appendix~\ref{app:resurgence}. 
It follows that the Borel summation of the asymptotic series of the identity block recovers the full block:
\begin{equation}\label{eq:V1-identity}
    \mathcal{F}_{1,1}^B = \mathcal{F}_{1,1} \, , \qquad 0 < z < 1 \, .
\end{equation}
The validity of~\eqref{eq:V1-identity} can be extended into the $z$ complex plane up to the points where the series $\widetilde{F}_1$ becomes non-Borel summable.
To find these points, observe that the function~\eqref{eq:Borel-transform-id-4h} has singularities, among others (see \eqref{eq:singularities-hatF1}), at
\begin{equation}\label{eq:tk-singularities}
    t_k(r) = \log\left(\frac{r}{r-1}\right)+2\pi i k \, , \qquad k\in \mathbb{Z} \, .
\end{equation}
When $r > 1$ we see that $t_0$ becomes positive and real turning the series non-Borel summable. 
The discontinuity of the lateral summation will then be a non-perturbative effect proportional to
\begin{equation}
     (s_{0+}-s_{0-})\widetilde{\mathcal{F}}_{1,1}(z;\epsilon) 
    \propto e^{-\frac{1}{\epsilon} t_0(r)} = \left(\frac{r}{r-1}\right)^{-\frac{1}{\epsilon}}.
    \label{eq:non-pert}
\end{equation}
By specializing~\eqref{eq:stokes-final} for $a=1/6$ and $b=5/6$, we get the full result
\begin{equation}\label{eq:stokes-hhhh}
    (s_{0+}-s_{0-})\widetilde{\mathcal{F}}_{1,1}(z;\epsilon) 
    = - i e^{\pm \frac{i\pi}{\epsilon}}\, s_0\left(\widetilde{\mathcal{F}}_{1,1}(1-z;\epsilon)\right) \, .
\end{equation}
Note how the non-perturbative factor \eqref{eq:non-pert} is naturally absorbed into $\widetilde{\mathcal{F}}_{1,1}$ evaluated at the crossing point $z\to 1-z$, up to a phase.
The sign ambiguity of the phase in~\eqref{eq:stokes-hhhh} depends on whether $z$ is in the upper or lower half-plane,
since
\begin{equation}\label{eq:r-phase}
    \left(\frac{r-1}{r}\right)^{\frac{1}{\epsilon} -1} 
    = - \left(\frac{1-r}{r}\right)^{\frac{1}{\epsilon} -1} 
    \begin{cases}
        e^{\frac{i\pi}{\epsilon}} & \im{z} > 0 \\
        e^{-\frac{i \pi}{\epsilon}} & \im{z} < 0
    \end{cases} \;,
\end{equation}
which is necessary to retrieve $\widetilde{\mathcal{F}}_{1,1}(1-z)$.
Observe that if $r$ is such that $\widetilde F_1(z)$ is not Borel summable, then $\widetilde F_1(1-z)$ certainly is, because $t_0\to -t_0$ when $z\to 1-z$. 
For convenience we define
\begin{equation}
    \widetilde{\mathcal{G}}(z;\epsilon) =    \widetilde{\mathcal{F}}_{1,1}(1-z;\epsilon)\,.
\end{equation}
We now determine in which regions of the $z$ plane Stokes jumps can occur.
At a generic point in the $z$ complex plane, both $\widetilde{\mathcal{F}}_{1,1}(z)$ and $\widetilde{\mathcal{G}}(z)$ are Borel summable to the functions $\mathcal{F}^B_{1,1}$ and $\mathcal{G}^B$. 
It is useful to combine them in a vector
\begin{equation}
    V(z;\epsilon) = 
    \begin{pmatrix}
        \mathcal{F}^B_{1,1}(z;\epsilon)\\
        \;\,\mathcal{G}^B(z;\epsilon)
    \end{pmatrix} \, .
\end{equation}
By symmetry, when $r<0$, $\widetilde{\mathcal{G}}$ is not Borel summable while $ \widetilde{\mathcal{F}}_{1,1}$ is. 
We find
\begin{equation}
    (s_{0+}-s_{0-})\widetilde{\mathcal{G}} 
    = - i e^{\mp \frac{i\pi}{\epsilon}} \, s_0\left(\widetilde{\mathcal{F}}_{1,1}\right) \, .
\end{equation}
These results determine the Stokes jumps at a fixed value of $r(z)$. 
As we move around the $z$ plane, the line $r\in(1,+\infty)$ is mapped to three lines in the $z$-plane.
We have 
\begin{equation}
    r\left(\frac{1}{2} + i a \right) = \frac{1}{2} + i \frac{a(9+4a^2)}{(3-4 a^2)^{\frac{3}{2}}} > 1 
    \qquad \text{if} \qquad 
    a < - \frac{\sqrt{3}}{2} \, ,
\end{equation}
leading to a line from $-i\infty$ to $z_-$. 
Moreover, $r>1$ also when $z= e^{i \theta}$ with ${\theta \in (-\pi/3, \pi/3)}$, which leads to two circle arcs connecting $1$ to $z_\pm$.
When $z$ crosses one of these lines, $t_0$ becomes positive and real.
Thus the Borel summations along the positive real axis before and after $z$ crossing the line differ, and they differ precisely by Stokes jump calculated when $z$ lies precisely at the line, i.e.~\eqref{eq:stokes-hhhh}. 
The Stokes jump in the Borel plane now appears as Stokes lines in the $z$ plane emanating from the forbidden singularities and ending at the actual singularities.
We can establish the region $D_0$ in which~\eqref{eq:V1-identity} holds, by continuing from $0 < z < 1$ until we reach either a Stokes line or a branch cut of the $r(z)$ map, and thus
\begin{equation}
\label{eq:D0-def}
    D_{0} = \left\{z \in \mathbb{C} \ \colon \re(z)<\frac{1}{2} \lor |z|<1\right\} \, .
\end{equation}
See appendix~\ref{app:2F1-expansion-4H-reduction} for further discussion.
We can also consider the Stokes lines where $\widetilde{\mathcal{G}}$ is not Borel summable which, from $z\rightarrow 1-z$, correspond to a vertical line from $z_+$ to $+i\infty$ and two circle arcs connecting $0$ to $z_\pm$. 
We then define the region
\begin{equation}
    E_{0} = \big\{z \in \mathbb{C} \ \colon |z|<1 \land |1-z|<1\big\} \,,
\end{equation}
between the two circle arcs, which is a result of continuing the interval $(0,1)$ up until we reach a Stokes line of $\widetilde{\mathcal{F}}_{1,1}$ or $\widetilde{\mathcal{G}}$, see figure~\ref{fig:4H-stokes-and-branch} (shaded region).

We report the whole web of Stokes lines in figure~\ref{fig:4H-stokes-and-branch}, where we also report the branch cuts of the $r(z)$ map itself, emanating from the singularities at $z_\pm$.
At the branch cut, the map $r(z)$ is discontinuous, obeying
\begin{equation}\label{eq:r-map-branch}
    r(z+\delta z) = 1-r(z-\delta z) = r\left(1-(z-\delta z)\right) \,,
\end{equation}
for $z$ in the branch cut and $\delta z$ an infinitesimal displacement orthogonal to it.
To avoid overlapping these cuts with the Stokes lines, we place them by convention slightly to the right/left, as in figure~\ref{fig:4H-stokes-and-branch}.\footnote{
    Alternatively, we could keep the branch cuts parallel to the imaginary axis and assign a small imaginary part to $\epsilon$ to slightly rotate the Stokes lines.
}  

An important feature which can be read from the Borel transform is the large-order behaviour of the asymptotic series. 
As recalled in~\eqref{eq:lboDer1} and~\eqref{eq:lboDer2}, the singularities of the Borel transform closest to the origin dictate the dominant asymptotic behaviour of the series. 
For real series, typically either a single singularity $t=t_0$ dominates, leading to a series such as $t_0^{-n} n!$, or two complex conjugate singularities, which lead to an oscillatory modulation of the factorial divergence. 
For most complex values of $z$, one of the $t_k$ singularities dominates. 
However, for real $z$ we have two equidistant $t_k$ singularities at $\log\!\big(\tfrac{r}{1-r}\big)\pm \pi i$. 
\eqref{eq:fn-LOB-cases} reduces in this case to
\begin{equation}\label{eq:LBO4H}
    f_n \approx -\frac{(n-1)!}{\pi}
    \left(\frac{r}{1-r}\right)\left|\log\frac{r}{r-1}\right|^{-n}\cos\left(n \arg\!\big(\!\log\tfrac{r}{r-1}\big)\right)\,.
\end{equation}
Notice that if one takes $z\rightarrow 0$, using~\eqref{eq:r-at-origin}, the oscillatory behaviour is damped out and we retrieve the scaling $(\log z^2)^{-n}n!$ which is implicit in the saddle points identified in~\cite{Benjamin:2023uib}. We refer to appendix~\ref{app:resurgence} for further details of the large order behaviour analysis. As we discuss below, we can already see that this series is always Borel summable for real $\epsilon$. Thus, non-Borel summability and resurgence must appear at complex $z$.

It is useful to look at the $z\rightarrow 0$ limit to identify what $\widetilde {\mathcal{G}}$ is. 
We get
\begin{equation}\label{eq:pc18}
    \lim_{z\to 0} \widetilde{\mathcal{G}}  \propto z^{-\frac{1}{2\epsilon}+\frac 34} (1+\ldots ) \,.
\end{equation}
Denoting by $h_X$ the conformal weight of the exchanged operator, we should have as $z\to 0$ $ \widetilde{\mathcal{G}}  \propto  z^{- 2h_{2,1}+h_X } (1+ \ldots)$.
By matching we get
\begin{equation}
    h_X = -\frac 2 \epsilon +2 = -2 b^2 -1 \qquad \Rightarrow \qquad h_{X} = h_{3,1}\,,
\end{equation}
and hence we see that $\widetilde{\mathcal{G}}$ includes the asymptotic expansion of the conformal block of the $\phi_{3,1}$ field.
We can obtain an exact expression for the resummation of $\mathcal{G}$ by taking~\eqref{eq:V1-identity} at $z\rightarrow 1-z$.
This holds for $0 < z < 1$, but now we also know we can extend it to the entire region contained between the circle arcs, see figure~\ref{fig:4H-stokes-and-branch}, where neither $\widetilde{\mathcal{F}}_{1,1}$ nor $\widetilde{\mathcal{G}}$ have a Stokes jump.
Then, using simple hypergeometric identities we get $\mathcal{G}^B$. Then we have
\begin{equation}\label{eq:V3Der}
\begin{aligned}
\mathcal{F}^B_{1,1}(z) &= \mathcal{F}_{1,1}(z),\\
     \mathcal{G}^B(z) &= \frac{1}{2\sin\!\big(\tfrac{\pi}{\epsilon}\big)} \mathcal{F}_{1,1}(z) + k(\epsilon)  \mathcal{F}_{3,1}(z) \,,\\
\end{aligned}\qquad
\epsilon > 0 \, , \quad z\in E_0,
\end{equation}
where 
\begin{equation}\label{eq:V3Def}
    k(\epsilon) = 
    \frac{\Gamma\big(\tfrac{2}{\epsilon} -2 \big) \Gamma\big(\tfrac{2}{\epsilon} -1\big)}{\Gamma\big(\tfrac {1}{\epsilon} -\tfrac{1}{2}\big) \Gamma\big(\tfrac{3}{\epsilon} -\tfrac{5}{2}\big)}\,.
\end{equation}
We recognize $\mathcal{F}_{3,1}$ as the block of the $\phi_{3,1}$ field, which we have then ``discovered'' from the Stokes discontinuity of the identity block.

Conversely, we can invert the relation above and state that, while $\mathcal{F}_{1,1}(z)$ is given by $\mathcal{F}_{1,1}^B(z)$  for $z \in D_0$, $\mathcal{F}_{3,1}(z)$ is given by the resummation of a trans-series containing both $\widetilde{\mathcal{G}}$ and $\widetilde{\mathcal{F}}_{1,1}$. 
Notice that since  $\mathcal{F}_{1,1}(z)$ is regular across Stokes jumps but the resummation of $\widetilde{\mathcal{F}}_{1,1}$ is not, once we cross a Stokes lines $\mathcal{F}_{1,1}(z)$ becomes itself also a trans-series. That is
\begin{equation}
\label{eq:f11-trans-series}
    \mathcal{F}_{1,1}(z-\delta z) 
    = {\mathcal{F}}^B_{1,1}(z-\delta z) 
    = {\mathcal{F}}^B_{1,1}(z+\delta z) \mp i e^{\pm \frac{i\pi}{\epsilon}} \mathcal{G}^B(z+\delta z) 
    = \mathcal{F}_{1,1}(z+\delta z) \, ,
\end{equation}
where $\delta z$ is some small displacement from the Stokes line such that $z-\delta z$ is in $E_0$ and $z+\delta z$ is to the right of the Stokes lines, while the signs are for upper/lower half plane, respectively. Thus, generically in the complex plane, both blocks are given by trans-series which include both $\widetilde{\mathcal{F}}_{1,1}$ and $\widetilde{\mathcal{G}}$.

\subsection{Stokes lines and monodromy matrices}
\label{subsec:StMono4H}

\begin{figure}[t!]
    \centering
    \renewcommand{\arraystretch}{2}
    \renewcommand{\tabcolsep}{1.5em}
    \begin{tabular}{lll}
        \begin{tikzpicture}[scale=1.3]
            \filldraw[fill=black, draw=black]  (0,0) circle (0.05 cm);
            \draw (2,0) -- (0,0) [postaction={decorate, decoration={
                    markings, mark=at position 0.5 with {\arrow{latex}}
                }}];
            \node[right] at (2,0) {\( = S_{-}\phantom{^{-1}} \)};
            \draw[-stealth,red] ([shift=(-20:1.5)]0,0) arc (-20:20:1.5);
        \end{tikzpicture}
        &   
        \begin{tikzpicture}[scale=1.3]
            \filldraw[fill=black, draw=black]  (0,0) circle (0.05 cm);
            \draw (0,0) -- (2,0) [postaction={decorate, decoration={
                    markings, mark=at position 0.5 with {\arrow{latex}}
                }}];
            \node[right] at (2,0) {\( = S_{+} \)};
            \draw[-stealth,red] ([shift=(-20:1.5)]0,0) arc (-20:20:1.5);
        \end{tikzpicture}
        &
        \begin{tikzpicture}[scale=1.3]
            \filldraw[fill=black, draw=black]  (0,0) circle (0.05 cm);
            \draw[decorate, decoration={snake,segment length=5pt}] (0,0) -- (2,0);
            \node[right] at (2,0) {\( = B \)};
            \draw[-stealth,red] ([shift=(-20:1.5)]0,0) arc (-20:20:1.5);
        \end{tikzpicture} \\
        \begin{tikzpicture}[scale=1.3]
            \filldraw[fill=black, draw=black]  (0,0) circle (0.05 cm);
            \draw (2,0) -- (0,0) [postaction={decorate, decoration={
                    markings, mark=at position 0.5 with {\arrow{latex}}
                }}];
            \node[right] at (2,0) {\( = S_{-}^{-1} \)};
            \draw[stealth-,red] ([shift=(-20:1.5)]0,0) arc (-20:20:1.5);
        \end{tikzpicture} 
        &
        \begin{tikzpicture}[scale=1.3]
            \filldraw[fill=black, draw=black]  (0,0) circle (0.05 cm);
            \draw (0,0) -- (2,0) [postaction={decorate, decoration={
                    markings, mark=at position 0.5 with {\arrow{latex}}
                }}];
            \node[right] at (2,0) {\( = S_{+}^{-1} \)};
            \draw[stealth-,red] ([shift=(-20:1.5)]0,0) arc (-20:20:1.5);
        \end{tikzpicture}
        &
        \begin{tikzpicture}[scale=1.3]
            \filldraw[fill=black, draw=black]  (0,0) circle (0.05 cm);
            \draw[decorate, decoration={snake,segment length=5pt}] (0,0) -- (2,0);
            \node[right] at (2,0) {\( = B^{-1} \)};
            \draw[stealth-,red] ([shift=(-20:1.5)]0,0) arc (-20:20:1.5);
        \end{tikzpicture}
    \end{tabular}
    \caption{Local connection matrices as we pass an oriented Stokes line (straight lines with arrows) or a branch cut (wavy lines) emanating from $z_\pm$  (adapted from figure 2 in~\cite{Bucciotti:2023trp}).}
     \label{fig:Airy2}
\end{figure}
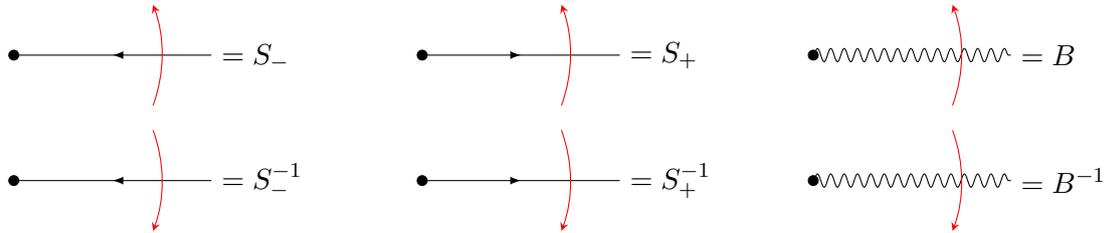

It is useful to organise the Stokes jumps in the graphical notation of figure~\ref{fig:Airy2} (adapted from~\cite{Bucciotti:2023trp}).
The Stokes lines become noted as ``incoming'' or ``outgoing'' lines, with respect to the turning point at $z_\pm$.
As discussed in~\eqref{eq:r-map-branch}, there are also branch cuts emanating from the forbidden singularities.
At these lines,
\begin{equation}
    V(z-\delta z) = S_{-} V(z+\delta z), \quad 
    V(z-\delta z) = S_{+} V(z+\delta z), \quad
    V(z-\delta z) = B V(z+\delta z)\,,
\end{equation}
for each line.
Here we define $\delta z$  such that in the orientation of figure~\ref{fig:Airy2}, $z\pm \delta z$ is above/below the line. 
In this convention, the matrices are
\begin{equation}\label{eq:monodromy}
    S_{+} = 
    \begin{pmatrix}
        1 & 0 \\
        -i e^{\mp\frac{i \pi }{\epsilon }} & 1
    \end{pmatrix} 
    \,, \quad 
    S_{-} = 
    \begin{pmatrix}
        1 & -i e^{\pm\frac{i \pi }{\epsilon }} \\
        0 & 1
    \end{pmatrix} 
    \,, \quad 
    B =
    \begin{pmatrix}
        0 & i e^{\pm\frac{i \pi }{\epsilon }} \\
     i e^{\mp\frac{i \pi }{\epsilon }} & 0
    \end{pmatrix}
    \,,\quad  \epsilon > 0\,,
\end{equation}
where in the phases we take the upper/lower sign according to whether we are in upper/lower half plane. 
There are two other branch cuts, at $(-\infty,0)$ and $(1,+\infty)$, due to the leading terms $S(z)$ and $A(z)$. 
These are simply
\begin{equation}\label{eq:AS-branch-cuts}
    \begin{aligned}
        V(z+i0) &=
        B_0
        V(z-i0) \, ,
        \quad
        B_0 = \begin{pmatrix}
            -i e^{\frac{3i \pi }{\epsilon }} & 0 \\
            0 & -i e^{-\frac{ i \pi }{\epsilon }}
        \end{pmatrix}
        \, ,
        \quad & z\in(-\infty,0) \, , \\[0.5em]
        V(z-i0) &=
        B_1
        V(z+i0) \, ,
        \quad
        B_1 =
        \begin{pmatrix}
            -i e^{-\frac{ i \pi }{\epsilon }} & 0 \\
            0 & -i e^{\frac{3i \pi }{\epsilon }}
        \end{pmatrix}
        \, ,
        \quad & z\in(1,+\infty) \, .
    \end{aligned}
\end{equation}
The final graph is figure~\ref{fig:4H-stokes-and-branch}, including the branch cuts~\eqref{eq:AS-branch-cuts}. 
We also include the anti-Stokes lines, which are 
the lines at which ${\mathcal{F}}_{1,1}^B$ and $\mathcal{G}^B$ are of comparable size, defined in appendix~\ref{app:hypgeom}. Their arrows are placed such that when you cross an outgoing line clockwise you go from a region where $\widetilde{\mathcal{F}}_{1,1}$ is exponentially enhanced to suppressed, and the opposite for clockwise crossing an ingoing line.\footnote{
    Briefly, the ``non-perturbative'' effect $\widetilde{\mathcal{G}}$ is suppressed (or enhanced) relative to $\widetilde{\mathcal{F}}_{1,1}$ as $e^{-\frac{1}{\epsilon} \Delta S(z)}$ when $\Delta S(z)$ has a positive (or negative) real part, with $\Delta S$ being the difference between the semi-classical parts of each. 
    When it is purely imaginary then $|\widetilde{\mathcal{F}}_{1,1}|\sim |\widetilde{\mathcal{G}}|$. The lines of $\re\left[\Delta S(z)\right]=0$ are the anti-Stokes lines. 
}

We now determine the monodromies around the singular points of $V$, $z=0,1,z_\pm$. We do not need to determine the monodromy around $z=\infty$ since this is 
fixed by the other ones. We start with $z=z_\pm$. Under a $2\pi$ rotation around $z_\pm$, the total monodromy is
\begin{equation}\label{eq:trivial-monodromy-1}
\begin{aligned}
    M_{z=z_+}&=S_{-} S_{+} B S_{+} = I \,,\\
    M_{z=z_-}&=S_{-} B S_{-} S_{+} = I \,.
\end{aligned}
\end{equation}
Trivial monodromy is a necessary condition for $z_\pm$ to be a regular point. We then see how concretely and globally Stokes discontinuities 
resolve forbidden singularities.
Note that, modulo the phase factors $-e^{\pm \frac{i\pi }{\epsilon}}$ the connection matrices~\eqref{eq:monodromy} are the same as those appearing in the Airy function, which are also the universal Stokes lines for simple turning points in quantum mechanics in exact WKB~\cite{voros-quartic}, see e.g.~\cite{Serone:2024uwz} for a pedagogical introduction.\footnote{
    The connection matrices appearing in~\cite{Bucciotti:2023trp,Serone:2024uwz} act on the coefficients of the functions, rather than the function themselves, as here. 
    For this reason they are given by the transpose of those in~\eqref{eq:monodromy} (without the phase factors), but the two actions are completely equivalent.
}
Even though forbidden singularities persist to all orders in perturbation theory, once we include all non-perturbative effects they are resolved. 
This is a crucial result from this analysis.
Furthermore, these non-perturbative effects can all be retrieved from the large order behaviour of the perturbative part.

We now compute the monodromy matrices around 1 and 0. The monodromy of the resumed functions can of course be derived using the known discontinuity formula of the hypergeometric function, but 
crucially we ``reconstruct'' them here from products of connection matrices. Moreover, this can be seen as a non-trivial check of the formalism.
Under a $2\pi i$ rotation around $z=1$, starting slightly above $(0,1)$,  leads to
\begin{equation}\label{eq:monodromy-1}
    V \rightarrow
    M_{z=1}
    V \,,\qquad M_{z=1}=
(S_{-})^{-1} B_1 (S_{-})^{-1}
    =
    \begin{pmatrix}
        -i e^{-\frac{i \pi }{\epsilon }} & 1+e^{\frac{2 i \pi }{\epsilon }} \\
        0 & -i e^{\frac{3 i \pi }{\epsilon }}
    \end{pmatrix}\,.
\end{equation}
The leftmost $S_{-}$ is in the lower half plane, and the rightmost in the upper one. 
Note that generically the monodromy matrix $M$ depends on the starting region, since matrices $M_i$ and $M_j$ starting in two different adjacent regions $i$ and $j$ are related by $M_j=S_{ij} M_i S_{ij}^{-1}$, with $S_{ij}$ the product of connection matrices $S_\pm$ and $B_i$ associated with lines separating regions $i$ and $j$.
Further, by replacing the asymptotic series with its summation in~\eqref{eq:monodromy-1}, one can check that
\begin{equation}
    \left(1+e^{\frac{2 i \pi }{\epsilon }}\right) \mathcal{G}^B
    = -2 i \pi e^{i \pi  b^2}  z^{\frac{3}{2}b^2+1} (1-z)^{\frac{3}{2} b^2+1} \sin(\pi  b^2)
    \hypgeo{2}{1}(b^2+1,3 b^2+2;2 b^2+2;1-z) \, ,
\end{equation}
which reproduces the exact discontinuity of the original hypergeometric representation of $\mathcal{F}_{1,1}$. 
The analysis at $z=0$ is symmetric under $z\rightarrow 1-z$,
\begin{equation}\label{eq:mono4Hz0}
    M_{z=0}=
    (S_{+})^{-1}B_0(S_{+})^{-1}
    =
    \begin{pmatrix}
        -i e^{\frac{3i \pi }{\epsilon }} & 0  \\
        1+e^{\frac{2 i \pi }{\epsilon }} & -i e^{\frac{-i \pi }{\epsilon }}
    \end{pmatrix}\,,
\end{equation}
where the first and second $S_{+}$ matrices in~\eqref{eq:mono4Hz0} are in the upper and lower half planes respectively.

\subsection{Fixing the four-point function}
\label{subsec:fixfourPT}

The above analysis applies verbatim to the anti-holomorphic sector starting from the asymptotic expansion of 
\begin{equation}
    \widetilde{\overline{\mathcal{F}}}_{1,1}(\bar z) = \Big(\widetilde{\mathcal{F}}_{1,1}(z)\Big)^*\,.
\end{equation} 
Putting together holomorphic and anti-holomorphic sectors, we finally show how to determine the entire four-point function by demanding its single-valuedness around $z=0,1$.
 
We write the four-point function as an \textit{a priori} unfixed quadratic form, up to a $z$-independent normalization factor,
\begin{equation}\label{eq:four-point-ansatz}
    {\cal A} = g(\epsilon) \overline V_i C_{ij} V_j \, , \qquad 
    C =  
    \begin{pmatrix}
        1 & c_1  \\
        c^{*}_{1} &  c_2
    \end{pmatrix} 
    \, , \qquad c_1 \in \mathbb{C}, \ c_2 \in \mathbb{R} \, .
\end{equation}
For the reader more familiar with resurgence, one can take the asymptotic form of the above expression and consider a trans-series
\begin{equation}
    \widetilde{\cal A} = g(\epsilon)\left( 
    \mathcal{C}_0\widetilde{\overline{\mathcal{F}}}_{1,1}\widetilde{\mathcal{F}}_{1,1}
    +\mathcal{C}_1 \widetilde{\overline{\mathcal{F}}}_{1,1}\widetilde{\mathcal{G}}
    +\overline{\mathcal{C}}_1 \widetilde{\overline{\mathcal{G}}}\mathcal{F}_{1,1}
    +\mathcal{C}_2 \widetilde{\overline{\mathcal{G}}}\widetilde{\mathcal{G}}
    \right) \, ,
\end{equation}
where the $\mathcal{C}_i$ are piece-wise constant in the $z$ complex plane (and also piece-wise constant in the $\epsilon$-plane). 
In $E_0$, we should have
\begin{equation}
    \mathcal{C}_0=1,\quad \mathcal{C}_1= c_1,\quad \mathcal{C}_2 = c_2.
\end{equation}
At Stokes lines, these parameters change to ``counter-act'' the discontinuities of Borel summation in $\widetilde{\mathcal{F}}_{1,1}$ and $\widetilde{\mathcal{G}}$, keeping $\mathcal{A}^B$ continuous and ensuring that
\begin{equation}
\mathcal{A}^B=\mathcal{A},
\end{equation}
in the entire complex $z$-plane.

Then, knowing the Stokes constants in the preceding section, we fix the matrix $C$, or trans-series parameters, with physical principles. Namely, we ensure that the sequence of Stokes jumps around the singularities should bring the trans-series back to itself.
This is a resurgence rephrasing of the fact that the monodromy around $z=0,1$ does not spoil the single-valuedness of the four-point function. After we determine the trans-series parameters in some region, they are automatically fixed in all regions, since they must change according to Stokes jumps, keeping $\mathcal{A}$ continuous. 

We determine the matrix $C$, by convention, for $z\in E_0$ which includes the interval $(0,1)$.
We impose
\begin{equation}
    \overline{V} M_{z=0}^\dagger C M_{z=0} V = \overline{V} M_{z=1}^\dagger C M_{z=1} V = \overline{V} C V,
\end{equation}
where $M_{z=0}$ and $M_{z=1}$ are the monodromy matrices~\eqref{eq:mono4Hz0} and~\eqref{eq:monodromy-1}.
We get as solution
\begin{equation}
    c_1 = -\frac{1}{2\sin\!\big(\tfrac{\pi}{\epsilon}\big)} \,, \quad c_2=1\,.
\end{equation}
We fix the factor $g(\epsilon)$ by demanding the correct normalization of the identity block.
This requires to consider values of $\epsilon>1$, in which case the identity becomes the leading contribution in the correlator. We then have
\begin{equation}\label{eq:A4HCorrez0}
    {\cal A} \approx z^{\frac{3}{\epsilon }-\frac{5}{2}} \, , \quad \epsilon>1 \, .
\end{equation}
Since $z=0$ is a regular point of $\mathcal{F}_{1,1}^B$ and the perturbative series trivializes to $1+\mathcal{O}(z)$ at small $z$, we can quickly derive $\mathcal{F}_{1,1}^B\approx z^{\frac{3}{2 \epsilon }-\frac{5}{4}}$ even for $\epsilon$ close to $1$.
For $\mathcal{G}$, however, the origin in $z$ is a singular point so we must first resum the asymptotic series and then use
\begin{equation}\label{eq:4PrecAux}
    \hypgeo{2}{1}\left(\frac{1}{6},\frac{5}{6};\frac{1}{\epsilon };1-r\right)
    = r^{\frac{1}{\epsilon }-1} \left(\frac{1}{2} \csc\left(\frac{\pi }{\epsilon }\right)+\mathcal{O}(r)\right)
    +\left(-\frac{\pi  \csc\!\big(\tfrac{\pi }{\epsilon }\big) \Gamma \big(\tfrac{1}{\epsilon }\big)}{\Gamma \big(2-\tfrac{1}{\epsilon }\big) \Gamma \big(\tfrac{1}{\epsilon }-\tfrac{5}{6}\big) \Gamma \big(\tfrac{1}{\epsilon }-\tfrac{1}{6}\big)}+\mathcal{O}(r)\right) \, .
\end{equation}
While the identity block dominates for $\epsilon>1$, it is not sufficient to look at $\mathcal{F}_{1,1}^B$ since, as shown in~\eqref{eq:V3Der}, $\mathcal{G}$ contains a contribution to the $|\mathcal{F}_{1,1}|^2$ part of the four-point function.
Using the $z\to 0$ limits as given by~\eqref{eq:4PrecAux} and matching with the behaviour~\eqref{eq:A4HCorrez0} allows us to fix $g(\epsilon)$ and the whole correlator~\eqref{eq:four-point-ansatz}. 
We finally get
\begin{equation}\label{eq:gepsCmat}
    \renewcommand{\arraystretch}{1.2}
    g(\epsilon)=\frac{4 \sin^2\!\big(\tfrac{\pi }{\epsilon }\big)}{1-2 \cos \!\big(\tfrac{2 \pi }{\epsilon }\big)} \, , \qquad 
    C =
    \begin{pmatrix}
        1 & -\frac{1}{2} \csc \left(\frac{\pi }{\epsilon }\right) \\
        -\frac{1}{2} \csc \left(\frac{\pi }{\epsilon }\right) & 1 \\
    \end{pmatrix}  
    \,, \qquad \epsilon>0.
\end{equation}
It is now easy to check that the correlator given by~\eqref{eq:four-point-ansatz} and~\eqref{eq:gepsCmat} reproduce the known correlator~\eqref{eq:4H-block-decomposition} and~\eqref{eq:4H-blocks} in the ordinary Virasoro block basis.
Note how the less ``physical'', but more ``mathematical'', $V$-basis gives rise to simpler formulas, where crossing symmetry is manifest.

\subsection{Negative central charge}\label{subsec:4H-eps-negative}

The analysis in the previous sections assumed $\epsilon > 0$. 
We started with this case because there are regions in the $z$ plane where the identity block is fully reconstructable from its asymptotic series, which makes the analysis more conventional and easier to follow.
Moreover, as we will see in section~\ref{subsec:MM}, the reconstruction extends up to $c<1$ and allows us to reach the unitary minimal models.
However, since for $\epsilon>0$ the conformal weight of the other exchanged highest weight state $h_{3,1} < 0$, 
the identity block is \emph{not} the leading contribution in any OPE channel. 
Physically speaking, one might then legitimately question the motivation to start the analysis from the identity block. 
For this reason, it is important to discuss the $\epsilon <0$ case, where $h_{3,1}>0$ and the identity operator provides the dominant contribution.
The above analysis valid for $\epsilon > 0$ readily applies for $\epsilon <0$ (and in fact can be extended for any complex value of $\epsilon$).

For $\epsilon<0$, the resummation of the asymptotic series in terms of hypergeometric functions is given by~\eqref{eq:resum-neg-eps}.
The difference with $\epsilon>0$ is not unexpected, since we cross multiple Stokes phenomena, due to the multitude of singularities~\eqref{eq:singularities-hatF1} that we encounter, as we move from the positive to the negative real $t$ axis.
As a consequence, the resummation of $\widetilde{\mathcal{F}}_{1,1}$ no longer corresponds to $\mathcal{F}_{1,1}$ in $E_0$, while the resummation of $\mathcal{G}$ does match $\mathcal{F}_{3,1}$, up to an overall factor. More precisely, we have
\begin{equation}\label{eq:4Hepsm0FG}
    \begin{aligned}
        \mathcal{F}_{1,1}^B(z;\epsilon)
        &=\mathcal{F}_{1,1}(z;\epsilon)+\frac{ 2 \sin\!\big(\tfrac{\pi }{\epsilon }\big) }{2 \cos \!\big(\tfrac{2 \pi }{\epsilon }\big)-1}k(\epsilon )\mathcal{F}_{3,1}(z;\epsilon)
        ,\\
        \mathcal{G}^B(z;\epsilon)
        &=\frac{4 \, {\sin\!\big(\tfrac{\pi }{\epsilon }\big)}^{2}}{1-2 \cos\!\big(\tfrac{2 \pi }{\epsilon }\big)}k(\epsilon )\mathcal{F}_{3,1}(z;\epsilon),
    \end{aligned}
    \qquad \epsilon<0,\quad z\in E_0,
\end{equation}
where $k(\epsilon)$ is defined in~\eqref{eq:V3Def}.
Although perhaps unfamiliar, it should not be surprising that the Borel resummation $f^B$ of the asymptotic series $\widetilde f$ of a function $f$ 
does not reproduce the function $f$. Equivalence classes of functions which differ by non-perturbative terms 
give rise to the same asymptotic expansion and the Borel resummed function $f^B$ picks up the unique representative in the equivalence class with given analyticities properties,
see the discussion at the beginning of appendix~\ref{app:resurgence}. In our case, this representative is given by the specific linear combination 
reported in the right-hand side of the first relation in~\eqref{eq:4Hepsm0FG}.\footnote{The fact that $\mathcal{F}_{1,1}^B\neq \mathcal{F}_{1,1}$ for $\epsilon <0$ is related to the $c$ parameter of the $\hypgeo{2}{1}(a,b,c;z)$ appearing in the identity block~\eqref{eq:4H-block-decomposition}, which turns negative. A similar comment applies to the resummation of the block $\mathcal{F}_{3,1}$ for $\epsilon >0$.}
Inverting this relation, we see that the identity block has a trans-series representation even for $0 < z < 1$, while $\mathcal{F}_{3,1}$ can be obtained from the resummation of a series. 

Like for $\epsilon >0$, the relations~\eqref{eq:4Hepsm0FG} are subject to Stokes jumps as we move out of $E_0$.
The analysis is identical to the one we performed for $\epsilon>0$, the only change being that the relevant axis in the Borel $t$ plane is now the negative, and not the positive one.
We need to avoid that the singularities $t_k$ of~\eqref{eq:tk-singularities} become negative and real, as discussed in appendix~\ref{app:resurgence}. This is 
at the origin of the different summability conditions with respect to the $\epsilon >0$ case, i.e.\ $\widetilde{\mathcal{F}}_{1,1}$ is non-Borel summable when $r<0$ and $\widetilde{\mathcal{G}}$ is non-Borel summable when $r>1$. The jumps themselves remain the same, as shown in the same appendix. Consequently, we need to switch the rules for the Stokes lines, while the monodromies due to branch cuts $B$, $B_0$ and $B_1$ remain the same as for $\epsilon >0$. We then have 
\begin{equation}
S_{+}  =  
\begin{pmatrix}
 1 & -i e^{\pm\frac{i \pi }{\epsilon }} \\
 0 & 1 \\
\end{pmatrix}, \qquad 
S_{-}=  
\begin{pmatrix}
 1 & 0 \\
 -i e^{\mp\frac{i \pi }{\epsilon }} & 1 \\
\end{pmatrix} \,, \quad 
B =
\begin{pmatrix}
 0 & i e^{\pm\frac{i \pi }{\epsilon }} \\
 i e^{\mp\frac{i \pi }{\epsilon }} & 0 \\
\end{pmatrix} \,,\quad  \epsilon < 0\,.
\label{eq:monodromy-neg}
\end{equation}
Since the Stokes graph is unaltered, the products of connection matrices that construct the monodromy matrices are unchanged. Then \eqref{eq:trivial-monodromy-1} applies, even with the new Stokes matrices, and the monodromy matrices $M_{z=z_\pm}$ are trivial.
Hence, forbidden singularities are resolved in the same way as in the $\epsilon>0$ case.
The monodromy matrices around $z=0$ and $z=1$ read
\begin{equation}
\begin{aligned}
M_{z=0} &= (S_{+})^{-1} B_0 (S_{+})^{-1} = \begin{pmatrix}
        -i e^{\frac{3i \pi }{\epsilon }} & 1+e^{\frac{2 i \pi }{\epsilon }}  \\
        0  & -i e^{-\frac{ i \pi }{\epsilon }}
    \end{pmatrix} \,,\\
M_{z=1} &= (S_{-})^{-1} B_1 (S_{-})^{-1} = \begin{pmatrix}
        -i e^{-\frac{i \pi }{\epsilon }} & 0  \\
        1+e^{\frac{2 i \pi }{\epsilon }} & -i e^{\frac{3 i \pi }{\epsilon }}
    \end{pmatrix}\,.
\end{aligned}
\end{equation}
Finally, we can determine the full correlator. As in the $\epsilon>0$ case, we write the amplitude as 
\begin{equation}
\label{eq:four-point-ansatz-neg}
{\cal A} = g(\epsilon) \overline V_i C_{ij} V_j \,, 
\end{equation}
and determine the matrix $C$ demanding trivial monodromy around $z=0$ and $z=1$.  
We get
\begin{equation}\label{eq:Cmatepsneg}
    \renewcommand{\arraystretch}{1.2}
    C =
    \begin{pmatrix}
        1 & \frac{1}{2} \csc \left(\frac{\pi }{\epsilon }\right) \\
        \frac{1}{2} \csc \left(\frac{\pi }{\epsilon }\right) & 1 \\
    \end{pmatrix} 
    \, ,\qquad \epsilon<0 \, .
\end{equation}
The determination of $g(\epsilon)$ is trivial for $\epsilon <0$. The identity block dominates at small $\epsilon$ and 
$\mathcal{G}$ corresponds only to $\mathcal{F}_{3,1}$, up to normalization. Since only $\mathcal{F}_{1,1}^B$ contributes to the identity block, we have
\begin{equation}\label{eq:gfunepsneg}
g(\epsilon)=1 \,, \qquad \epsilon<0\,.
\end{equation}
One can check that~\eqref{eq:four-point-ansatz-neg},~\eqref{eq:Cmatepsneg} and~\eqref{eq:gfunepsneg} reproduce 
the correct result~\eqref{eq:4H-block-decomposition}, which holds for both signs of $\epsilon$.

\begin{figure}[t!]
    \centering
    \hspace{0.5em}
    \begin{subfigure}[c]{0.48\textwidth}
        \includegraphics[width=\textwidth]{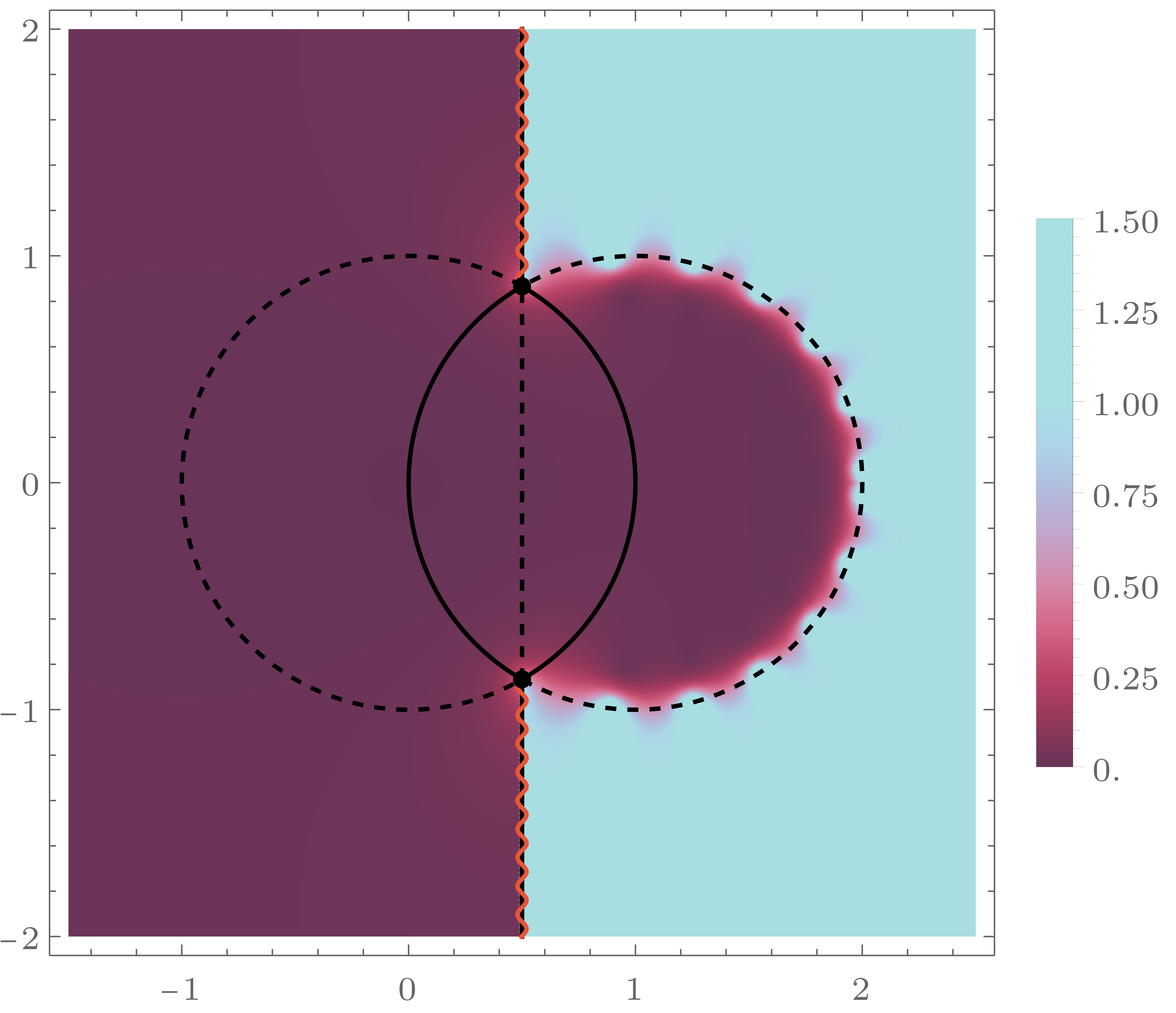}
        \caption{$ \epsilon > 0$}
    \end{subfigure}
    \hfill
    \begin{subfigure}[c]{0.48\textwidth}
        \includegraphics[width=\textwidth]{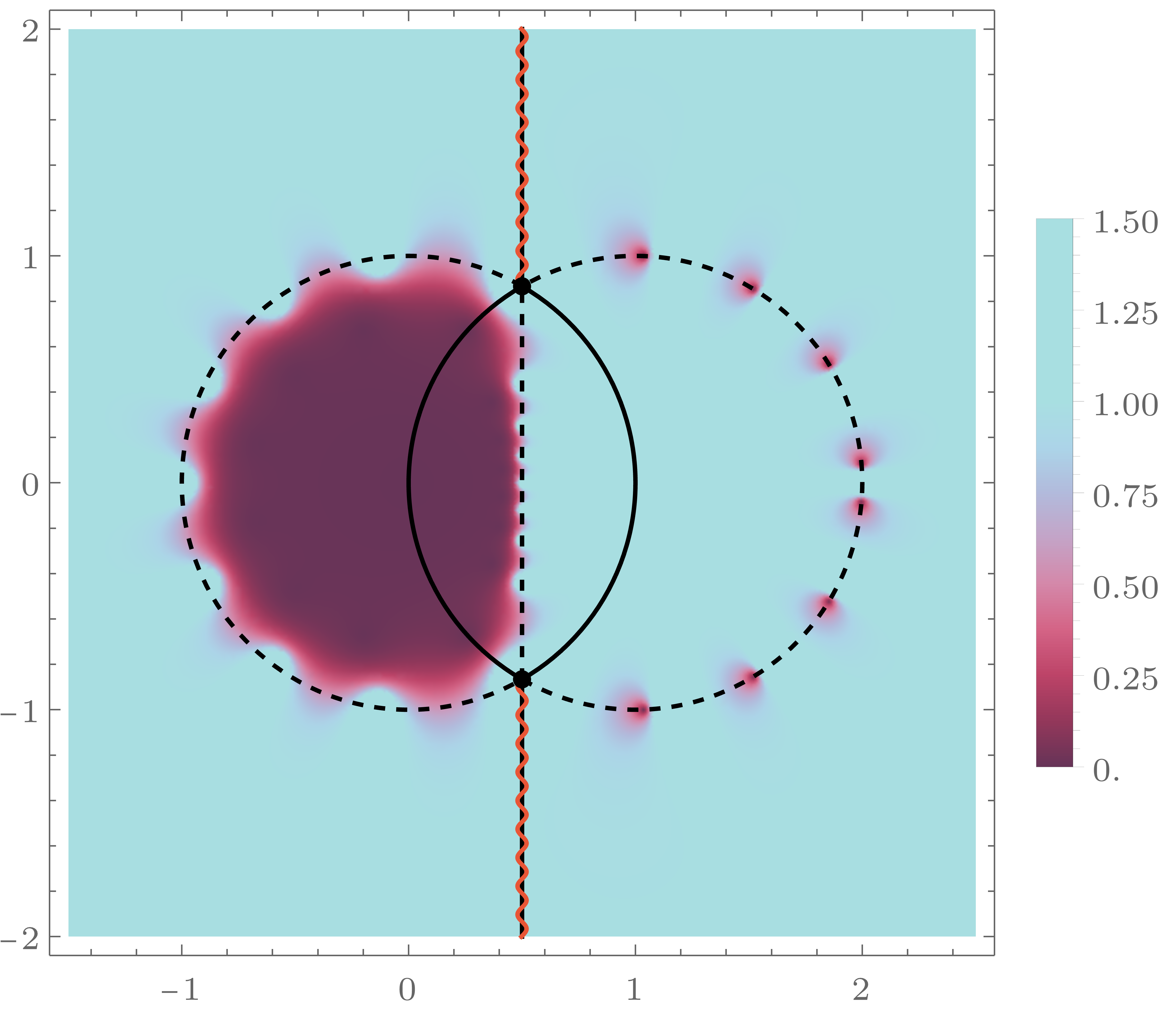}
        \caption{$ \epsilon < 0$}
    \end{subfigure}
    \hspace{0.5em}
    \caption{Density plots of $|1-\mathcal{F}_{1,1}^{\, {\rm sc}}/\mathcal{F}_{1,1}|$ as a function of $z$.
    The black lines are the Stokes (solid) and the anti-Stokes (dashed) lines. 
    Wiggled lines are branch cuts of the semi-classical approximation emanating from $z = z_{\pm}$ (black dots). 
    The density plot was generated with $\epsilon = \pm 0.10137$ (corresponding to central charge $c \simeq 82 $ and $c \simeq -34$, respectively).\label{fig:4H-exact-vs-semiclassic}}
\end{figure}

In figure~\ref{fig:4H-exact-vs-semiclassic}, we compare  $\mathcal{F}_{1,1}$ with its semi-classical approximation 
\begin{equation}\label{eq:F11sc}
    \mathcal{F}_{1,1}^{\, \rm sc} = e^{-\frac{1}{\epsilon}S_1} A_1 \, ,
\end{equation}
for $\epsilon> 0$ (left panel) and $\epsilon<0$ (right panel).  

For $\epsilon >0$, $\mathcal{F}_{1,1}=\mathcal{F}_{1,1}^B$ for $z\in D_0$ and $\mathcal{F}_{1,1}^{\,\rm sc}$ provides a good approximation. Observe that anti-Stokes lines in $D_0$ are ``deactivated'', as there is no other saddle contribution.
After we cross the Stokes lines and $z\notin D_0$, $\mathcal{F}_{1,1}$ becomes a trans-series but $\mathcal{F}_{1,1}^{\,\rm sc}$ remains the leading contribution in a disc around $z=1$ until we cross the anti-Stokes line, in which case $\widetilde{\mathcal{G}}$ becomes the dominant contribution and the approximation breaks down.

For $\epsilon <0$, $\mathcal{F}_{1,1}$ is always a combination of $\mathcal{F}_{1,1}^B$ and $\mathcal{G}^B$. We see that $\mathcal{F}_{1,1}^{\,\rm sc}$ provides a good approximation when $z$ 
is in the left half of region $E_0$ and beyond, in a disc around $z=0$. However, as we cross anti-Stokes lines (both sides), $\mathcal{G}^B$ dominates and the semi-classical approximation breaks down.
Interestingly enough, the rightmost Stokes line triggers a Stokes jump of $\widetilde{\mathcal{G}}$ given by $S_{-}$ in~\eqref{eq:monodromy-neg}. 
As a consequence, when 
we cross the outer anti-Stokes line on the right, $\widetilde{\mathcal{F}}_{1,1}$ becomes dominant again but the semi-classical approximation remains poor since the prefactor has changed.

\section{2H2L correlator with \texorpdfstring{$\phi_{2,1}$}{phi21}}
\label{sec:2H2L}

Having established the analysis in the 4H case, the 2H2L case will follow a very similar path.
We use the results of appendix~\ref{app:resurgence} to derive the asymptotic expansion of the identity block, study in detail its Borel summability and Stokes phenomena, verify that the non-perturbative Stokes jump resolve forbidden singularities, and, lastly, fix the four-point function. 
As in the previous section, we find it useful to first start with $\epsilon>0$ and then discuss the case $\epsilon < 0$.

By using the analysis of appendix~\ref{app:2F1-expansion-2H2L-reduction}, we can write the identity block in~\eqref{eq:2H2L-blocks} in the form~\eqref{eq:pc1}, but now with prefactors\footnote{
    The exponential of $S_{1}$ is just a re-writing of the kinematical factor $z^{-2 h_{2,1}}$ in~\eqref{eq:Int2}.
}
\begin{equation}\label{eq:hhll-SA}
  S_1(z)  = -\frac{3}{2}\log(z) \, , \qquad 
  A_1(z) = z^{-\frac{5}{4}}(1-v(z))^\gamma \, ,
\end{equation}
where the map $v$ is given by
\begin{equation}\label{eq:vmapSec}
    v(z) = \frac{z^2}{(2-z)^2}\,.
\end{equation}
See appendix~\ref{app:2F1-expansion-2H2L-reduction} for a detailed derivation of how the map $v(z)$ is found.
The same $v(z)$ governs the form of the coefficients in the asymptotic series:
\begin{equation}\label{eq:asym-hhll}
    \widetilde{F}_1 = f_0 + \sum_{n=1}^{+\infty} f_{n}(v) \epsilon^{n} \qquad
    f_{0} = 1 \, , \quad 
    f_{n}(v) = \sum_{k=1}^{n} (-1)^{n+k} \stirlingII{n-1}{k-1} \frac{(\gamma)_k (\gamma + 1/2)_k }{k!} v^k \,,
\end{equation}
which is obtained from~\eqref{eq:2F1-expansion-large-c-no-shift} with the identification $a\to \gamma$, $b\to \gamma+\tfrac{1}{2}$, $z\to v(z)$.
Like in the 4H case, it is remarkable that the coefficients $f_n$ have a simple form when expressed in terms of the variable $v(z)$ in~\eqref{eq:vmapSec}.

It is useful to match our results for $S_1$ and $A_1$ with the form of the semi-classical blocks for a generic 2H2L correlator found in~\cite{Fitzpatrick:2014vua,Fitzpatrick:2015zha}.
For two heavy operators with conformal weight $h\ped{H} = \eta c\,(1+\mathcal{O}(c^{-1}))$ and two light operators with weight $h\ped{L}$, the block associated with identity exchange has the semi-classical limit
\begin{equation}\label{eq:SCblockFK}
    \lim_{c\rightarrow\infty}\mathcal{F}(h_i,0,c;z) 
    = z^{-2h\ped{H}} (1-w)^{h\ped{L}\left(1-\frac{1}{a}\right)}\left(\frac{w}{a z}\right)^{-2h\ped{L}} \, ,
\end{equation}
where
\begin{equation}
    w = 1-(1-z)^a \, , \qquad a = \sqrt{1-24 \eta} \, .
\end{equation}
This block has forbidden singularities at
\begin{equation}\label{eq:ForbSingFK}
    z_{n} = 1 - e^{\frac{2\pi i}{a} n} \, , \qquad n \in \mathbb{Z}_{\neq 0} \, .
\end{equation}
From~\eqref{eq:hhll-SA} we have
\begin{equation}\label{eq:hhll-SCblock}
    \widetilde{\mathcal{F}}_{1,1}(z;\epsilon) = e^{\frac{1}{4} c \log{z}} \, \left(\frac{4(1-z)}{(2-z)^2}\right)^{h\ped{L}} \Big(1 + O(c^{-1})\Big)\,,
\end{equation}
where we use $\gamma = h\ped{L} + O(c^{-1})$ and have rewritten the expression in terms of the central charge $c$. The expression~\eqref{eq:hhll-SCblock} matches~\eqref{eq:SCblockFK} with 
$\eta = -1/8$ and $a=2$.  For $a=2$, we get from~\eqref{eq:ForbSingFK} a single forbidden singularity at $z=2$, which is precisely the singular point of the map $v(z)$. Like in the 4H case discussed before, in addition to the semi-classical block, every single order in perturbation theory is more and more singular at $z=2$. Unlike the map $r(z)$, $v(z)$ does not have branch cuts. The singular behaviour of perturbation theory had already been observed in~\cite{Fitzpatrick:2016ive}, where it was conjectured a form for the $f_n$ as a polynomial in $z$ divided by a power of $\left(z(2-z)\right)^2$. We see in~\eqref{eq:asym-hhll} that the $f_n$ are polynomials in $v$.

\subsection{Borel summation and discontinuity}

The Borel transform of~\eqref{eq:asym-hhll} is given by
\begin{equation}
    \hatF_1(v;t) = \gamma\left(\gamma+\tfrac{1}{2}\right) v \hypgeo{2}{1}\left(\gamma+1,\gamma+\tfrac{3}{2},2;v(1-e^{-t})\right)\,,
\end{equation}
with $z$-dependent singularities given by
\begin{equation}
t_k(v) = \log\left(\frac{v}{v-1}\right)+2\pi i k, \quad k\in \mathbb{Z},
\label{eq:tk-singularities-hhll}
\end{equation}
which become real and positive for $v>1$. We can read from the Borel transform the large order behaviour, which, for example, for $0 < z < 1$ is given by
\begin{equation}\label{eq:LBO2H2L}
    f_n\approx  \frac{\Gamma(n+2\gamma-\tfrac{1}{2})}{\Gamma\left(\gamma\right)\Gamma\left(\gamma+\tfrac{1}{2}\right)} \frac{v}{(1-v)^{2\gamma+\tfrac{1}{2}}}\left|\log\frac{v}{v-1}\right|^{\frac{1}{2}-2\gamma-n}2\sin\left(\big(n+2\gamma-\tfrac{1}{2}\big)\arg\!\big(\!\log\tfrac{v}{v-1}\big)-2\pi\gamma\right),
\end{equation}
as follows from~\eqref{eq:fn-LOB-cases}.

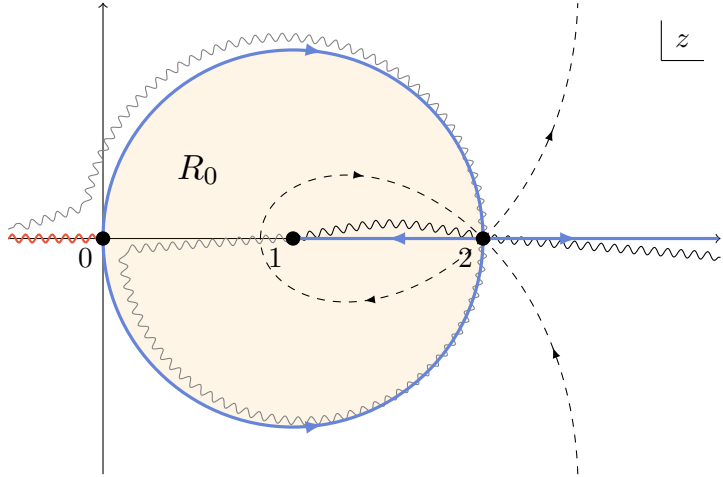
\begin{figure}
    \centering
    \begin{tikzpicture}[scale=2.5]
        \draw[->] (-0.5,0)--(3.25,0);
        \draw[->] (0,-1.25)--(0,1.25);

        \node[draw=none, inner sep=5pt, append after command={
            \pgfextra{\draw (\tikzlastnode.south west) -- (\tikzlastnode.south east);}
            \pgfextra{\draw (\tikzlastnode.south west) -- (\tikzlastnode.north west);}
        }] at (3.05, 1.05) {$z$};
        
        \fill[region, opacity=0.1] (2,0) arc[start angle=0, end angle=360, radius=1]; 
        \node[draw=none, scale=1.2] at (1/2, 0.365) {$R_{0}$};

        \draw[branch, thick, decorate, decoration={snake, segment length=5, amplitude=1.5}] (-0.5,0) -- (0,0);
        \draw[black, decorate, decoration={snake, segment length=5, amplitude=1.5}] (1,0) .. controls (3/2, 1/10) .. (2,0) -- (3.25,-0.1);
        \draw[gray, decorate, decoration={snake, segment length=5, amplitude=1.5}] (2,0) arc (0:170:1.05) -- (-1/2,1/20);
        \draw[gray, decorate, decoration={snake, segment length=5, amplitude=1.5}] (2,0) arc (0:-176:0.95) -- (1,0);

        \draw[stokes, very thick] (1,1) arc[start angle=90, end angle=0, radius=1] [postaction={decorate, decoration={
                markings, mark=at position 0.1 with {\arrow{latex}}
            }}];
        \draw[stokes, very thick] (1,1) arc[start angle=90, end angle=180, radius=1]; 
        \draw[stokes, very thick] (1,-1) arc[start angle=270, end angle=180, radius=1];
        \draw[stokes, very thick] (1,-1) arc[start angle=-90, end angle=0, radius=1] [postaction={decorate, decoration={
                markings, mark=at position 0.1 with {\arrow{latex}}
            }}];
        \draw[stokes, very thick] (2, 0) -- (1,0) [postaction={decorate, decoration={
                markings, mark=at position 0.5 with {\arrow{latex}}
            }}];
        \draw[stokes, very thick] (2, 0) -- (3.24,0) [postaction={decorate, decoration={
                markings, mark=at position 0.4 with {\arrow{latex}}
            }}];
        \draw[dashed, samples=50, domain=-1.1911:-0.785398, variable=\x, postaction={decorate, decoration={markings, mark=at position 0.5 with {\arrow{latex}}}}] 
        plot ({2+(2.82843-4*cos(deg(\x)))*cos(deg(\x))}, {(2.82843-4*cos(deg(\x)))*sin(deg(\x))});
        \draw[dashed, samples=50, domain=0:-0.785398, variable=\x, postaction={decorate, decoration={markings, mark=at position 0.5 with {\arrow{latex}}}}] 
        plot ({2+(2.82843-4*cos(deg(\x)))*cos(deg(\x))}, {(2.82843-4*cos(deg(\x)))*sin(deg(\x))});
        \draw[dashed, samples=50, domain=0.785398:0, variable=\x, postaction={decorate, decoration={markings, mark=at position 0.5 with {\arrow{latex}}}}] 
        plot ({2+(2.82843-4*cos(deg(\x)))*cos(deg(\x))}, {(2.82843-4*cos(deg(\x)))*sin(deg(\x))});
        \draw[dashed, samples=50, domain=0.785398:1.1911, variable=\x, postaction={decorate, decoration={markings, mark=at position 0.5 with {\arrow{latex}}}}] 
        plot ({2+(2.82843-4*cos(deg(\x)))*cos(deg(\x))}, {(2.82843-4*cos(deg(\x)))*sin(deg(\x))});
        \filldraw[black, fill=black] (2,0) circle (1pt) node[below left,black]{$2$};
        \filldraw[black, fill=black] (1,0) circle (1pt) node[below left,black]{$1$};
        \filldraw[black, fill=black] (0,0) circle (1pt) node[below left,black]{$0$};
    \end{tikzpicture}
    \caption{Branch cuts (wiggly lines), Stokes lines (blue) and anti-Stokes lines (dashed) in the $z$-plane for the 2H2L asymptotic series $(\widetilde{\mathcal{F}}_{1,1},\widetilde{\mathcal{G}})$.
    Starting from $z = 2$, outgoing arrows denote lines where $\widetilde{\mathcal{F}}_{1,1}$ has a Stokes discontinuity (and $S_2$ has a branch cut), while ongoing arrows denote lines where $\widetilde{\mathcal{G}}$ jumps.
    The wiggled black and grey lines respectively denote branch cuts $A_{1,2}(z)$, while the red one denotes an additional branch cut from $S_{1,2}(z)$. 
    Refined version of figure~\ref{fig:Fig2Draft}.\label{fig:2H2L-stokes}}
\end{figure}

We see that $z(v)$ maps $v\in(1, +\infty)$ to two intervals, $(1,2)$ and $(2, +\infty)$. Outside of these intervals, $\widetilde{\mathcal{F}}_{1,1}$ is Borel summable and~\eqref{eq:V1-identity} also holds for $\epsilon>0$. We then have
\begin{equation}
  \mathcal{F}_{1,1}^B =  \mathcal{F}_{1,1} ,\qquad z\notin(1,+\infty)\,.
    \label{eq:hhll-BorelOk}
\end{equation}
When $z\in (1,+\infty)$, there is a Stokes jump because $t_0(v)$ lies in the positive real axis.
Thus the jump is non-perturbative in $1/c$, again with the form
\begin{equation}
     (s_{0+}-s_{0-})\widetilde{\mathcal{F}}_{1,1}(z;\epsilon) 
    \propto e^{-\frac{1}{\epsilon} t_0(v)} = \left(\frac{v}{v-1}\right)^{-\frac{1}{\epsilon}}.
    \label{eq:non-pert-hhll}
\end{equation}
Its full expression can be derived using the results of appendix~\ref{app:resurgence}, which give
\begin{equation}
    (s_{0+}-s_{0-}) \widetilde{\mathcal{F}}_{1,1}(z;\epsilon) = i \sqrt{2} e^{i \pi  \gamma \pm \frac{i \pi }{\epsilon }} \sqrt{\sin (2 \pi  \gamma )} \, s_0\Big(\widetilde{\mathcal{G}}(z+ i0;\epsilon)\Big),\quad z\in(1,+\infty),
    \label{eq:hhll-pure-stokes}
\end{equation}
where $\pm$ in the phase is taken depending on whether $z\lessgtr 2$. 
Unlike the 4H case, the series $\widetilde{\mathcal{G}}$ is related to the genuinely new series $\widetilde{F}_2$, described in detail in appendix~\ref{app:resurgence}.
The prefactors are given by
\begin{equation}
\begin{aligned}
    \widetilde{\mathcal{G}}(\epsilon; z) &= e^{\pm \left(i\pi\gamma+\frac{i \pi}{\epsilon}\right)}  e^{-\frac{1}{\epsilon} S_{1}(z)} A_{1}(z) \widetilde{F}_{2}(\epsilon; v(z))
\approx 
    e^{-\frac{1}{\epsilon} S_{2}(z)}
    \epsilon^{\frac{1}{2}-2 \gamma } 
    A_2(z)
    \Big(1+\mathcal{O}(\epsilon)\Big),\\[0.5em]
    S_2(z) &= S_1(z) +\log\left(\frac{v}{1-v}\right),\quad
    A_2(z) = z^{-5/4}
    \frac{v  }{(v-1)^{\gamma +\frac{1}{2}}}
    \, 4^{ \gamma } 
    \sqrt{\frac{\Gamma (1-2 \gamma )}{2\Gamma (2 \gamma )}},
\end{aligned}
\end{equation}
where we choose $\pm$ in the phase according to whether $z$ is in the upper/lower half plane. 
After combining with the prefactors in $\widetilde{F}_2$ this ambiguity is resolved, as can be seen in the second line. This is analogous to the point discussed in~\eqref{eq:r-phase}.
$\widetilde{\mathcal{G}}$ has Stokes jumps when $v\in(-\infty,0)$, which in the $z$-plane maps to two half unit circles centred around $z=1$, as in figure~\ref{fig:2H2L-stokes}. Then the maximal continuation of $(0,1)$ until a Stokes line of either $\widetilde{\mathcal{F}}_{1,1}$ or $\widetilde{\mathcal{G}}$ is the region
\begin{equation}
  R_0 =  \big\{\, z\in \mathbb{C}\,:\, |z-1|<1 \wedge z\notin (1,2) \, \big\}.
\end{equation}

Repeating the exercise of extracting the conformal weight by taking $z\rightarrow 0$, we discover once again that $\widetilde{\mathcal{G}}$ is the perturbative part of the $(3,1)$ block. However, $F_2^B(\epsilon;v)$ is a sum of both solutions of the hypergeometric equation, as discussed in appendix~\ref{app:resurgence}, meaning that $\mathcal{G}^B$ is a linear combination of both blocks, much like in the 4H case. We find
\begin{equation}
\begin{aligned}
\mathcal{F}^B_{1,1}&=\mathcal{F}_{1,1},\\
\mathcal{G}^B &= 
-\frac{i e^{-i \pi  \gamma } }{\sqrt{\sin (2 \pi  \gamma )}}
\left(
h(\epsilon)
\mathcal{F}_{3,1}
+ 
\frac{1}{\sqrt{2}}\frac{ \sin (2 \pi  \gamma )}{ \sin \left(\frac{\pi }{\epsilon }\right)}
\mathcal{F}_{1,1}
\right),
\end{aligned}
\qquad \epsilon>0,\quad z\in R_0,
\end{equation}
where
\begin{equation}
    h(\epsilon) = \frac{16^{\frac{1}{\epsilon }-1} \Gamma \left(\frac{1}{\epsilon }-1\right) \Gamma \left(\frac{1}{\epsilon }\right)}{\sqrt{2} \Gamma (2 \gamma ) \Gamma \left(-2 \gamma +\frac{2}{\epsilon }-1\right)}.
    \label{eq:h-coeff}
\end{equation}
$\widetilde{\mathcal{G}}$ becomes non-Borel summable when $v<0$, and its Stokes jump is given by $\widetilde{\mathcal{F}}_{1,1}$ up to some overall constants, as we discuss in detail in the next section.

\subsection{Stokes lines and monodromy matrices}

It is convenient to present the Stokes jumps using a graphical notation for Stokes lines similar to the 4H case.
This case requires some additional care since $S_{2}(z)$  has a branch along the Stokes line of $\widetilde{\mathcal{F}}_{1,1}$, while $A_1(z)$ and $A_2(z)$ have branch cuts which overlap with both the real axis and the Stokes lines. While the former is easy to incorporate in the Stokes jump matrices, for the latter we place them as in figure~\ref{fig:2H2L-stokes}.\footnote{This is equivalent to redefining the one-loop factors with a small angle $\alpha$ such that
\begin{equation}
A_1(z) \propto z^{-5/4} e^{i \pi  \alpha  \gamma } \left(e^{-i \pi  \alpha } (1-v(z))\right)^{\gamma },
\qquad
A_2 \propto \frac{z^{-5/4} v(z)}{e^{i \pi  \alpha  \left(\gamma +\frac{1}{2}\right)} \left(e^{-i \pi  \alpha } (v(z)-1)\right)^{\gamma +\frac{1}{2}}}.
\end{equation}}
Applying the inverse map $z(v)$ to $(1,\infty)$ and $(-\infty,0)$ results in figure~\ref{fig:2H2L-stokes}. In the convention of figure~\ref{fig:Airy2}, the Stokes jumps in figure~\ref{fig:2H2L-stokes} are
\begin{equation}
    S_{+} = \begin{pmatrix}
 1 & 0 \\
 - e^{\frac{i \pi }{\epsilon }-i \pi  \gamma } \sqrt{2\sin (2 \pi  \gamma )} & e^{\frac{2 i \pi }{\epsilon }} \\
\end{pmatrix}, 
    \qquad 
    S_{-} = \begin{pmatrix}
 1 & i e^{i \pi  \gamma -\frac{i \pi }{\epsilon }} \sqrt{2\sin (2 \pi  \gamma )} \\
 0 & e^{-\frac{2 i \pi }{\epsilon }} \\
\end{pmatrix},
\end{equation}
which can be derived from the Stokes jump of $\widetilde{F}_{1,2}$ discussed in appendix~\ref{app:resurgence}.
There are two types of branch cuts emanating from $z=2$, those of $A_1(z)$ (drawn in black in figure~\ref{fig:2H2L-stokes}) and those of $A_2(z)$ (drawn in gray). They are given by, respectively,
\begin{equation}
B_1 = \begin{pmatrix}
 e^{-2 i \pi  \gamma } & 0 \\
 0 & 1 \\
\end{pmatrix}\,,
\qquad 
B_2 = \begin{pmatrix}
 1 & 0 \\
 0 & -e^{2 i \pi  \gamma } \\
\end{pmatrix},
\end{equation}
when crossing the branch cut anti-clockwise with respect to $z=2$.
Lastly there is a branch cut at $(-\infty,0)$ due to the $\log(z)$ factor in the origin,
\begin{equation}
    B_3= -i e^{\frac{3 i \pi }{\epsilon }} I.
\end{equation}
In contrast to the singular points $z_\pm$ in the 4H correlator, where three Stokes lines meet, at $z=2$ four Stokes lines meet.
The point $z=2$ corresponds to a double turning point, where two simple turning points merge. This is evident from the analysis in appendix~\ref{app:wkb}, where
one can see that $\sqrt{Q_0}$ in~\eqref{eq:derivative_action_diff} has a double zero at $z=2$ for the values $a_{1}=0$, $b_{1}=1$, $c_{1}=2$. As is known in exact WKB, $q+2$ Stokes lines end at a turning point of order $q$. 

By taking care of all branch cuts and Stokes lines, we determine the total monodromy in making a $2\pi$ rotation around $z=2$. We get
\begin{equation}\label{eq:trivial-monodromy-2}
    M_{z=2} = B_2 S_{+} B_1 S_{-} B_2 S_{+} B_1 S_{-}= I\,,
\end{equation}
as it should be for a regular point, and importantly establishes how non-perturbative effects resolve the forbidden singularity at $z=2$. 
The monodromies around $z=0,1$, starting from $z\in(0,1)$, are
\begin{equation}\label{eq:2h2lM01}
\begin{aligned}
M_{z=0}
&=  
S_{+} B_3^{-1} S_{+}
=
\begin{pmatrix}
 i e^{-\frac{3 i \pi }{\epsilon }} & 0 \\
 -2 i  e^{-\frac{i \pi }{\epsilon }-i \pi  \gamma } \sqrt{2\sin (2 \pi  \gamma )} \cos \left(\frac{\pi }{\epsilon }\right) & i e^{\frac{i \pi }{\epsilon }} \\
\end{pmatrix}
, \\ 
M_{z=1} &= 
B_1 S_{-} B_2
=
\begin{pmatrix}
 e^{-2 i \pi  \gamma } & -i e^{i \pi  \gamma -\frac{i \pi }{\epsilon }} \sqrt{2\sin (2 \pi  \gamma )} \\
 0 & -e^{-\frac{2 i \pi }{\epsilon }+2 i \pi  \gamma } \\
\end{pmatrix} .
\end{aligned}
\end{equation}
It can be checked that, if we replace $\widetilde{\mathcal{F}}_{1,1}$ and $\widetilde{\mathcal{G}}$ by their resummed hypergeometric forms using~\eqref{eq:tildeF1-borel-resummation-deriv} and~\eqref{eq:saddle2}, the matrix $M_{z=1}$ rederives the correct discontinuity of the exact block in~\eqref{eq:2H2L-blocks}.

\subsection{Fixing the four-point function}

Repeating the steps done in the 4H case, we can fix the four-point function by combining holomorphic and anti-holomorphic terms. We write the four-point function as in~\eqref{eq:four-point-ansatz} and determine the matrix $C_{ij}$
by demanding the single-valuedness of the correlator around $z=0$ and $z=1$. Using the monodromy matrices $M_{z=0,1}$ in~\eqref{eq:2h2lM01}, we get
\begin{equation}
C = \begin{pmatrix}
 1 & -\frac{i}{\sqrt{2}} e^{i \pi  \gamma } \sqrt{\sin (2 \pi  \gamma )} \sec \left(\pi  \left(\frac{1}{\epsilon }-2 \gamma \right)\right) \\
 \frac{i}{\sqrt{2}} e^{-i \pi  \gamma } \sqrt{\sin (2 \pi  \gamma )} \sec \left(\pi  \left(\frac{1}{\epsilon }-2 \gamma \right)\right) & \sin \left(\frac{\pi }{\epsilon }\right) \sec \left(\pi  \left(\frac{1}{\epsilon }-2 \gamma \right)\right) \\
\end{pmatrix} .
\end{equation}
We determine the normalization factor $g(\epsilon)$ in~\eqref{eq:four-point-ansatz} by demanding the correct normalization of the identity block exchange in the $s$-channel as $z\to 0$.
As before, this requires to consider $\epsilon>1$, in which case the identity becomes the leading contribution in the correlator. 
Proceeding as before, we then have
\begin{equation}
    g(\epsilon)=1+\sin (2 \pi  \gamma ) \csc \left(2 \pi  \left(\frac{1}{\epsilon }-\gamma \right)\right).
\end{equation}
In region $R_0$, this matches~\eqref{eq:2H2L-blocks}, rederiving the OPE coefficient therein. For the full complex plane one must adjust $C$ whenever we cross a branch cut or Stokes line, as discussed in section~\ref{sec:4H}.

We emphasize that we reconstructed the entire correlator starting just from the asymptotic series of the identity block in the $s$-channel.

\subsection{Negative central charge}
\label{subsec:2H2L-eps-negative}

When $\epsilon<0$ the conformal weight $h_{3,1}$ becomes positive and the identity operator provides the dominant contribution in the $s$-channel OPE.
The situation is analogous to the 4H case, so we will be brief. 
The identity block is now a trans-series while the $(3,1)$ block is, up to $z$-independent factors, given by resummation of $\widetilde{\mathcal{G}}$, as follows from~\eqref{eq:resum-neg-eps}.
We have
\begin{equation}\label{eq:2H2Lepsm0FG}
    \begin{aligned}
        \mathcal{F}_{1,1}^B(z;\epsilon)
        &=\mathcal{F}_{1,1}(z;\epsilon)
        +
         \frac{\sqrt{2}\sin \left(\frac{\pi }{\epsilon }\right)}{ \sin \left(2 \pi  \left(\gamma -\frac{1}{\epsilon }\right)\right)} h(\epsilon )
        \mathcal{F}_{3,1}(z;\epsilon),\\
        \mathcal{G}^B(z;\epsilon)
        &=-
        e^{-3 i \pi  \gamma }\left(\frac{
        2  \sin ^2\left(\frac{\pi }{\epsilon }\right) }
        {\sin \left(2 \pi  \left(\gamma -\frac{1}{\epsilon }\right)\right)}\right)
        \frac{h(\epsilon)}{\sqrt{\sin (2 \pi  \gamma )}}
        \mathcal{F}_{3,1}(z;\epsilon),
    \end{aligned}
    \qquad\epsilon<0,\quad z\in R_0,
\end{equation}
where $h(\epsilon)$ is defined in~\eqref{eq:h-coeff}.
Furthermore, the Stokes jumps change, becoming
\begin{equation}
    S_{+} = \begin{pmatrix}
 1 & -i  e^{3 i \pi  \gamma +\frac{i \pi }{\epsilon }} \sqrt{2\sin (2 \pi  \gamma )} \\
 0 & e^{\frac{2 i \pi }{\epsilon }} \\
\end{pmatrix}, 
    \qquad 
    S_{-} =\begin{pmatrix}
 1 & 0 \\
 -e^{-\frac{i \pi }{\epsilon }-3 i \pi  \gamma } \sqrt{2\sin (2 \pi  \gamma )} & e^{-\frac{2 i \pi }{\epsilon }} \\
\end{pmatrix}.
\end{equation}
Here we have to be careful that not only the lines swap, but we also pick up additional phases due to~\eqref{eq:phi1-phi2-system-stokes-jumps-neg}.

As expected,~\eqref{eq:trivial-monodromy-2} still applies and the monodromy matrix around $z=2$ is again trivial. The forbidden singularity is resolved in the same way as in the $\epsilon>0$ case.
The monodromy matrices around $z=0$ and $z=1$ read
\begin{equation}\label{eq:2h2lM0M1}
\begin{aligned}
M_{z=0} &= 
\begin{pmatrix}
 i e^{-\frac{3 i \pi }{\epsilon }} & 2 e^{3 \pi  i \gamma } e^{-\frac{i \pi }{\epsilon }} \sqrt{2\sin (2 \pi  \gamma )} \cos \left(\frac{\pi }{\epsilon }\right) \\
 0 & i e^{\frac{i \pi }{\epsilon }} \\
\end{pmatrix}
\,, \\ 
M_{z=1} &= 
\begin{pmatrix}
 e^{-2 i \pi  \gamma } & 0 \\
 - e^{-3 i \pi  \gamma -\frac{i \pi }{\epsilon }} \sqrt{2\sin (2 \pi  \gamma )} & -e^{2 i \pi  \gamma -\frac{2 i \pi }{\epsilon }} \\
\end{pmatrix}.
\end{aligned}
\end{equation}
We can fix the four-point function, as discussed in the previous cases, by starting
from the ansatz~\eqref{eq:four-point-ansatz-neg}. From the monodromy matrices~\eqref{eq:2h2lM0M1} we can determine the matrix $C$ entering~\eqref{eq:four-point-ansatz-neg}.
Like in the 4H correlator, for $\epsilon<0$ the identity block dominates in the $s$-channel $z\rightarrow 0$  and this makes the
determination of $g(\epsilon)$ trivial. Eventually we get 
\begin{equation}
 g(\epsilon)=1\,, \quad    C = 
 \begin{pmatrix}
 1 & \frac{1}{\sqrt{2}}e^{3 i \pi  \gamma } \sqrt{\sin (2 \pi  \gamma )} \csc \left(\frac{\pi }{\epsilon }\right) \\
 \frac{1}{\sqrt{2}}e^{-3 i \pi  \gamma } \sqrt{\sin (2 \pi  \gamma )} \csc \left(\frac{\pi }{\epsilon }\right) & \csc \left(\frac{\pi }{\epsilon }\right) \cos \left(\pi  \left(\frac{1}{\epsilon }-2 \gamma \right)\right) \\
\end{pmatrix} \,, \quad \epsilon <0\,.
\end{equation}

\begin{figure}[t!]
    \centering
    \hspace{0.5em}
    \begin{subfigure}[c]{0.48\textwidth}
        \includegraphics[width=\textwidth]{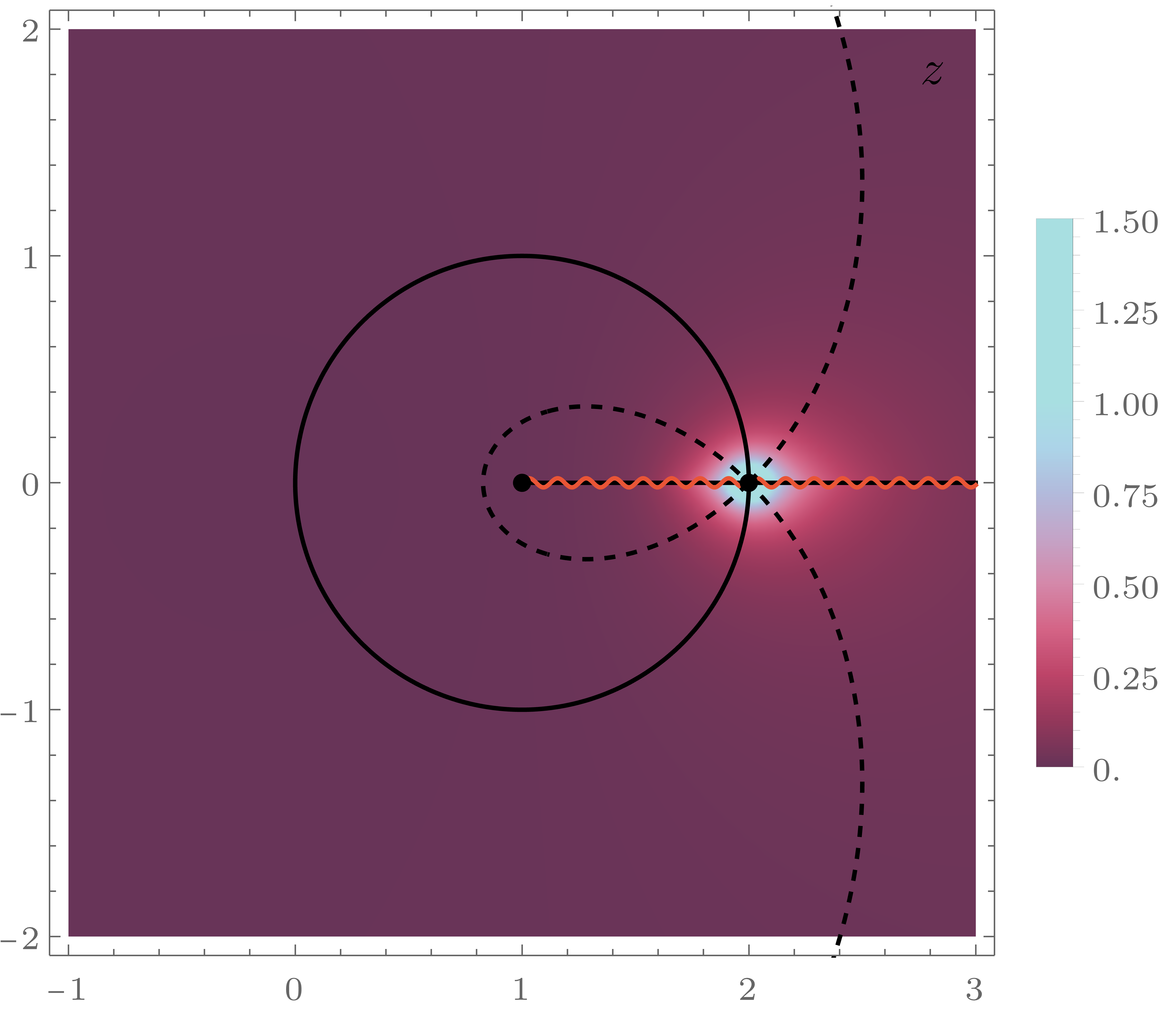}
        \caption{$ \epsilon > 0$}
    \end{subfigure}
    \hfill
    \begin{subfigure}[c]{0.48\textwidth}
        \includegraphics[width=\textwidth]{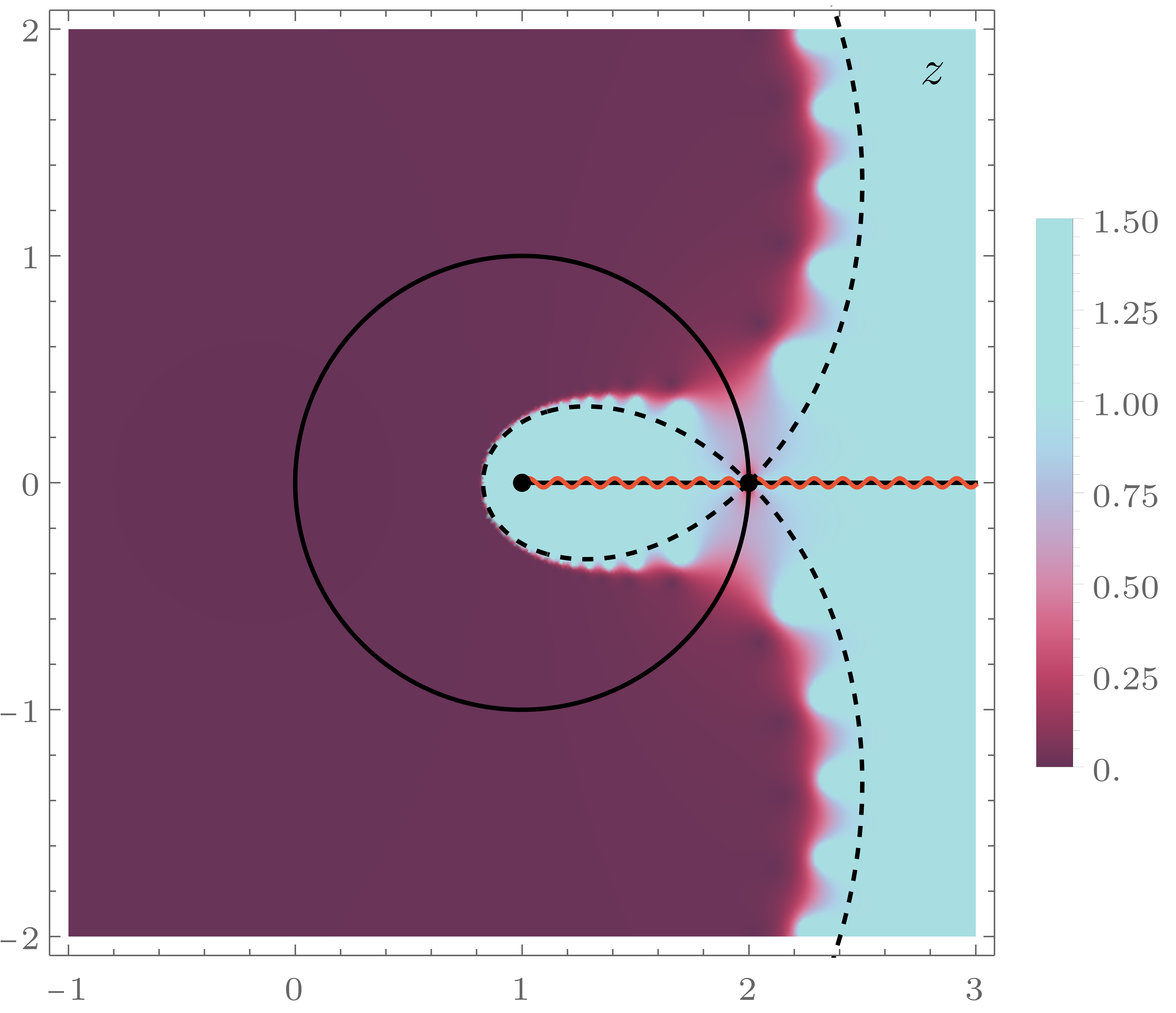}
        \caption{$ \epsilon < 0$}
    \end{subfigure}
    \hspace{0.5em}
    \caption{
    Density plots of $|1-\mathcal{F}_{1,1}^{\, {\rm sc}}/\mathcal{F}_{1,1}|$ as function of $z$.
    The black lines are the Stokes (solid) and anti-Stokes (dashed) lines.
    Wiggled lines are branch cuts of the semi-classical approximation emanating from $z = 1, 2$ (black dots). 
    The density plot was generated with $\epsilon\simeq  \pm 0.01447$ (corresponding to central charge $c \simeq 436 $ and $c \simeq -393$, respectively).
    }
\label{fig:2H2L-exact-vs-semiclassic}
\end{figure}

We compare in figure~\ref{fig:2H2L-exact-vs-semiclassic} $\mathcal{F}_{1,1}$ with its semi-classical approximation 
$\mathcal{F}_{1,1}^{\, \rm sc}$ defined as in~\eqref{eq:F11sc},
for $\epsilon> 0$ (left panel) and $\epsilon<0$ (right panel).  
For $\epsilon >0$, $\mathcal{F}_{1,1}=\mathcal{F}_{1,1}^B$ for $z\notin (1,\infty)$ and $\mathcal{F}_{1,1}^{\,\rm sc}$ provides a good approximation all over the complex plane, but a small region around the forbidden singularity.
For $\epsilon <0$, $\mathcal{F}_{1,1}$ is always a combination of $\mathcal{F}_{1,1}^B$ and $\mathcal{G}^B$. $\mathcal{F}_{1,1}^{\,\rm sc}$ provides a good approximation in the purple 
region in panel (b) of figure~\ref{fig:2H2L-exact-vs-semiclassic},\footnote{This purple region does not extend to infinity, but is enclosed by the anti-Stokes line which has a ``lima\c{c}on'' shape \cite{wiki:Limaçon}.} until we cross anti-Stokes lines.

\subsection{Unitary minimal models from a large \texorpdfstring{$c$}{c} expansion}
\label{subsec:MM}

We have seen in the previous sections how we can determine the whole correlator of 4H and 2H2L starting from the asymptotic expansion
of the identity block, both for $\epsilon>0$ and $\epsilon<0$. The result applies also for finite values of $\epsilon$, and hence at finite central charge.
In fact, starting from $\epsilon>0$, we see that as $\epsilon$ increases, at some point $b^2$ turns negative and the 
conformal weights~\eqref{eq:hDegConfWeights} turn positive. In particular, the values
\begin{equation}
    \epsilon = \frac{2(p+1)}{p+3}\,, \qquad p\geq 3\,,
\end{equation}
correspond to the {\it unitary} minimal models ${\cal M}_{p+1,p}$ with 
\begin{equation} 
    c = 1 - \frac{6}{p(p+1)}\,,  \qquad h_{r,s} = \frac{ (p r - (p+1) s)^2-1}{4p(p+1)}\,.
\end{equation}
Interestingly enough, the Borel summability of $\widetilde{\mathcal{F}}_{1,1}$ extends all the way to these values.
Consider for example the 4-$\sigma$ correlator, with $h_\sigma= \overline{h}_\sigma = 1/16$, in the Ising model $p=3$.
The identity block exchanged in this correlator is known to be given by 
\begin{equation}
    \mathcal{F}_{1,1}^\text{Ising}(z)  =\frac{1}{\left(z(1-z)\right)^{\frac 18}} \sqrt{\frac{1+\sqrt{1-z}}{2}} \,. 
\end{equation}
We can match this formula in two ways. We can take the $\widetilde{\mathcal{F}}_{1,1}$ series of the 4H case and resum it for $\epsilon=4/3$ to get
\begin{equation} 
    \mathcal{F}_{1,1}^\text{Ising}(z)  = \mathcal{F}_{1,1}^B\left(z;\epsilon=\frac 43\right) \,, \qquad z \in E_0 \qquad (\text{4H})  \,.
\end{equation}
By tracking the Stokes jumps across the $z$-plane, the above formula can be extended to retrieve $\mathcal{F}_{1,1}^\text{Ising}(z)$ everywhere in terms of $\mathcal{F}_{1,1}^B$ and $\mathcal{G}^{B}$, using \eqref{eq:f11-trans-series}. 
Alternatively, we can consider the 2H2L correlator, where the light field has $\gamma=3/8$, which corresponds to the conformal weight $h_L=1/16$ at $c=1/2$,
and resum it for $\epsilon=4/3$. Again we get
\begin{equation} 
    \mathcal{F}_{1,1}^\text{Ising}(z)  = \mathcal{F}_{1,1}^B\left(z;\epsilon=\frac 43\right)_{\gamma = 3/8} \,,  \qquad z \in R_0 \, , \qquad (\text{2H2L}) \,.
\end{equation}
As in the 4H approach, we can use our results to continue this formula into the full complex $z$-plane by incorporating the Stokes phenomena of 2H2L. 
For both 4H and 2H2L we can extend the outcome of the discussion about discovering $\mathcal{F}_{3,1}$, resumming its trans-series, and fixing the four-point correlator to the Ising model.
The above analysis applies of course to correlators involving the fields $\phi_{2,1}$ in all ${\cal M}_{p+1,p}$ unitary minimal models.

From a $3d$ ``dual'' theory point of view, we are interpolating the 4-$\sigma$ Ising correlator to a family of four-point correlators in holographic non-unitary CFTs, which would correspond 
to a pair of negative energy sources interacting with each other (4H) or to a small perturbation in the presence of a negative energy source (2H2L). 
Correlators of unitary models can then be seen as the ``deep quantum gravity'' regime of amplitudes in exotic $3d$ theories with negative energy sources.

\section{Outlook}
\label{sec:outlook}

To conclude, we point out connections among our results and related topics. In particular, we highlight how to approach unitary theories. As a general remark, our analysis can be useful in the countless contexts involving large parameter expansions of $\hypgeo{2}{1}$.

\subsection{Related topics}
\paragraph{AdS picture} In this paper we have discussed the systematics of the large $c$ expansion of Virasoro blocks associated with the identity operator for a class of correlators involving degenerate operators. 
Due to the identification of the central charge with the $3d$ Newton constant through ${c=3 \ell/2 G_{N}}$~\cite{Brown:1986nw}, such an expansion is then mimicked by the small $G_N$ ones in the semi-classical gravitational theory on $\text{AdS}_{3}$. 
Depending on the nature of the external operators, e.g.\ how their conformal dimensions behave as $c$ becomes large, we expect the Virasoro block of the identity to encapsulate the contribution of the propagation of gravitons in empty AdS, in the defect/excess conical defect geometry or in the BTZ black hole background. 
For the first few orders in the large $c$ expansion, this has been proven using the techniques of the geodesics Witten diagrams in~\cite{Hijano:2015zsa,Hijano:2015qja,CarneirodaCunha:2016zmi,Nishida:2016vds,Chen:2017yia,Beccaria:2015shq} and while challenging to explicitly check, especially in the 4H case, we envision that our analysis can be a testing ground for further holographic explorations.\footnote{
    Even though the CFT that we consider is non-unitary, thus generically not suited for a straightforward holographic description, this identification should still hold.
}
In this context, it would be interesting to understand the role of the maps~\eqref{eq:rmapSec} and~\eqref{eq:vmapSec} in the holographic computations.  An interesting observation is that the presence of forbidden singularities, reflected in the gravitational setup by singularities of the bulk to boundary propagators~\cite{Keski-Vakkuri:1998gmz,Fitzpatrick:2016ive,Collier:2018exn} is not exclusive to the 2H2L case but extends also to the 4H case, predicting that this should persist in the holographic counterpart and that they would be higher order singularities as probing larger and larger orders in the large $c$ expansion. Our findings provide hints on characterizing the geometries dual to 2H2L and 4H correlators in the dual CFT. In particular, they enter in the set of saddle points that we need to sum over to reconstruct the gravitational partition function. This is generically a complicated task and we envision that our results can shed light on some aspects of it.

\paragraph{Chern-Simons description}

Due to its natural split into a holomorphic and anti-holomorphic part, the $\text{SL}(2,\mathbb{R}) \times \text{SL}(2,\mathbb{R})$ Chern-Simons description of $\text{AdS}_{3}$ gravity is particularly useful to directly compute Virasoro blocks in a $1/c$ expansion~\cite{Achucarro:1986uwr, Witten:1988hc,Verlinde:1989ua}. In particular, Wilson lines anchored at the boundary have been recognized to be the correct gauge-invariant observables. 
In this framework, one can construct Virasoro blocks in the $1/c$ expansion using Wilson lines networks~\cite{Fitzpatrick:2016mtp}  making order-by-order computations technically easier than $\text{AdS}_{3}$ Witten diagrams, provided one finds a good regularization prescription for the lines. As discussed in~\cite{Fitzpatrick:2016mtp}, this turns out to be quite a non-trivial task. It would be interesting to explore whether our all-order results in section~\ref{sec:2H2L} might help determine such regularization procedure at higher order in $1/c$ for Wilson line correlators corresponding to a 2H2L configuration. 

\paragraph{More general degenerate operators}
The four-point function of the degenerate $\phi_{2,1}$ satisfies a differential equation which is hypergeometric, leading to the analytical results contained in this work. One might ask what can be concluded concerning resurgent structures for four-point functions of higher degenerates of the form $\phi_{r,1}$ appearing in different minimal models. Even though they satisfy ordinary differential equations of order $r$, we still expect an ansatz of the form~\eqref{eq:2F1-expansion-asymptotic-ansatz} to be valid. In these cases, Stokes jumps are expected to mix multiple solutions of the underlying differential equation. This mixing pattern
can be deduced from knowledge of the saddle actions $S(z)$ corresponding to each Virasoro block. Generalizing the discussion in appendix~\ref{app:2F1-expansion-semiclassic}, it follows that $S'(z)$ solves a polynomial equation of order $r$, suggesting that the cases $r=3,4$ might be still treatable semi-analytically. 

\paragraph{Higher dimensional CFTs} While the structure of CFTs defined in spacetime $d>2$ is different from the $2d$ case, the explicit form of global conformal blocks still involves hypergeometric functions in even spacetime dimensions. It would be interesting to see if the expansions that we found for generic hypergeometric functions are useful in some limiting cases, such as the large dimension expansion. A potential arena for such explorations could be the analytic bootstrap. In particular, the fact that the identity block, either global or Virasoro, is present when expanding a four-point function of (at least pairwise) identical operators in the $s$-channel lead to conclude that specific composite operators (dubbed double twist or double trace) need to be exchanged in the $t$-channel, and the correction to the conformal dimension of such operators for large spin is completely determined by the behaviour of the identity block, together with unitarity~\cite{Fitzpatrick:2012yx, Komargodski:2012ek,Caron-Huot:2017vep} (see~\cite{Fitzpatrick:2014vua,Kusuki:2018wpa,Collier:2018exn} for studies in $d=2$). There has been a plethora of results generalising, extending and complementing these findings (see for instance~\cite{Poland:2018epd,Bissi:2022mrs}), and we envision that the manipulations of the hypergeometric functions that we present in this paper can be handy also in that context. 
\subsection{Unitary case: challenges and directions}
 
Virasoro blocks associated with degenerate operators do not appear in unitary CFTs aside from minimal models.
The most compelling question is whether there are general lessons from our analysis that extend to generic correlators and blocks in \emph{unitary} $2d$ CFTs.

A partial list of key questions to answer are the following: 
does the Stokes phenomenon consistently occur as we move $z$ over the complex plane?
Is there a region in cross-ratio space ($z$) where the asymptotic series of the identity block reconstructs the full block or not? In other words,
does the unitary case resemble the $\epsilon>0$ or the $\epsilon<0$ scenarios we discussed? 
Can we always resolve forbidden singularities with non-perturbative effects discovered through Stokes jumps?
Can resurgence techniques provide predictions on the spectrum of unitary $2d$ CFTs at large central charge, such as in holographic CFTs?
How would they compare with similar bounds found using the modular bootstrap~\cite{Hellerman:2009bu}?\footnote{See~\cite{Afkhami-Jeddi:2019zci} for the state-of-the-art bound.}
Is our analysis generalizable to averages of product of correlators in CFT ensembles, e.g.\ those discussed in~\cite{Maloney:2020nni,Afkhami-Jeddi:2020ezh,Chandra:2022bqq}?

The same Stokes phenomena which resolve the appearance of forbidden singularities are expected to play a role in the problem of information loss when we consider
the correlator in Lorentzian signature. It would be extremely interesting to extend our study to this regime and possibly shed light on this issue.
However, the analysis of the large-order behaviour in unitary cases is challenging.
In~\cite{Benjamin:2023uib}, the authors analysed the leading asymptotic behaviour in the generic unitary case using Zamoldochikov's recursion for Virasoro blocks with four light or four heavy operators in the regime $z\ll 1$. In these cases, the leading singularity in the Borel plane lies away from the positive real axis in this limit. With the insight from our analysis, we know that finite $z$ is crucial to discern the non-perturbative effects.

A promising setup where we might be able to answer these questions is 2H2L unitary correlators. 
The Virasoro blocks entering such correlators have been shown to have an infinite number of forbidden singularities in the leading semi-classical limit, whose position is known analytically~\cite{Fitzpatrick:2014vua,Fitzpatrick:2016ive}, see \eqref{eq:ForbSingFK}. 
The behaviour of generic and unitary heavy-light Virasoro blocks near forbidden singularities was investigated in the numerical work~\cite{Chen:2017yze}. The exact block was computed using Zamolodchikov's $h$-recursion at very high precision and compared with the semi-classical block at large central charge (an analogue of what we reported in figure~\ref{fig:4H-exact-vs-semiclassic} and~\ref{fig:2H2L-exact-vs-semiclassic} for blocks involving degenerate operators). It was shown that the exact Virasoro block deviates drastically from its semi-classical approximation in a neighbourhood of the forbidden singularity, the former being regular there. It was speculated that this deviation could be explained by a web of Stokes and anti-Stokes lines emanating from the forbidden singularities (see also~\cite{Faulkner:2017hll}).

A systematic $1/c$ expansion of heavy-light Virasoro blocks can be in principle obtained from a Zamolodchikov-like recursion relation using the semi-classical heavy-light block as the seed of the recursion~\cite{Fitzpatrick:2015zha,Perlmutter:2015iya}.
It then follows that we should be able to build upon the analysis of~\cite{Chen:2017yze} to explore the resurgence structure of generic and unitary 2H2L correlators, by studying the large order behaviour of the asymptotic series of the Virasoro identity blocks and its Borel transform.
A limitation of this approach is that we don't have access to the full analytic form of each $1/c$ term, but only to a truncated power series in the cross-ratio $z$, thus requiring us to resort to numerical techniques.
Nevertheless, preliminary investigations in this direction have shown evidence that forbidden singularities occur not only in the leading semi-classical term but in each order in the $1/c$ expansion. 
Moreover, by studying the large-order behaviour of the asymptotic series of the identity block, we have gathered numerical evidence that Stokes phenomenon occurs (at least in a neighbourhood of the forbidden singularity closest to $z = 0$). 
From a resurgence point of view, this would imply that the asymptotic series of the identity block cannot be Borel summable over the entire complex $z$-plane, and hence that non-perturbative terms, associated with heavy operator exchange, must be included in unitary heavy-light correlator. 
We aim to report and refine our findings in this direction in a separate publication.

\section*{Acknowledgements}
We thank In$\hat{\rm e}$s Aniceto and Kohei Iwaki for discussions.
We also thank Per Kraus and Adri Olde Daalhuis for correspondence, and Bruno Bucciotti for the preliminary version of the code used to generate figure \ref{fig:appC-figs}.
A.B.\ is partially supported by the Knut and Alice Wallenberg Foundation grant KAW 2021.0170 and Olle Engkvists Stiftelse grant 2180108.
T.R.\ is supported by the ERC-COG grant NP-QFT No.\ 864583 ``Non-perturbative dynamics of quantum fields: from new deconfined phases of matter to quantum black holes'', by the MUR-FARE2020 grant No.\ R20E8NR3HX ``The Emergence of Quantum Gravity from Strong Coupling Dynamic''.
N.D.\ acknowledges the receipt of the joint grant from the Abdus Salam International Centre for Theoretical Physics (ICTP), Trieste, Italy and INFN, sede centrale Frascati, Italy.
Work partially supported by INFN Iniziativa Specifica ST\&FI. 

\appendix
\addtocontents{toc}{\protect\vskip1.5em}

\section{Asymptotic expansion of \texorpdfstring{$\hypgeo{2}{1}$}{2F1}}
\label{app:hypgeom}

The theory of hypergeometric functions is a well-known subject in mathematics, see e.g.\ section 15 of~\cite{NIST:DLMF} for a collection of results.
Since the early work of Watson~\cite{watson1918} several asymptotic expressions of $\hypgeo{2}{1}$ when different combinations of parameters are large have been discussed in the literature, see e.g.~\cite{2003OLDEDAALHUIS,Paris2013AsymptoticsOT,Paris:2013ugv}, and~\cite{TEMME2003441,NIST:DLMF} for a collection of results.\footnote{
    See also~\cite{AokiPub} for a WKB-like approach. This is useful to quickly get the Stokes lines associated with the resulting asymptotic series (see appendix~\ref{app:wkb}), but we did not find it convenient to get the whole explicit asymptotic series.
}
In these works the coefficients of the resulting asymptotic expansion are either given in implicit form or only the first few terms of the expansion are explicitly reported. To the best of our knowledge, a completely explicit form of the coefficients of the whole asymptotic expansion of the $\hypgeo{2}{1}$ entering~\eqref{eq:4H-block-decomposition} and~\eqref{eq:2H2L-block-decomposition} is missing.
This appendix aims to close this gap and discuss some generalizations.

The hypergeometric function $\hypgeo{2}{1}(a,b,c;z)$ is one of the solutions of the second order differential equation 
\begin{equation}\label{eq:2F1-ode}
    z(1-z)f''(z) + \big(c - (a+b+1)z\big)f'(z) - a b f(z) = 0 \,.
\end{equation}
This equation has three regular-singular points at $z=0,1,\infty$. The solution $\hypgeo{2}{1}(a,b,c;z)$ is analytic at $z=0$ in the principal sheet,
and has a branch cut singularity at $z=1$, with a branch cut over the real positive axis from $z=1$ up to infinity. 
Its series expansion in $z$, known as the Gauss series, is convergent in the open disc $|z|<1$.
We are here interested in an expansion for large parameters $a$, $b$ and $c$.

The general problem is to determine, at least in some domain $z \in D \subseteq \mathbb{C}$, the asymptotic series in $\lambda \to +\infty$ of a $\hypgeo{2}{1}(a,b,c;z)$ where the parameters scale as
\begin{equation}\label{eq:2F1-expansion-large-parameters}
    a = a_{0} + a_{1} \lambda \,, \qquad b = b_{0} + b_{1} \lambda \,, \qquad c = c_{0} + c_{1} \lambda \,.
\end{equation}
When $c_{1} \neq 0$ the series is expected to take the semi-classical form
\begin{equation}\label{eq:2F1-expansion-asymptotic-ansatz}
    \hypgeo{2}{1}(a_{0} + a_{1} \lambda,b_{0} + b_{1} \lambda,c_{0} + c_{1} \lambda; z) \sim e^{-\lambda S(z)}A(z) \left(1 + \sum_{n=1}^{+\infty} f_{n}(z)
    \lambda^{-n}\right)\,, \qquad z \in D\,.
\end{equation}
The case $c_{1} = 0$ will not be considered. This case could be in principle worked out with a similar approach to what follows, but needs a slight modification of the semi-classical ansatz. 

This appendix is quite articulated and is then useful to summarize its structure. 
We discuss in sections~\ref{app:thimbles} and~\ref{app:wkb} how to determine the domain $z \in D \subseteq \mathbb{C}$ where the asymptotic expansion~\eqref{eq:2F1-expansion-asymptotic-ansatz} is Borel resummable and where~\eqref{eq:2F1-expansion-asymptotic-ansatz} is a good approximation of the function.
This requires to determine the so-called Stokes and anti-Stokes lines associated to the differential equation~\eqref{eq:2F1-ode}, which are governed by the factor $S(z)$ in~\eqref{eq:2F1-expansion-asymptotic-ansatz}. The explicit form of the factors of $S$ and $A$ is derived in section~\ref{app:2F1-expansion-semiclassic}.
The remaining sections are dedicated to the derivation of the coefficients $f_{n}$ in~\eqref{eq:2F1-expansion-asymptotic-ansatz}.
It is hard to determine a closed analytic form of $f_n$ when all parameters scale with $\lambda$. The case where
\begin{equation}
    a_{1} = 0\,, \qquad b_{1} = 0 \,, \qquad c_{1} \neq 0 \,,
\end{equation}
is instead particularly simple and will be discussed first in section~\ref{app:2F1-expansion-c-large}. 
Fortunately enough, the hypergeometric functions with special parameters in~\eqref{eq:4H-block-decomposition} and~\eqref{eq:2H2L-block-decomposition} entering our $2d$ CFT problem can be reduced to this case thanks to hypergeometric identities that we discuss in sections~\ref{app:2F1-expansion-4H-reduction} and~\ref{app:2F1-expansion-2H2L-reduction}. 
As a by-product, we present in section~\ref{app:2F1-expansion-b-and-c-large} a sketch of the derivation and a closed formula for the coefficients $f_{n}$ in the case
\begin{equation}
    a_{1} = 0 \,, \qquad b_{1} \neq 0\,,  \qquad c_{1} \neq 0  \,.
\end{equation}
For the convenience of the reader, we summarize all the hypergeometric expansions we found in appendix~\ref{app:summary2F1}.

\subsection{Determining Stokes lines I: Steepest descent contours}
\label{app:thimbles}

\begin{figure}[t!]
\centering
\hspace{0.5em}
\begin{subfigure}[c]{0.3\textwidth}
    \includegraphics[width=\textwidth]{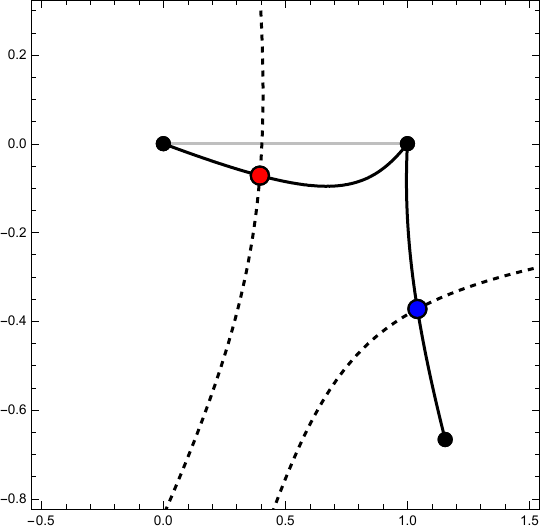}
    \caption{$\rho = 3/4$\label{fig:Thimbles-1}}
\end{subfigure}
\hfill
\begin{subfigure}[c]{0.3\textwidth}
    \includegraphics[width=\textwidth]{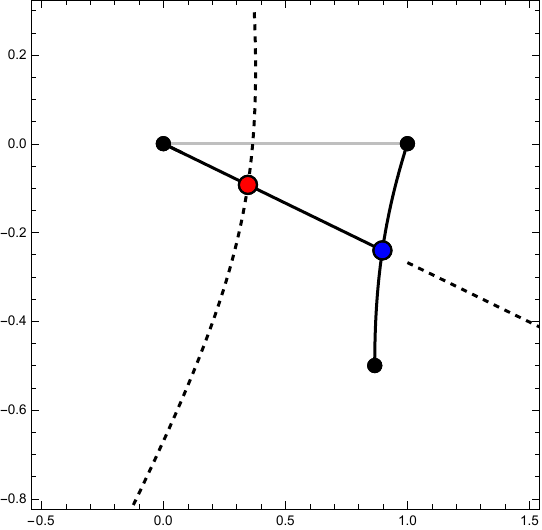}
    \caption{$\rho = 1$\label{fig:Thimbles-2}}
\end{subfigure}
\hfill
\begin{subfigure}[c]{0.3\textwidth}
    \includegraphics[width=\textwidth]{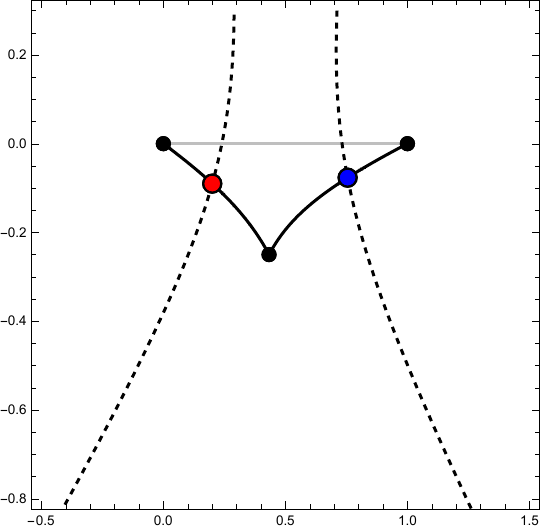}
    \caption{$\rho = 2$\label{fig:Thimbles-3}}
\end{subfigure}
\hspace{0.5em}
\caption{Thimble decomposition for $\mathcal{I}(\lambda, z)$ in~\eqref{eq:2F1_int_rep} in the complex $x$-plane with $a_1=-1$, $b_1=2$, $c_{1}=2$.
We take $\lambda =100$ and $z = \rho e^{i \pi/6}$ with different values of $\rho$, 
corresponding to three points in the trajectory crossing the Stokes line depicted in orange in figure~\ref{fig:Fig12Draft}, with $\rho=1$ being on the line itself.
Black dots correspond to the singularities of $S(x,z)$ at $x=0,1,1/z$ while the red and blue dots are the saddles $x_-$ and $x_+$ respectively.
Solid and dashed lines are the steepest descent contours and steepest ascent contours respectively.
The grey solid line is the original contour of integration.
In~\ref{fig:Thimbles-1}, only the saddle point $x_-$ contributes.
In~\ref{fig:Thimbles-2}, the thimble decomposition is degenerate.
In~\ref{fig:Thimbles-3}, both saddles $x_-$ and $x_+$ contribute.\label{fig:Thimbles}} 
\end{figure}

Whenever a function can be recast into a definite integral form, a steepest descent contour analysis can be used to develop its asymptotic expansions, classify the possible non-perturbative corrections and reveal possible Stokes lines.
See e.g.~\cite{Witten:2010cx} for a detailed review of the method applied to the Airy function.
Hypergeometric functions admit such a representation:
\begin{equation}
    \hypgeo{2}{1}\left( a,b,c ; z \right) = \frac{\Gamma\left( c \right)}{\Gamma\left( b \right)\Gamma\left( b-c \right)} \mathcal{I}(a,b,c,z)\,, \quad \mathcal{I}(a,b,c,z) = \int_0^1 \dif x \, x^{b-1}(1-x)^{c-b-1}(1-z x)^{- a}  \,,
\end{equation}
where the integral converges for $\re(b) > 0$, $\re(b) < \re(c)$ and $z \in \mathbb{C}\setminus[1,+\infty)$.
The integral can be recast in exponential form as follows:
\begin{align}\label{eq:2F1_int_rep}
    \mathcal{I}(\lambda,z) &= \int_{0}^1 \dif x \, x^{b_0-1}(1-x)^{c_0-b_0-1}(1-z x)^{- a_0} \, e^{-\lambda S(x,z)} \,, \nonumber\\
S(x,z) &= - \log \left[   x^{b_1}(1-x)^{c_1-b_1} (1-zx)^{-a_1}  \right]  \,.
\end{align}
The original integration contour $x \in [0,1]$ can be freely deformed in the complex $x$-plane, as long as no integrand singularities are crossed.
In this form, the $\lambda \rightarrow \infty$ expansion takes the form of a saddle-point expansion. 
The saddles of $S(x,z)$ are at
\begin{equation}
    x_{\pm} = \frac{(a_1-b_1)z - c_1 \pm \sqrt{[(a_1-b_1)z - c_1]^2 + 4 b_1(a_1-c_1) z}}{2(a_1-c_1)z}\, .
\end{equation}
From the saddle $x_{\pm}$ emanate steepest descent contours $\mathcal{C}_{\pm}$ where $\re(S)$ monotonically decreases moving away from the saddle.
Not all saddles can contribute simultaneously to the asymptotic integral $\mathcal{I}(\lambda,z)$ at given values of $z$ (for simplicity we assume $z\in \mathbb{C}$ to be the only parameter we vary, with $a_1,b_1,c_1$ kept fixed).
Which saddle contributes is determined by how the original contour of integration can be decomposed into a sum of the steepest descent contours $\mathcal{C}_{\pm}$ (Lefschetz thimbles). 
An example of thimble decomposition as $z$ varies is shown in figure~\ref{fig:Thimbles}. 

Since the imaginary part of the action remains constant on steepest descent contours, this decomposition becomes degenerate when $\mathcal{C}_{\pm}$ intersects, or equivalently:
\begin{equation}\label{eq:StokesCond}
    \im \!\Big(\lambda(S _{-} - S_{+})\Big) = 0, 
 \, \end{equation}
where $S_\pm(z) = S(x_\pm(z),z)$. This is the condition which determines where Stokes lines are in the $z$-plane. We can write this condition more explicitly noticing that 
\begin{equation}\label{eq:derivative_action_diff}
    \frac{\dif}{\dif z} (S_{-}-S_{+}) = \sqrt{Q_0(z)} , \quad  Q_0 = \frac{z^2(a_1-b_1)^2+2z\Big(2 a_1 b_1 - c_1 (a_1+b_1)\Big) +c_1^2}{4z^2(1-z)^2} \, .
\end{equation}
The function $Q_0$ vanishes at $z_{\pm}$, the two roots of its numerator:
\begin{equation}\label{eq:zpmExp}
    z_{\pm} = \frac{c_1(a_1+b_1)- 2a_1 b_1 \pm 2 \sqrt{a_1 b_1(a_1-c_1)(b_1-c_1)}}{(a_1-b_1)^2}\,.
\end{equation}
Noticing that $(S_{-} - S_{+})(z_{\pm})=0$ we can integrate the above equation and obtain the following condition for $z \in \mathbb{C}$ to be a Stokes point for generic $\lambda \in \mathbb{C}$:
\begin{equation}\label{eq:Stokes}
    \im\left\{\lambda \int_{z_{\pm}}^{z} \dif w \sqrt{Q_0(w)} \right\} = 0 \, , \qquad (\text{Stokes lines}) \,.
\end{equation}
For the specific cases of identity blocks in~\eqref{eq:4H-blocks} and~\eqref{eq:2H2L-blocks}, the above condition gives the Stokes line depicted in figure~\ref{fig:Fig12Draft} and figure~\ref{fig:Fig2Draft} respectively.
Anti-Stokes lines are instead defined as
\begin{equation}\label{eq:Anti-Stokes}
    \re\left\{\lambda \int_{z_\pm}^z \dif{w} \sqrt{Q_0(w) }\right\} = 0\,, \qquad (\text{anti-Stokes lines}) \,.
\end{equation}
Stokes lines are lines at which non-perturbative effects ``appear'' and anti-Stokes lines are the lines at which non-perturbative effects become of the same order as perturbative effects.

\subsection{Determining Stokes lines II:  exact WKB}
\label{app:wkb}

\begin{figure}[p]
\centering
\hspace{0.5em}
\begin{subfigure}[c]{0.4\linewidth}
\includegraphics[width=\linewidth]{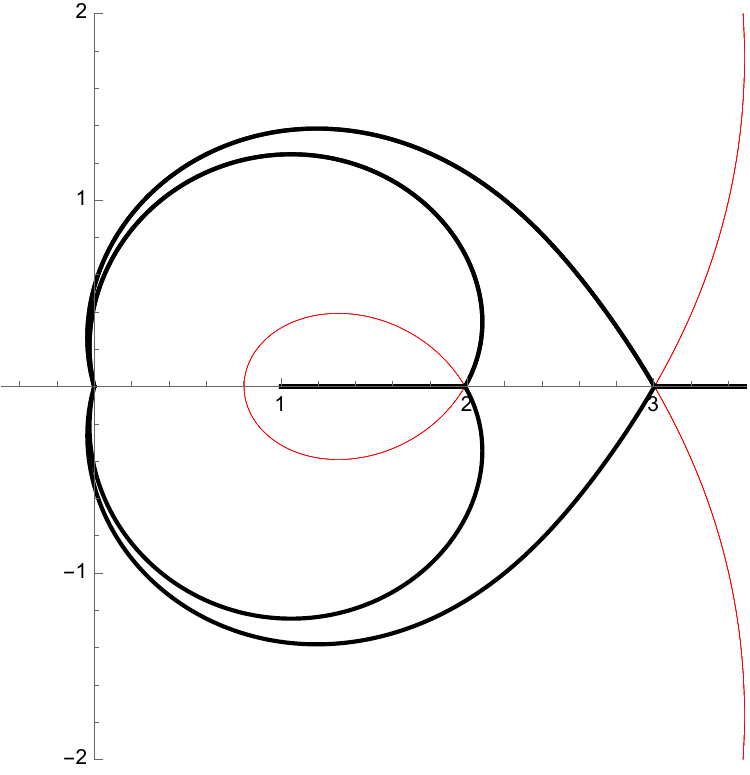}
\caption{$a_1=1$, $b_1=0.018$, $c_1 = 2.4$}
\end{subfigure}
\hfill
\begin{subfigure}[c]{0.4\linewidth}
\includegraphics[width=\linewidth]{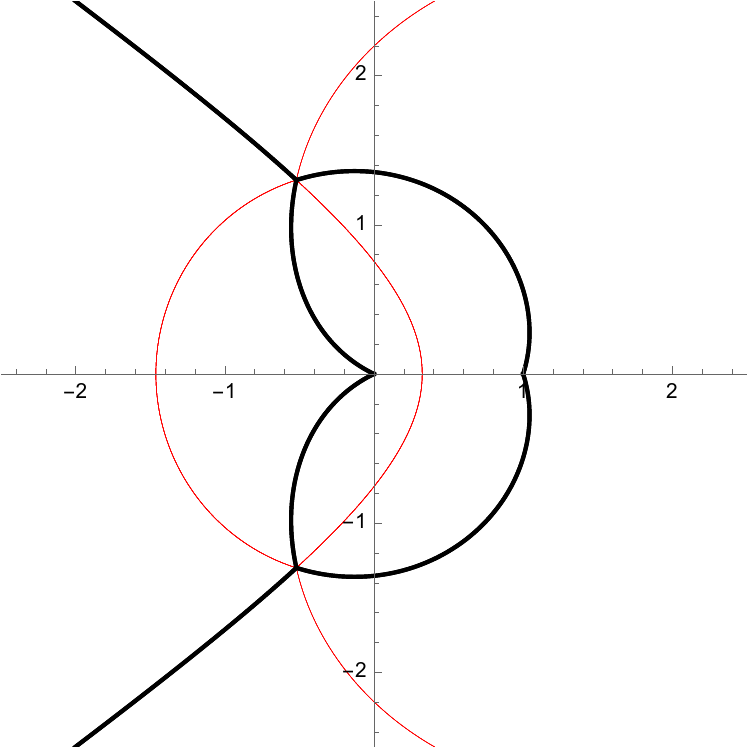}
\caption{$a_1=1.2$, $b_1=2.2$, $c_1=1.4$}
\end{subfigure}
\hspace{0.5em}

\vspace{1em}

\hspace{0.5em}
\begin{subfigure}[c]{0.4\linewidth}
\includegraphics[width=\linewidth]{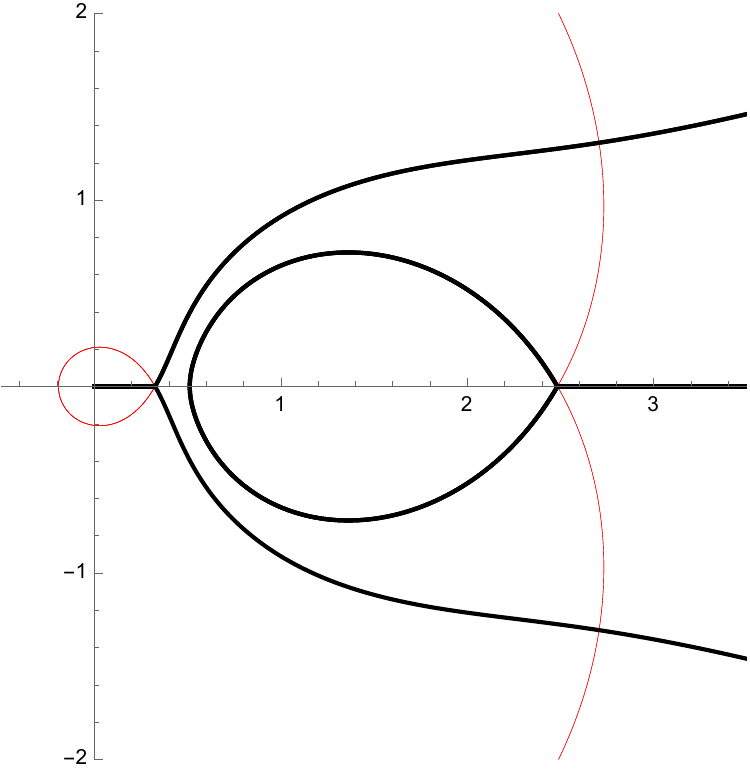}
\caption{$\begin{aligned}
a_1&=-0.95-0.5 i,\\ b_1&=0.05-0.5i,\\ c_1&= -0.9
\end{aligned}$}
\end{subfigure}
\hfill
\begin{subfigure}[c]{0.4\linewidth}
\includegraphics[width=\linewidth]{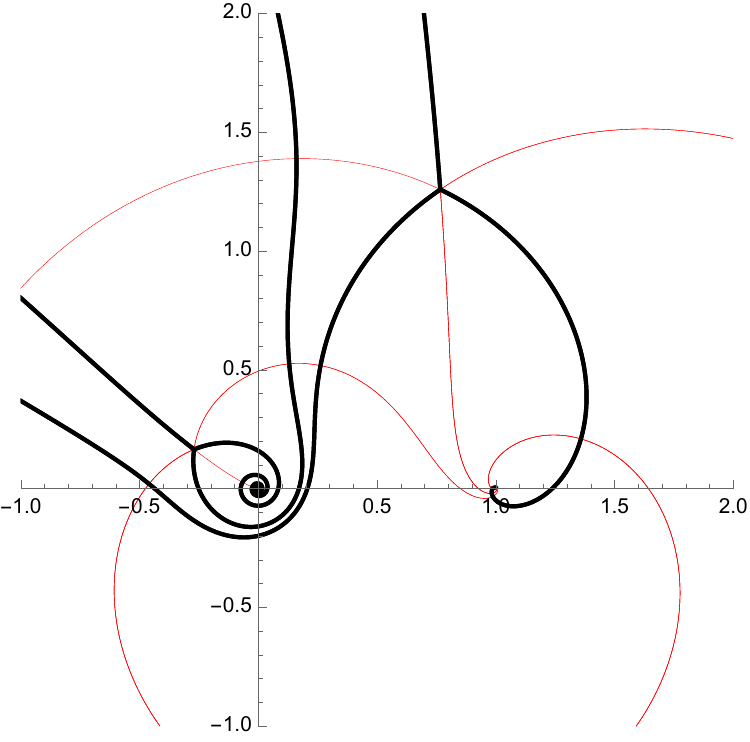}
\caption{$\begin{aligned}
a_1&=-1+0.9i,\\ b_1&=0.03 +0.8 i,\\ c_1&= -0.1+0.7 i
\end{aligned}$}
\end{subfigure}
\hspace{0.5em}
\caption{Four examples of Stokes (black) and anti-Stokes (red) lines for potentials of the form~\eqref{eq:derivative_action_diff}. While it is not visible in the figures, anti-Stokes lines in these examples are closed.\label{fig:appC-figs}} 
\end{figure}

The large parameter expansion of the hypergeometric function and its relation to exact WKB\footnote{
    See e.g.~\cite{INcluster} for a comprehensive math-oriented or~\cite{Bucciotti:2023trp} for a more basic and physics-oriented introductions to exact WKB.
} has been worked out in a series of papers, see e.g.~\cite{aoki2017hypergeometric}.\footnote{
    We thank Kohei Iwaki for drawing our attention to these papers.
} 

The hypergeometric differential equation~\eqref{eq:2F1-ode} can be rewritten in a Schr\"odinger-like form by defining a new function $g$ as
\begin{equation}\label{eq:app2}
    \hypgeo{2}{1}(a,b,c,z) = z^{-\frac c2}(1-z)^{-\frac{1}{2} (a+b-c+1)} g(z)\,,
\end{equation}
such that the linear derivatives terms in~\eqref{eq:2F1-ode} cancel. We then get
\begin{equation}\label{eq:app3}
    - g'' + Q(z) g = 0,
\end{equation}
where 
\begin{equation}\label{eq:app4}
    Q(z) = \frac{z^2\Big((a-b)^2-1\Big)+ 2 z\Big(2ab-c(a+b-1) \Big)+c (c-2) }{4z^2(1-z)^2}
\end{equation}
is a ``potential''.
If we take the limit of large $a,b,c\sim \lambda \rightarrow \infty$, as in~\eqref{eq:2F1-expansion-large-parameters} then~\eqref{eq:app3} turns into a Schr\"odinger equation containing classical and quantum (one and two-loop) potentials, where $1/\lambda$ is identified with $\hbar$.
Then one finds
\begin{equation}\label{eq:app6}
    - \lambda^{-2} g'' + \Big(Q_0(z) + \lambda^{-1} Q_1(z) + \lambda^{-2} Q_2(z)\Big) g = 0\,,
\end{equation}
where the leading potential $Q_0$ is given in~\eqref{eq:derivative_action_diff}.

It is well-known that Stokes lines emanate from the zeros $z_\pm$ of $Q_0$, which are turning points of the potential $Q_0$. 
For each $z_\pm$, Stokes lines and anti-Stokes lines are given by~\eqref{eq:Stokes} and~\eqref{eq:Anti-Stokes}.
The Stokes lines of simple zeros are locally those of the Airy function, with three Stokes lines emanating from the point with a $2\pi/3$ angle between them. 
When the $z_+$ and $z_-$ collide to form a double zero, four Stokes lines emanate from the point with a $\pi/2$ angle between them. 
For non-singular $Q_0$, Stokes lines emanating from turning points can go to infinity or end at another turning point.
A Stokes line ending at another turning point can be resolved by assigning a small phase to the parameter $\lambda$.
If $Q_0$ has singularities, Stokes lines emanating from turning points can also end at the singular points. 
The potential~\eqref{eq:derivative_action_diff} has two order-two poles at $z=0$ and $z=1$.
The way Stokes lines approach a singular order-two pole depends on the value of the residue of $\sqrt{Q_0}$~\cite{INcluster}.
The Stokes lines approach the singular point radially if the residue is real, form closed trajectories around the singular point if it is purely imaginary, and spiral around it for a generic complex residue.
For illustration we report in figure~\ref{fig:appC-figs} the Stokes and anti-Stokes lines as given by~\eqref{eq:Stokes},~\eqref{eq:Anti-Stokes} for a sample of choices of parameters $(a_1,b_1,c_1)$.

\subsection{Saddle and prefactor}
\label{app:2F1-expansion-semiclassic}

The form of the two factors $S_\pm$ (from now on loosely dubbed ``saddles'') can easily be derived from the results in sections~\ref{app:thimbles} or~\ref{app:wkb}.
From now on we consider for simplicity only the case of \emph{real} parameters and focus on the solution which is regular as $z\to 0$ and corresponds to $\hypgeo{2}{1}(a,b,c;z)$, the other solution corresponding instead to the solution $z^{1-c}\hypgeo{2}{1}(a-c+1,b-c+1,2-c;z)$. For $c_1>0$ 
the value of $S(z)$ for $\hypgeo{2}{1}(a,b,c;z)$ is\footnote{Without loss of generality, we can assume $a_{1} \leq b_{1}$ since the problem is symmetric in $(a,b)$ exchange.}
\begin{equation}\label{eq:2F1-expansion-Sminus-generic-params}
    \begin{aligned}
        S(z) &= \log \left[
        \frac{
        ((a_{1}-b_{1})z + c_{1}-2a_{1} +2 u(z))^{a_{1} +b_{1} -c_{1} } 
        ((a_{1}-b_{1})z + c_{1} +2 u(z))^{c_{1} } 
        }{
        \left((a_{1} -b_{1} )^2 z + 2 a_{1}  b_{1} -c_{1} 
        (a_{1} +b_{1} ) +2 (a_{1} -b_{1} ) u(z)\right)^{b_{1} }
        }\right] - S_{0}\, , \\[0.5em]
        S_{0} &= \log\left[2^{a_{1} } c_{1} ^{c_{1} } (b_{1}  (a_{1} -c_{1} ))^{-b_{1} } (c_{1} -a_{1} )^{a_{1} +b_{1} -c_{1} }\right] \, ,
    \end{aligned}
\end{equation}
with $S_{0}$ fixed by the boundary condition $S(0) = 0$, and 
\begin{equation}
    u(z) = \sqrt{\frac{c_{1}^2}{4} + \left(a_{1} b_{1}-\frac{1}{2}c_{1}(a_{1}+b_{1}) \right)z + \frac{(a_{1}-b_{1})^2}{4} z^2} \,.
\end{equation}
At next-to-leading order in $\lambda$ we can determine the factor $A(z)$, given by 
\begin{equation}\label{eq:2F1-expansion-Aminus-generic-params}
    \begin{aligned}
        A(z) 
        &= 
        A_{0} \frac{(1-z)^{c_{0}-a_{0}-b_{0}}}{\sqrt{u(z)}} 
        \left((a_{1} -b_{1})^2 z + 2 a_{1}  b_{1}-c_{1}  (a_{1} +b_{1} ) +2 (a_{1} -b_{1} ) u(z)\right)^{\frac{b_{0}-a_{0}}{2}} \\
        &\quad \times \left((a_{1} ^2 + b_{1} ^2-c_{1}  (a_{1} +b_{1} )) z +2 a_{1}  b_{1} +c_{1} ^2 -c_{1}  (a_{1} +b_{1} ) + 2 (a_{1} +b_{1} -c_{1} ) u(z)\right)^{\frac{a_{0}+b_{0}-c_{0}}{2}} \\
        &\quad \times \left((2a_{1}b_{1} - c_{1}(a_{1}+b_{1}))z + c_{1}^{2} + 2 c_{1} u(z)\right)^{\frac{1-c_{0}}{2}} \, , \\[0.5em]
        A_{0} &= 2^{c_{0}-b_{0}-1} c_{1}^{c_{0}-\frac{1}{2}} a_{1}^{\frac{c_{0}-a_{0}-b_{0}}{2}} b_{1}^{b_{0}-\frac{c_{0}}{2}} (a_{1}-c_{1})^{\frac{a_{0}-b_{0}}{2}} \, ,
    \end{aligned}
\end{equation}
with $A_{0}$ fixed by the boundary condition $A(0) = 1$.

When $a_{1} = 0$ the previous formulas appear singular but the solutions are still well-defined and dramatically simplify:
\begin{equation}\label{eq:2F1-expansion-semiclassic-a1-zero}
    S(z) = 0 \,, \qquad A(z) = \left(1-\frac{b_{1} z}{c_{1}}\right)^{-a_{0}}\,.
\end{equation}
When both $a_{1} = 0$ and $b_{1} = 0$, both the saddle and the prefactor become trivial:
\begin{equation}
    S(z) = 0 \,, \qquad A(z) = 1 \,.
\end{equation}

\subsection{Expansion of \texorpdfstring{$\hypgeo{2}{1}(a,b,c + \lambda; z)$}{2F1(a,b,c+lambda;z)}}
\label{app:2F1-expansion-c-large}

In this section we provide a formula for the coefficients $f_{n}$ of the asymptotic expansion~\eqref{eq:2F1-expansion-asymptotic-ansatz} in the case where only the third parameter scales with $\lambda$, namely $a_{1} = b_{1} = 0$ in~\eqref{eq:2F1-expansion-large-parameters}. 
Without loss of generality, we can set $c_{1} = 1$. To lighten the notation we will drop the subscript ``0'' from the coefficients in~\eqref{eq:2F1-expansion-asymptotic-ansatz}.

An implicit form of this expansion is known, see e.g.\ section 15.12(iii) of~\cite{NIST:DLMF}, but the coefficients~\eqref{eq:2F1-expansion-asymptotic-ansatz} are not given in a closed form, but rather in terms of a generating function.

An explicit form of the coefficients can be obtained instead by manipulating the Gauss series and expanding the Pochhammer symbols.
To this end, we can use the identity
\begin{equation}
    \frac{1}{(\lambda)_{k}} = \sum_{n=k}^{+\infty} (-1)^{n+k} \stirlingII{n-1}{k-1} \lambda^{-n} \,, \qquad |\lambda|> k-1 \,, \qquad k \geq 1\,,
\end{equation}
where $\stirlingII{n}{k}$ are the Stirling numbers of the second kind, or rather the following generalization
\begin{equation}\label{eq:2F1-expansion-Pochhammer-expansion}
    \frac{1}{(c + \lambda)_{k}} = \sum_{n=k}^{+\infty} (-1)^{n+k} \stirlingIIpoly{n-1}{k-1}{c} \lambda^{-n} \, , \qquad |\lambda| > \max_{0 \leq p \leq k-1}|c+p|\, , \qquad k \geq 1 \, ,
\end{equation}
where we have introduced the Stirling polynomials of the second kind
\begin{equation}
    \stirlingIIpoly{n}{k}{x} = \sum_{m=k}^{n} \stirlingII{m}{k} \binom{n}{m} x^{n-m} \,.
\end{equation}
Plugging~\eqref{eq:2F1-expansion-Pochhammer-expansion} into the Gauss series gives
\begin{equation}\label{eq:2F1-expansion-gauss-series-expand-Pochhammer}
    \hypgeo{2}{1}\left(a,b,c+\lambda; z\right) 
    \sim 1 + \sum_{k=1}^{+\infty}\sum_{n=k}^{+\infty} \frac{(a)_{k} (b)_{k}}{k!} z^{k} \, (-1)^{n+k} \stirlingIIpoly{n-1}{k-1}{c} \lambda^{-n} \,.
\end{equation}
Exchanging the order of the two sums, we get, as $\lambda \to +\infty$:
\begin{equation}\label{eq:2F1-expansion-large-c-asymptotic}
    \hypgeo{2}{1}\left(a,b,c+\lambda; z\right) \sim 1 + \sum_{n=1}^{+\infty} f_{n}(a,b,c;z) \lambda^{-n}\,, \qquad z \in \mathbb{C} \setminus [1,+\infty)\,,
\end{equation}
where
\begin{equation}\label{eq:2F1-expansion-large-c-asymptotic-coefficients}
    f_{n}(a,b,c;z)= \sum_{k=1}^{n} (-1)^{n+k} \stirlingIIpoly{n-1}{k-1}{c} \frac{(a)_{k}(b)_{k}}{k!} z^{k} \, .
\end{equation}
The sums in~\eqref{eq:2F1-expansion-gauss-series-expand-Pochhammer} are both convergent, but the exchange of the two is merely formal and makes the sum over $n$ in~\eqref{eq:2F1-expansion-large-c-asymptotic} asymptotic. 
In fact, applying the identity~\eqref{eq:2F1-expansion-Pochhammer-expansion} for all $k \in \mathbb{N}$ would imply a zero radius of convergence in $1/\lambda$.
The matching of the expansions~\eqref{eq:2F1-expansion-large-c-asymptotic} and the known one, e.g.\ equation 15.12.3 of~\cite{NIST:DLMF}, can be numerically verified to any order.
Note that this asymptotic series actually holds in a much larger domain in $z$ than the Gauss series we started from.
This explicit form of the expansion in $\lambda$ is particularly useful since each term of the asymptotic series is a degree $n$ polynomial in $z$ with coefficients known in closed form.
This will make it feasible to carry out the resurgence analysis of such asymptotic series in appendix~\ref{app:resurgence}.

In the case where $c = 0$, the Stirling polynomials reduce to ordinary Stirling numbers and thus we have the following simpler asymptotic expansion as $\lambda \to +\infty$:
\begin{equation}\label{eq:2F1-expansion-large-c-no-shift}
    \hypgeo{2}{1}\left(a,b,\lambda; z\right)\sim 1 + \sum_{n=1}^{+\infty} f_{n}(a,b;z) \lambda^{-n} \, , \qquad
    f_{n}(a, b; z) = \sum_{k=1}^{n} (-1)^{n+k} \stirlingII{n-1}{k-1} \frac{(a)_{k}(b)_{k}}{k!} z^{k} \, .
\end{equation}
The formula holds for $z \in \mathbb{C} \setminus [1,+\infty)$ as can be deduced by analyzing the web of Stokes and anti-Stokes lines that can be obtained as in~\ref{app:wkb}.
This asymptotic expansion also holds for $\lambda \to -\infty$, in the domain $\{z \in \mathbb{C} \setminus (-\infty,0] \ \colon \re{(z)} < 1/2\}$, as can be understood by studying the Stokes and anti-Stokes lines of the associated differential equation studied in~\ref{app:wkb}.
See appendix~\ref{app:resurgence-eps-negative} for further discussion.

\subsection{Reduction of \texorpdfstring{$\hypgeo{2}{1}(-\lambda,1+\lambda,2+2\lambda;z)$}{2F1(-lambda,1+lambda,2+2lambda;z)}}
\label{app:2F1-expansion-4H-reduction}

The asymptotic expansion as $\lambda \to +\infty$ of $\hypgeo{2}{1}(-\lambda,1+\lambda,2+2\lambda;z)$, corresponding to the $\hypgeo{2}{1}$ entering the identity block in the 4H correlator~\eqref{eq:4H-block-decomposition}, can be reduced to the case discussed in section~\ref{app:2F1-expansion-c-large} with a non-trivial change of variable.

The saddle $S(z)$ and the prefactor $A(z)$ are obtained from~\eqref{eq:2F1-expansion-Sminus-generic-params} and~\eqref{eq:2F1-expansion-Aminus-generic-params} respectively, with $a_{0} = 0$, $a_{1} = -1$, $b_{0} = 1$, $b_{1} = 1$, $c_{0} = 2$ and $c_{1} = 2$.

The idea is to solve the following auxiliary problem. $\hypgeo{2}{1}(-\lambda,1+\lambda,2+2\lambda;z)$ is a solution of the hypergeometric differential equation~\eqref{eq:2F1-ode} with $a = -\lambda,b=1+\lambda,c=2+2\lambda$.
We factor out the leading asymptotic and define
\begin{equation}
    g(w(z)) = \frac{f(z)}{e^{-\lambda S(z)} A(z)},
\end{equation}
where $w$ is a map to be determined. The function $g$ satisfies a second-order differential equation which can be put in the the form
\begin{equation}\label{eq:2F1-ode-g}
    w(1-w) \partial_{w}^{2} g + H_{1}(z) \partial_{w} g - H_{0}(z) g = 0 \,,
\end{equation}
where $H_{1}, H_{0}$ are functions of $z$ and the map $w(z)$, given by
\begin{align}
    H_{1}(z) &= \frac{w(z)(1-w(z))}{z(1-z) w'(z)^{2}}\left[2 \lambda u(z) w'(z) + \left(\frac{1}{2}-z + \frac{1-2z}{2u(z)^2} + u(z)\right)w'(z)+z(1-z)w''(z)\right] \, ,\nn \\[0.5em]
    H_{0}(z) &= \frac{15}{16} \frac{w(z)(1-w(z))}{u(z)^{4} w'(z)^{2}} \, ,
\end{align}
with
\begin{equation}\label{eq:uDef}
u(z) = \sqrt{1-z+z^2}\,.
\end{equation}
The goal is to find a map $w(z)$ such that~\eqref{eq:2F1-ode-g} is again a hypergeometric differential equation where now only the ``third'' parameter scales with $\lambda$ (namely $a_{1} = b_{1} = 0$ and $c_{1} \neq 0$ in~\eqref{eq:2F1-expansion-large-parameters}).
More precisely we have to find $w(z)$ and constants $(A, B, C_{0}, C_{1})$ such that
\begin{align}
    H_{1}(z) &= C_{0} + C_{1} \lambda - (A+B+1) w(z) \label{eq:2F1-expansion-H1-constraint}\,, \\
    H_{0}(z) &= A B \,. \label{eq:2F1-expansion-H0-constraint} 
\end{align}
It is highly non-trivial that this map exists. If a solution exists, we will have derived the identity
\begin{equation}
    \hypgeo{2}{1}(-\lambda,1+\lambda,2+2\lambda;z) = e^{-\lambda S(z)}A(z)\hypgeo{2}{1}(A, B, C_{0} + C_{1} \lambda; w(z))\,, \qquad z \in D\,,
\end{equation}
at least in some domain $D \subseteq \mathbb{C}$ where the asymptotic ansatz holds.
We have also assumed that the map $w$ behaves as $w(z) \sim \alpha z + O(z^2)$ for small $z$, so that we can correctly match the behaviour as $z \to 0$ of the $\hypgeo{2}{1}$ on both sides.
Otherwise one needs to pick another appropriate solution of the hypergeometric differential equation.

The problem is in principle overconstrained and the existence of a solution is related to the particular way in which the parameters in the $\hypgeo{2}{1}$ scale with $\lambda$. 
Remarkably, for the case at hand a map $w(z)$ does exist and its form can be found in the following way.
Isolating from~\eqref{eq:2F1-expansion-H1-constraint} the term proportional to $\lambda$ gives the following first order differential equation for $w(z)$:
\begin{equation}\label{eq:2F1-expansion-4H-wprime}
    w'(z) =\frac{2 u(z)}{C_{1} z(1-z)} w(z)(1-w(z)) \, .
\end{equation}
We don't solve this equation directly but rather use this relation in $H_{0}$ to remove the $w'(z)$ term.
Then the constraint~\eqref{eq:2F1-expansion-H0-constraint} reduces to a simple second-degree polynomial equation in $w$
\begin{equation}\label{eq:2F1-expansion-4H-w-map-wpm}
    w(z)^{2} - w(z) + \frac{15}{64} \frac{C_{1}^{2}}{A B} \frac{z^{2}(1-z)^{2}}{u(z)^{6}} = 0 \quad \Rightarrow \quad
    w_{\pm}(z) = \frac{1}{2}\left(1\pm\sqrt{1-\frac{15}{16} \frac{C_{1}^{2}}{A B} \frac{z^{2}(1-z)^{2}}{u(z)^{6}}}\right) \, .
\end{equation}
Plugging these solutions into the differential equation~\eqref{eq:2F1-expansion-4H-wprime} and demanding that the constraint is solved for all $z$ implies\footnote{
    Note that while the solutions $w_{\pm}$ in~\eqref{eq:2F1-expansion-4H-w-map-wpm} depend on $C_{1}^{2}$, the constraint~\eqref{eq:2F1-expansion-4H-wprime} depends on $C_{1}$.
    For this reason, picking either $w_{+}$ or $w_{-}$ requires a different sign of $C_{1}$ to solve the constraint.\label{foot:wpm}
}
\begin{equation}
    w_{\pm} \quad \colon \quad A B = \frac{5}{36} \, , \qquad  C_{1} = \mp 1 \, .
\end{equation}
Finally, plugging the relations~\eqref{eq:2F1-expansion-4H-wprime} and~\eqref{eq:2F1-expansion-4H-w-map-wpm} in~\eqref{eq:2F1-expansion-H1-constraint} results in a simple equation:
\begin{equation}
    (A+B-1)w(z) + 1 - C_{0} + \frac{1}{2} C_{1} = 0 \quad \Rightarrow \quad A+B = 1 \, , \quad C_{0} = 1 +  \frac{1}{2} C_{1}\, .
\end{equation}
We have therefore found two solutions to the constraints~\eqref{eq:2F1-expansion-H1-constraint} and~\eqref{eq:2F1-expansion-H0-constraint}.
Demanding that $w(z) \sim \alpha z + O(z^2)$ for small $z$ selects the solution\footnote{
    The other solution is given by $w_{+}(z) = 1-r(z)$, with $C_1=-1$, $C_0=1/2$, and the same values of $A$ and $B$.
    In both cases the solutions are symmetric under $(A,B)$ exchange.
}
\begin{equation}
    r(z) = w_{-}(z) = \frac{1}{2}+ \frac{(z+1)(2z-1)(2-z)}{4 u(z)^{3}}\, , \quad  (A, B) = \left(\frac{1}{6},\frac{5}{6}\right) \, , \ \  C_{0} = \frac{3}{2} \, , \ \   C_{1} = 1 \,.
\end{equation}
We have thus established the identity
\begin{equation}\label{eq:2F1-expansion-4H-identity}
    \hypgeo{2}{1}(-\lambda, 1+\lambda, 2+2\lambda) = e^{-\lambda S(z)} A(z) g(r(z)) \, ,
    \quad g(r) = \hypgeo{2}{1}\left(\frac{1}{6},\frac{5}{6},\frac{3}{2}+\lambda; r\right) \, ,
    \quad z \in D_{0}\,.
\end{equation}
Note that the saddle and the prefactor can be written more conveniently in terms of $r(z)$ as
\begin{equation}
    S(z) = \frac{1}{2} \log \left(\frac{27}{16} \frac{z^2}{(1-z)^2}\frac{1-r(z)}{r(z)}\right) \, , \qquad
    A(z) = \frac{e^{-\frac{1}{2}S(z)}}{\sqrt{u(z)}} \, .
\end{equation}

\begin{figure}[t!]
    \centering
    \begin{tikzpicture}[scale=2.5]
    \draw[->] (-1.5,0) -- (2.5,0);
    \draw[->] (0,-1.7) -- (0,1.7);
    \foreach \x in {-1,-0.5,0.5,1.5,2} {
        \draw (\x,0) -- (\x,-0.03) node[below,scale=0.5] {$\x$} ;
    }
    \foreach \y in {-1.5,-1,-0.5,0.5,1,1.5} {
        \draw (0,\y) -- (-0.03,\y) node[left,scale=0.5] {$\y$} ;
    }
    \draw[branch, thick, decorate, decoration={snake, segment length=5, amplitude=1.5}] (0.5,{-sqrt(3)/2}) arc[start angle=-60, end angle=60, radius=1];
    \draw[path, thick, decorate, decoration={snake, segment length=5, amplitude=1.5}] (1,0) -- (2.45,0);
    \draw[stokes, thick, decorate, decoration={snake, segment length=5, amplitude=1.5}] (0.5,{sqrt(3)/2}) -- (0.5,1.7); 
    \draw[stokes, thick, decorate, decoration={snake, segment length=5, amplitude=1.5}] (0.5,{-sqrt(3)/2}) -- (0.5,-1.7); 
    \filldraw[black] (0.5,{sqrt(3)/2}) circle (0.02) node[above left] {$z_+$};
    \filldraw[black] (0.5,{-sqrt(3)/2}) circle (0.02) node[below left] {$z_-$};
    \filldraw[black] (1,0) circle (0.02) node[below right] {$1$};
    \node at (0.5,0.25) {$D_0$};
    \node at (1.5,1) {$D_+$};
    \node at (1.5,-1) {$D_-$};
    \node[draw=none, inner sep=5pt, append after command={
        \pgfextra{\draw (\tikzlastnode.south west) -- (\tikzlastnode.south east);}
        \pgfextra{\draw (\tikzlastnode.south west) -- (\tikzlastnode.north west);}
    }] at (2.3, 1.5) {$z$};
    \end{tikzpicture}
    \caption{Branch cuts of the hypergeometric function when written using the variable $r(z)$.
    The points $z_\pm$ turn out to be analytic points and we are only left with the discontinuity at $z=1$ (orange branch cut).\label{fig:hypgeo-identity-largeb2-branchcuts}}
\end{figure}
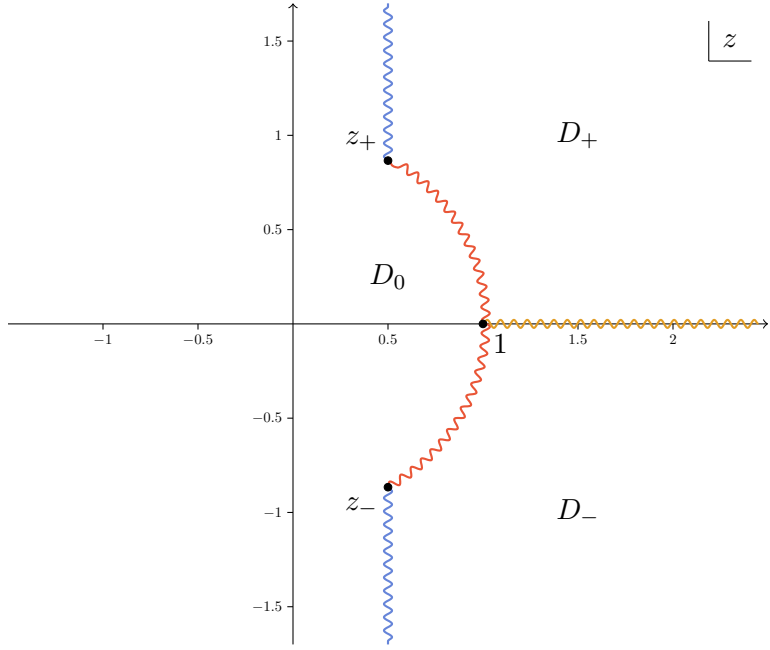

\paragraph{Comments on the analytic structure}
Let us discuss in more detail the analytic structure stemming from~\eqref{eq:2F1-expansion-4H-identity}, and specify in particular the domain $D_{0}$ in which it is valid.
We restrict our analysis to the principal sheet in the $z$-plane.
We have possible branching points where $u(z)$ vanishes, at $z=z_\pm = \exp(i \pi/3)$,
branching points of the logarithm in $S$, and branching points of $g(r)$ when $r>1$.
We report in figure~\ref{fig:hypgeo-identity-largeb2-branchcuts} the branching points and branch cuts which result in the complex $z$-plane. From $z=1$, where $r=1$, the two red branch cuts represent the points where $r$ is real and greater than one, reaching infinity at $z_\pm$.
The two blue branch cuts are instead square root branch cuts of the function $u(z)$, which induce a transformation $r\to 1-r$. 
Despite the two branch cuts looking quite different, they act equally on the whole function, implying that $z_\pm$ are actually analytic regular points, as expected from the left-hand side of~\eqref{eq:2F1-expansion-4H-identity}.
Finally, we have the orange branch cut from $z=1$ up to infinity, which is invisible in the principal sheet of $g(r)$, but is a branch cut of its continuation to the nearby Riemann sheets. 
The $z$-plane is then split into three regions, denoted $D_p$ in the figure, $p=0,\pm$. 
One can verify that over the whole principal sheet in the $z$-plane, we have
\begin{equation}\label{eq:equivGlobal}
    \hypgeo{2}{1}(-\lambda,1+\lambda,2+2\lambda; z) = e^{-\lambda S(z)} A(z) g^{(p)}(r(z)) \, , \qquad z\in D_p\,,
\end{equation}
where $g^{(0)}(r) = g(r)$ represents the $\hypgeo{2}{1}$ on the right-hand side of~\eqref{eq:2F1-expansion-4H-identity} on the principal Riemann sheet in the $r$-plane, while
\begin{equation}\label{eq:gRsheets}
g^{(\pm)}(r) = g(r) \pm \operatorname{Disc}[g(r)]\,,\qquad \qquad \operatorname{Disc}[g(r)] = i \left(\frac{r}{r-1}\right)^{-\frac{1}{2}-\lambda}g(1-r)\,,
\end{equation}
is the function $g$ in the two Riemann sheets adjacent to the principal one. 
One can see from~\eqref{eq:gRsheets} the appearance of the orange branch cut for $g^{(\pm)}$, which eventually maps to the ordinary branch cut of the $\hypgeo{2}{1}$ in $z$ in the left-hand side of~\eqref{eq:equivGlobal}.
Region $D_0$ and \eqref{eq:gRsheets} reappear in the resurgence analysis of the asymptotic series, as seen in the main text around~\eqref{eq:D0-def}.

\subsection{Reduction of \texorpdfstring{$\hypgeo{2}{1}(2\gamma,1+\lambda,2+2\lambda;z)$}{2F1(2 gamma,1+lambda,2+2lambda;z)}}
\label{app:2F1-expansion-2H2L-reduction}

The asymptotic expansion as $\lambda \to +\infty$ of $\hypgeo{2}{1}(2\gamma,1+\lambda,2+2\lambda;z)$, the $\hypgeo{2}{1}$ entering the identity block in the 2H2L correlator~\eqref{eq:2H2L-block-decomposition}, can also be reduced to the case discussed in section~\ref{app:2F1-expansion-c-large} with a non-trivial change of variable.
The procedure is identical to what we described in section~\ref{app:2F1-expansion-4H-reduction} so we will be more brief.

The saddle and prefactor can be read from~\eqref{eq:2F1-expansion-semiclassic-a1-zero} with $a_{0} = 2\gamma$, $b_{1} = 1$ and $c_{1} = 2$.
The functions $H_{0}, H_{1}$ can be computed straightforwardly and will not be reported.
Isolating from the $H_{1}$ equation the term proportional to $\lambda$ gives the following first order differential equation for $w(z)$:
\begin{equation}\label{eq:2F1-expansion-2H2L-wprime}
    w'(z) =\frac{(2-z)}{C_{1} z(1-z)} w(z)(1-w(z)) \, .
\end{equation}
Using this relation into the constraint~\eqref{eq:2F1-expansion-H0-constraint}, we reduce to an algebraic equation for $w$:
\begin{equation}\label{eq:2F1-expansion-2H2L-w-map-wpm}
    \resizebox{\textwidth}{!}{\(\displaystyle
    w(z)^{2} - w(z) + 2\gamma(2\gamma+1)\frac{C_{1}^{2}}{A B} \frac{z^{2}(1-z)}{(2-z)^{4}} = 0 \quad \Rightarrow \quad
    w_{\pm}(z) = \frac{1}{2}\left(1\pm\sqrt{1-8\gamma(2\gamma+1)\frac{C_{1}^{2}}{A B} \frac{z^{2}(1-z)}{(2-z)^{4}}}\right) \, .
    \)}
\end{equation}
Plugging these solutions into the differential equation~\eqref{eq:2F1-expansion-2H2L-wprime} and asking for the constraint to be solved for all $z$ and generic $\gamma$ implies\ap{\ref{foot:wpm}}
\begin{equation}
    w_{\pm} \quad \colon \quad  A B = \frac{\gamma(2\gamma+1)}{2} \, , \qquad   C_{1} = \mp 1 \, .
\end{equation}
Finally, we can use the relation~\eqref{eq:2F1-expansion-2H2L-wprime} and~\eqref{eq:2F1-expansion-2H2L-w-map-wpm} in~\eqref{eq:2F1-expansion-H1-constraint} to fix $A$, $B$, $C_{0}$ and $C_{1}$.
Demanding that $w(z) \sim \alpha z + O(z^2)$ for small $z$ selects the solution\footnote{
    The other solution is given by $w_{+}(z) = 1-v(z)$, with $C_1=-1$, $C_0=2\gamma$, and the same values of $A$ and $B$.
}
\begin{equation}
        v(z) = w_{-}(z) = \frac{z^{2}}{(2-z)^{2}} \, , \quad
         (A, B) = \left(\gamma,\gamma+\frac{1}{2}\right) \, , \ \
         C_{0} = \frac{3}{2} \, , \ \
         C_{1} = 1 \, .
\end{equation}
We have thus established the identity
\begin{equation}\label{eq:2F1-expansion-2H2L-identity-lambda}
    \hypgeo{2}{1}\left(2\gamma, 1+\lambda, 2+2\lambda\right) = A(z) g\big(v(z)\big) \, ,
    \quad g(v) = \hypgeo{2}{1}\left(\gamma,\frac{1}{2}+\gamma,\frac{3}{2}+\lambda; v\right) \, ,
    \quad z \in D\,,
\end{equation}
where
\begin{equation}
    A(z) = \left(1-\frac{z}{2}\right)^{-2\gamma} \, .
\end{equation}
In this case, the analytic structure is much simpler than the one discussed in section~\ref{app:2F1-expansion-4H-reduction}.
The region where~\eqref{eq:2F1-expansion-2H2L-identity-lambda} holds is the entire cut complex plane $D = \mathbb{C} \setminus[1,+\infty)$.

Note that the relation~\eqref{eq:2F1-expansion-2H2L-identity-lambda} can also be obtained using identities between hypergeometric functions, see e.g.\ equation 15.8.13 of~\cite{NIST:DLMF}.

\subsection{Expansion of \texorpdfstring{$\hypgeo{2}{1}(a, b_{0} + b_{1} \lambda , c + \lambda; z)$}{2F1(a,b0 + b1 lambda,c + lambda;z)}}
\label{app:2F1-expansion-b-and-c-large}

In this section, we provide a formula for the coefficients $f_{n}$ of the asymptotic expansion~\eqref{eq:2F1-expansion-asymptotic-ansatz} in the case where $a_{1} = 0$ in~\eqref{eq:2F1-expansion-large-parameters} while $b_{1},c_{1} \neq 0$.
Without loss of generality, we can set $c_{1} = 1$.
To lighten the notation we will drop the subscript ``0'' from the coefficients in $a_{0}$ and $c_{0}$.
This generalization is not used in the main text.

In the case where $a,b_{0},b_{1},c$ are generic, the strategy devised in section~\ref{app:2F1-expansion-4H-reduction} does not work, namely, the map $w(z)$ does not exist unless the parameters satisfy particular relations. 

Nevertheless, it's possible to plug the asymptotic ansatz (the saddle and prefactor are given in~\eqref{eq:2F1-expansion-semiclassic-a1-zero})
\begin{equation}\label{eq:2F1-expansion-b-and-c-large-asymptotic-form}
    f(z) = \hypgeo{2}{1}(a, b_{0} + b_{1} \lambda , c + \lambda; z) \sim (1-b_{1} z)^{-a} \left(1+\sum_{n=1}^{+\infty} f_{n}(z) \lambda^{-n}\right),
\end{equation}
into the differential equation~\eqref{eq:2F1-ode} to obtain a differential recursion formula for the coefficients $f_{n}$.
The recursion gives $f_{n}'(z)$ in terms of $f_{n-1}(z)$, $ f_{n-1}'(z)$ and $f_{n-1}''(z)$. Starting from the seed $f_{0}(z) = 1$ and imposing the boundary condition $f_{n}(0) = 0$ it's straightforward to obtain the first few coefficients of the asymptotic series.
It's easy to realize (and prove inductively) that the coefficient $f_{n}$ is always a polynomial of degree $2n$ in the variable
\begin{equation}
    w(z) = \frac{z}{1-b_{1} z} \, .
\end{equation}
Thus, using the ansatz $f_{n}(z) = \sum_{l=0}^{2n} d_{n,l} w(z)^{l}$ into the recursion formula for $f_{n}$ we can obtain an algebraic recursion formula for the coefficients $d_{n,k}$. 

The explicit form of the coefficients for all $n,k$ is difficult to obtain from the recursion, but we can leverage the knowledge we obtained so far to compute the coefficients $d_{n,k}$ by directly expanding Pochhammers in the Gauss series.
Given the previous discussion, it's clear that a better starting point is obtained after absorbing the prefactor $(1-b_{1} z)^{-a}$ and changing variable $z \to z(w) = w/(1+b_{1} w)$:
\begin{equation}
    g(w) = (1+b_{1} w)^{-a} f\left(\frac{w}{1+b_{1} w}\right) = \sum_{l=0}^{+\infty} \sum_{k=0}^{l} \binom{l}{k} \frac{(b_{0}+b_{1} \lambda)_{k}}{(c+\lambda)_{m}} (-b_{1})^{l-k} \frac{(a)_{l}}{l!} w^{l} \,.
\end{equation}
The Pochhammer in the denominator can be expanded in terms of Stirling polynomials of the second kind as in~\eqref{eq:2F1-expansion-Pochhammer-expansion}, while the one in the numerator in terms of Stirling polynomials of the first kind
\begin{equation}
    (b_{0} + b_{1} \lambda)_{n} = \sum_{k=0}^{n} \stirlingIpoly{n}{k}{b_{0}} (b_{1} \lambda)^{k} \,, \quad (n \geq 0) \,, \qquad \stirlingIpoly{n}{k}{x} = \sum_{m=k}^{n} \stirlingI{n}{m}\binom{m}{k} x^{m-k} \,,
\end{equation}
and $\stirlingI{n}{k}$ are \emph{unsigned} Stirling numbers of the first kind. This produces an expression for $g$ with four nested summations.
After performing several sum exchanges and index re-labeling, we can collect the inverse powers of $\lambda$ and obtain the asymptotic expansion of $g(w)$ as $\lambda \to +\infty$:
\begin{equation}
    g(w) \sim 1 + \sum_{n=1}^{+\infty} g_{n}(w) \lambda^{-n} \,.
\end{equation}
The coefficients $g_{n}$ are then explicitly written as polynomials in $w$ of degree $2n$.
We omit the details of the computation.
We find:
\begin{align}\label{eq:2F1-expansion-b-and-c-large-gn}
    g_{n}(w) 
    &= \sum_{l=1}^{2n} d_{n,l}(a,b_{0},b_{1},c) \, w^{l} \,, \\[0.5em]
    d_{n,l}(a,b_{0},b_{1},c)
    &= \frac{(a)_{l}}{l!}\sum_{r=1}^{l}\sum_{k=0}^{\min\{n,r\}}(-1)^{n+l+r-k}\binom{l}{r} \stirlingIIpoly{r+n-k-1}{r-1}{c} \stirlingIpoly{r}{r-k}{b_{0}}  b_{1}^{l-k} \,. \nn
\end{align}
Combining~\eqref{eq:2F1-expansion-b-and-c-large-gn} with~\eqref{eq:2F1-expansion-b-and-c-large-asymptotic-form}, where $f_{n}(z) = g_{n}(z/(1-b_{1}z))$, provides the full asymptotic expansion of $\hypgeo{2}{1}(a, b_{0} + b_{1} \lambda , c + \lambda; z)$ as $\lambda \to +\infty$.

\section{Resurgence of \texorpdfstring{$\hypgeo{2}{1}$}{2F1}}
\label{app:resurgence}

In this section, we study the resurgence properties of the asymptotic series~\eqref{eq:2F1-expansion-large-c-no-shift}.
Before proceeding, we review the few basic facts needed to follow the analysis.
The interested reader can consult~\cite{Marino:2012zq,sauzin2014,Dorigoni:2014hea,abs} for 
reviews on resurgence from different perspectives and~\cite{Serone:2024uwz} for a quick overview.

Given an asymptotic series expansion $\widetilde f(\epsilon)$, we can determine its Borel function $\hat f(t)$ in a disc around the origin, which we assume can analytically be continued over the whole complex plane.
If $\hat f$ has no singularities over the real axis, and is bounded in such a way that its Laplace transform~\eqref{eq:laplace-transform} is finite, its Borel resummation $s(\widetilde f)$ defines a well-defined function $f^B$.

Asymptotic expansions are not one-to-one with functions. 
Equivalence classes of functions which differ by non-perturbative terms of the form $\exp(-1/\epsilon)$ give rise to the same asymptotic expansion. 
The Borel resummed function $f^B$ picks up the unique representative in the equivalence class which is analytic in an open disc $D_R$ of radius $R$ with centre at $\epsilon=R$~\cite{Nevanlinna,Sokal}. 
Therefore the function $f^B$ does not necessarily coincide with the original function $f$.

\begin{figure}[t!]
    \centering
    \begin{tikzpicture}[scale=2]
    \draw[->] (0,-1.2) -- (0,1.2);
    \draw[->] (-1,0) -- (3,0);

    \node[draw=none, inner sep=5pt, append after command={
        \pgfextra{\draw (\tikzlastnode.south west) -- (\tikzlastnode.south east);}
        \pgfextra{\draw (\tikzlastnode.south west) -- (\tikzlastnode.north west);}
    }] at (2.8, 1) {$t$};

    \draw[path, thick] (0,0) -- (3,0.5) node[midway, above, text=black]  {${\cal C}_+$} [postaction={decorate, decoration={
        markings, mark=at position 0.75 with {\arrow{latex}}
    }}];
    \draw[path, thick] (0,0) -- (3,-0.5) node[midway, below, text=black]  {${\cal C}_-$} [postaction={decorate, decoration={
        markings, mark=at position 0.75 with {\arrow{latex}}
    }}];

    \draw[branch, thick, decorate, decoration={snake, segment length=3, amplitude=1.5}] (3/2,0) -- (3-0.03,0);
    \filldraw[branch] (3/2,0) circle (0.05);
    
    \end{tikzpicture}
    \caption{Contours associated to lateral Borel resummations around the real positive axis. The red dot indicates a branch cut singularity of the Borel function.\label{fig:lateral}}
\end{figure}
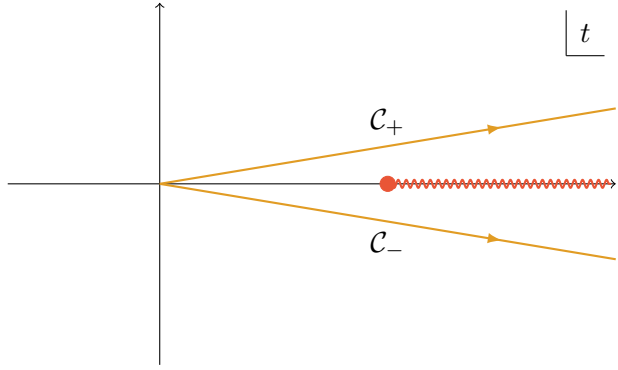 

An asymptotic series $\widetilde f(\epsilon)$ is not Borel summable if singularities occur along the positive real axis in $\hat f(t)$. 
We can deform the original contour into contours ${\cal C}_\pm$ which go in the upper or lower plane, as in figure~\ref{fig:lateral}.
The non-Borel summability is quantified by the difference of Borel resummation along ${\cal C}_\pm$: 
\begin{equation}\label{eq:resurgence2a}
    (s_{0+}-s_{0-})(\widetilde f) \neq 0\,,
\end{equation}
which define so-called lateral Borel resummations. 
The ambiguity~\eqref{eq:resurgence2a}, also called Stokes discontinuity, is quantified by the location $S$ of the closest singularity to the origin and is of order $\exp(- S/\epsilon)$. 
This is a useful way to discover non-perturbative effects. 

Discontinuities of the Borel function away from the real axis are also useful because they are associated with the large-order behaviour of the asymptotic series. 
In general, if the Borel function has a singularity at the point $S=\rho e^{i\phi}$ in the Borel plane, the discontinuity of the Borel sum at the same angle $\phi$ will contain terms of the form
\begin{equation}\label{eq:lboDer1}
    (s_{\phi+}-s_{\phi-})(\widetilde{f}) \approx 2\pi i C_0 \left(\frac{\epsilon}{S}\right)^\beta e^{-S/\epsilon} \, , \qquad
    |\epsilon|\ll 1\,.
\end{equation}
For each such term, the asymptotic behaviour of the coefficients $f_n$ of $\widetilde f$ has a term of the form
\begin{equation}\label{eq:lboDer2}
    f_n \approx C_0 S^{-n} \Gamma\left(n-\beta\right).
\end{equation}
The leading large order behaviour of $\widetilde f$ is then determined by the singularity closest to the origin, wherever located in the complex plane. 

The right-hand side of~\eqref{eq:resurgence2a} can often be rewritten as $s(\widetilde g)$, the Borel resummation of another
asymptotic series $\widetilde g$.
In turn, $\widetilde g$ might in turn also suffers of ambiguities, and so on. 
In the best case scenario, one hopes that by replacing the original asymptotic series $\widetilde f$ with a trans-series, a collection of asymptotic series, weighted by non-perturbative factors, $\widetilde f + \exp(-S/\epsilon) \widetilde g, \ldots$, the whole result can be non-ambiguous and reconstructs the full function.  
In a nutshell, resurgence is the theory explaining how to systematically implement this procedure.

We can now discuss the case at hand. The asymptotic series~\eqref{eq:2F1-expansion-large-c-no-shift} is the solution of a second-order differential equation and it has simple resurgent properties involving just two series, the original one and another obtained using~\eqref{eq:resurgence2a}.
The presence of just two asymptotic series is related to the second order of the hypergeometric differential equation~\eqref{eq:2F1-ode} which admits two independent solutions. 
The key point of the analysis that follows is to see how the resurgence properties of 
\begin{equation}
    \widetilde F_{1}(\epsilon; w) \sim \hypgeo{2}{1}\left(a,b,\frac{1}{\epsilon}; w\right),
\end{equation}
vary as we vary $w$.
We find it more convenient for the resurgence analysis to work in terms of 
\begin{equation}
    \epsilon = \frac{1}{\lambda} \, ,
\end{equation} 
so that the asymptotic limit is $\epsilon \to 0^{\pm}$ instead of $\lambda \to \pm \infty$.
We take $\epsilon$ real and discuss the cases $\epsilon>0$ and $\epsilon<0$ separately.

\subsection{Stokes jumps for large positive parameter}

We determine the Borel function \( \hatF_{1} \) associated to the \( \widetilde{F}_{1} \) series.
Using~\eqref{eq:borel-function-def} and~\eqref{eq:2F1-expansion-large-c-no-shift} we have
\begin{equation}\label{eq:hatF1-derivation}
    \resizebox{\textwidth}{!}{\(
    \begin{aligned}
        \hatF{}_{1}(w;t) 
        &= \sum_{n=1}^{+\infty} \sum_{k=1}^{n} (-1)^{n+k} \stirlingII{n-1}{k-1} \frac{(a)_{k}(b)_{k}}{k!} w^{k} \frac{t^{n-1}}{(n-1)!} 
        = \sum_{k=1}^{+\infty} \frac{(a)_{k}(b)_{k}}{k!} (-w)^{k} \sum_{n=k}^{+\infty} \stirlingII{n-1}{k-1} \frac{-(-t)^{n-1}}{(n-1)!} \\
       & = a b w  \sum_{k=0}^{+\infty} \frac{(a+1)_{k}(b+1)_{k}}{(k+1)!} \frac{(w(1-e^{-t}))^{k}}{k!} = a b w  \hypgeo{2}{1}(a+1,b+1,2; u)\, ,
    \end{aligned}
    \)}
\end{equation}
where
\begin{equation}
    u = w(1-e^{-t})\,,
\end{equation}
and from the first to the second row of~\eqref{eq:hatF1-derivation} we have used the relation
\begin{equation}\label{eq:stirlingII-generating-function-k}
    \sum_{n=k}^\infty \stirlingII{n}{k}\frac{t^n}{n!} = \frac{(e^t-1)^k}{k!}\,.
\end{equation}
We can determine the singularities of $\hatF_1$ in the Borel $t$-plane analytically as a function of $w \in \mathbb{C}$.
In a generic Riemann sheet $\hypgeo{2}{1}(A,B,C; u)$ is singular for $u=0,1,\infty$. 
We get a rich spectrum of singularities: 
\begin{equation}\label{eq:singularities-hatF1}
    \begin{aligned}
        u & = 0 
        && \Longrightarrow && 
        w=0, \, \forall t \, , \text{ or } t = 2i \pi k , \,  k\in \mathbb{Z}_{\neq 0} , \, \forall w \, , \\
        u & = 1 
        && \Longrightarrow && 
        t=t_k(w) = \log\left(\frac{w}{w-1}\right) + 2i \pi k\,, \quad  k\in \mathbb{Z} \, , \\
        u & = \infty 
        && \Longrightarrow && 
        w=\infty, \, \forall t, \text{ or } t=-\infty, \, \forall w \,. 
    \end{aligned}
\end{equation}
The most interesting ones are those which depend on $w$, in the second-row of~\eqref{eq:singularities-hatF1}, because as we vary $w \in \mathbb{C}$ they can cross the positive real axis and lead to the Stokes discontinuity~\eqref{eq:resurgence2a}.
When $w \notin [1,+\infty)$, the positive real line in the Borel $t$-plane is free of singularities.
In this case $\widetilde{F}_{1}$ is Borel summable and $F_{1}^B = s_{0}(\widetilde F_1)$ coincides with the original function:\footnote{
    This is expected, because $F_1$ is analytic in the appropriate disc in the $1/\epsilon$ plane.\label{foot:Sokal1}
}
\begin{align}
    s_{0}\big(\widetilde{F}_{1}\big)(w;\epsilon) 
    &= 1 + a b w \int_{0}^{+\infty} \dif{t} \, e^{-\frac{t}{\epsilon}} \hypgeo{2}{1}(a+1,b+1,2;u) \label{eq:tildeF1-borel-resummation-deriv}  \\
    &\overset{\mathclap{s = 1 - e^{-t}}}{=} \quad \ 1 + a b w \int_{0}^{1}\dif{s} (1-s)^{\frac{1}{\epsilon}-1} \hypgeo{2}{1}(a+1,b+1,2; w s) \\
    & =\hypgeo{2}{1}\left(a, b, \frac{1}{\epsilon}; w\right) 
    = F_1(w;\epsilon)\,.  \nn
\end{align}
When $w \in (1,+\infty)$, the singularity $t_0(w)$ in the second row of~\eqref{eq:singularities-hatF1} lie on the positive real axis of the Borel $t$-plane and gives rise to a Stokes discontinuity.
This is easily computed.
We have
\begin{equation}\label{eq:tildeF1-stokes-jump-deriv-1}
    \begin{aligned}
        (s_{0+} - s_{0-})[\widetilde{F}_{1}]
        &=         \lim_{\delta \to 0^{+}} \left(\int_{0}^{e^{i \delta} \infty} - \int_{0}^{e^{-i \delta} \infty}\right) \dif{t}\, e^{-\frac{t}{\epsilon}} \hatF_{1}(w; t)
        = \int_{\mathcal{H}} \dif{t}\,  e^{-\frac{t}{\epsilon}} \hatF_{1}(w; t) \\
        &= a b w \int_{t_0(w)}^{+\infty}\!\! \dif{t} \, e^{-\frac{t}{\epsilon}} \operatorname{Disc}\left[\hypgeo{2}{1}(a+1,b+1,2; u)\right] \, ,
    \end{aligned}
\end{equation}
where $\mathcal{H}$ is a Hankel contour starting around $t_0(w)$ and extending over the positive real axis. The contour integral picks up the discontinuity of the hypergeometric $\hypgeo{2}{1}(A, B, C; u)$ across the branch cut starting at $u = 1$. For reference the explicit formula is
\begin{equation}
\label{eq:2F1-discontinuity}
    \begin{aligned}
        &\operatorname{Disc}\left[\hypgeo{2}{1}(A,B,C;u)\right] = 2 \pi i \frac{\Gamma(C)}{\Gamma(A)\Gamma(B)\Gamma(C-A-B+1)} \\
        &\qquad\qquad  \times (u-1)^{C-A-B} \hypgeo{2}{1}(C-A,C-B,C-A-B+1;1-u) \, , \quad u \in (1,+\infty) \, .
    \end{aligned}
\end{equation}
Applying this formula to~\eqref{eq:tildeF1-stokes-jump-deriv-1} and shifting $t \to t_0(w) + t$ gives the integral
\begin{equation}\label{eq:2F1-derivation}
\begin{aligned}
    (s_{0+}-s_{0-})(\widetilde{F}_{1}) 
 &= \frac{2\pi i}{\Gamma(a)\Gamma(b)\Gamma(1-a-b)}  \left(\frac{w}{w-1}\right)^{-\frac{1}{\epsilon}} w (w-1)^{-a-b}   \\
    & \times \int_{0}^{+\infty} \dif{t} \, e^{-\frac{t}{\epsilon}}(1-e^{-t})^{-a-b} \hypgeo{2}{1}(1-a,1-b,1-a-b,(1-w)(1-e^{-t}))\,,
\end{aligned}
\end{equation}
leading to
\begin{equation}\label{eq:stokes-final}
    (s_{0+}-s_{0-})(\widetilde{F}_{1}) 
     =  \frac{ 2\pi i \, \Gamma\big(\tfrac{1}{\epsilon}\big) \, w^{-\frac{1}{\epsilon}+1} (w-1)^{\frac{1}{\epsilon}-a-b}}{\Gamma(a)\Gamma(b)\Gamma\big(1-a-b+\tfrac{1}{\epsilon}\big)}  \hypgeo{2}{1}\left(1-a,1-b,1-a-b+\frac{1}{\epsilon}, 1-w\right) \, . 
\end{equation}
Note that while~\eqref{eq:stokes-final} applies also when $a+b=1$, its derivation needs to be adjusted.
In such cases,~\eqref{eq:2F1-discontinuity} has to be adapted for $A+B=C+1$ and the singularity of $\widehat{F}_1$ changes nature from a rational branch cut to a pole and a logarithmic cut.
We rewrite~\eqref{eq:stokes-final} as
\begin{equation}\label{eq:tildeF1-stokes-jump}
    (s_{0+}-s_{0-})(\widetilde{F}_{1}) = i \eta F_{2}(\epsilon; w) \, , \qquad \quad w\in (1,+\infty) \,,
\end{equation}
where
\begin{align}
    F_{2}(w;\epsilon)  & = \frac{1}{\eta}  e^{-\frac{1}{\epsilon} S(w)} A(w)
    \mathcal{C}(\epsilon) \hypgeo{2}{1}\left(1-a,1-b,1-a-b+\frac{1}{\epsilon}, 1-w\right) \,, \label{eq:2F1F1jump} \\
    S(w) & = \log\left(\frac{w}{w-1}\right) \, , \quad 
    A(w) = \frac{w}{(w-1)^{a+b}} \, , \quad  \mathcal{C}(\epsilon) = \frac{2\pi\Gamma\big(\tfrac{1}{\epsilon}\big)}{\Gamma(a)\Gamma(b)\Gamma\big(1-a-b+\tfrac{1}{\epsilon}\big)} \,, \label{eq:saddle2}
\end{align}
and we have multiplied and divided by the factor
\begin{equation}\label{eq:eta-definition}
    \eta  = 2\sqrt{\sin(\pi a)\sin(\pi b) } \, ,
\end{equation}
to simplify some of the formulas which will follow. 
Note that $F_2$ in the right-hand side of~\eqref{eq:tildeF1-stokes-jump} is proportional to 
\begin{equation}
    (1-w)^{\frac{1}{\epsilon} - a - b} \hypgeo{2}{1}\left(\frac{1}{\epsilon} -a,\frac{1}{\epsilon} -b,\frac{1}{\epsilon}+1-a-b, 1-w\right)\,,
\end{equation}
which is one of the two independent solutions of~\eqref{eq:2F1-ode} when expanded around $w=1$.
Using Kummer's connection formulas, see e.g.\ equation 15.10.21 of~\cite{NIST:DLMF}, the Stokes discontinuity~\eqref{eq:tildeF1-stokes-jump} can be rewritten as a linear combination of the two independent solutions around $w=0$, $\hypgeo{2}{1}(a,b,1/\epsilon; w)$ and $w^{1-1/\epsilon}\hypgeo{2}{1}(1+a-1/\epsilon,1+b-1/\epsilon,2-1/\epsilon; w)$. 
From the Stokes jump of the asymptotic series of one solution, we have then ``discovered'' the other one.
The formulation in terms of the hypergeometric~\eqref{eq:2F1F1jump} is useful, as we have a closed-form expression for the asymptotic series of $\hypgeo{2}{1}$ when one parameter is large.

We now compute the Borel transform $\hatF_2(t)$, determine its Stokes discontinuities, and show that they give back $F_{1}$.
The asymptotic series $\widetilde F_{2}$ in $\epsilon$ is the product of the ones coming from $\mathcal{C}(\epsilon)$ and $\hypgeo{2}{1}$.\footnote{
See \eqref{eq:2F1-expansion-large-c-asymptotic-coefficients}
for the asymptotic series of the $\hypgeo{2}{1}$ part, and equation 5.11.13 of \cite{NIST:DLMF} for $\mathcal{C}(\epsilon)$.}
Its explicit expression is not needed, because we can immediately deduce $\hatF_2$ from the integrand appearing in~\eqref{eq:2F1-derivation}.
We have
\begin{equation}
    \hatF_{2}(w;t) = e^{-\frac{1}{\epsilon} S(w)} \frac{A(w)}{\eta}\frac{2\pi \left(1-e^{-t}\right)^{-a-b}}{\Gamma(a)\Gamma(b)\Gamma(1-a-b)} \, \hypgeo{2}{1}\left(1-a,1-b;1-a-b;(1-w) (1-e^{-t})\right).
\end{equation}

The singularities of $\hatF_{2}$ in the Borel $t$-plane are identical to the ones of $\hatF_{1}$ in~\eqref{eq:singularities-hatF1}, provided $w \to 1-w$.
As for $\hatF_{1}$, the most interesting ones are those which depend on $w$.
When $w \in \mathbb{C} \setminus (-\infty,0]$, the positive real line in the Borel $t$-plane is free of singularities.
In this case, $\widetilde{F}_{2}$ is Borel summable and we recover the original function $F_{2}$.
When $w \in (-\infty,0)$, instead, the singularities $t_0(1-w) = \log\!\big(\tfrac{w-1}{w}\big)$ lie on the positive real axis of the Borel $t$-plane.
The computation of the Stokes discontinuity is very similar to the one in~\eqref{eq:2F1-derivation} and will not be reported.
We have
\begin{equation}\label{eq:tildeF2-stokes-jump}
    (s_{0+}-s_{0-})(\widetilde{F}_{2})= i \eta A(w) A'(w) F_{1} \, , \qquad \quad w \in (-\infty,0) \, ,
\end{equation}
where
\begin{equation}
    A'(w) = -\frac{(1-w)^{a+b}}{w}\,.
\end{equation}
The factor of $A$ in~\eqref{eq:saddle2} has a branch cut on the line $(-\infty,1)$ and thus the product $A A'$ is ambiguous for $w \in (-\infty,0)$. 
The ambiguity amounts to the choice of the Riemann sheet of $w \in (-\infty,0)$.
We take $w-1 = \exp(i \pi) (1-w)$ in the principal sheet of $A$, so that
\begin{equation}
    A(w) A'(w) = - e^{-i\pi (a+b)} \,.
\end{equation}

\begin{figure}[t!]
    \centering
    \begin{tikzpicture}[scale=3]
    \draw[->] (-0.75,0) -- (1.75,0);
    \draw[->] (0,-0.75) -- (0,0.75);
    \draw (0,0) node[below right] {$0$};
    \draw (1,0) node[below left] {$1$};
    \draw[stokes, very thick] (1,0) -- (1.73,0) [postaction={decorate, decoration={
        markings, mark=at position 0.4 with {\arrow{latex}}
    }}];
    \draw[stokes, very thick] (-0.75,0) -- (0,0) [postaction={decorate, decoration={
        markings, mark=at position 0.7 with {\arrow{latex}}
    }}];

    \draw[dashed] (0.5,-0.75) -- (0.5,0.75) [postaction={decorate, decoration={
                markings, mark=at position 0.75 with {\arrow{latex}}
            }}];
    \filldraw[black] (0,0) circle (0.02);
    \filldraw[black] (1,0) circle (0.02);
    \node[draw=none, inner sep=4, append after command={
        \pgfextra{\draw (\tikzlastnode.south west) -- (\tikzlastnode.south east);}
        \pgfextra{\draw (\tikzlastnode.south west) -- (\tikzlastnode.north west);}
    }] at (1.62, 0.65) {$w$};
    \end{tikzpicture}
    \caption{Stokes lines (blue) and anti-Stokes lines (dashed) in the $w$-plane for the $(\widetilde{F}_{1},\widetilde{F}_{2})$ system.
    $w = 0, 1$ are singular points.
    Outgoing arrows denote lines where $\widetilde{F}_{1}$ has a Stokes jump, while ongoing denote lines where $\widetilde{F}_{2}$ jumps.\label{fig:simple-2F1-stokes}}
\end{figure}
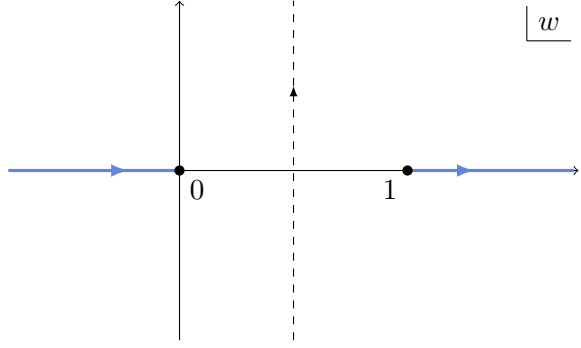

Summarizing, we have the following relations:
\begin{equation}\label{eq:phi1-phi2-system-stokes-jumps}
    \left\{ \
    \begin{aligned}
        s_{0}(\widetilde{F}_{1}) & = F_{1} \, , &\qquad w \in \mathbb{C} \setminus [1,+\infty) \, , \\
        s_{0}(\widetilde{F}_{2}) & = F_{2} \, , &\qquad w \in \mathbb{C} \setminus (-\infty,0] \, , \\
        (s_{0+}-s_{0-})(\widetilde{F}_{1}) &= i \eta \, s_{0}(\widetilde{F}_{2}) \, , &\qquad w \in (1,+\infty) \, , \\
        (s_{0+}-s_{0-})(\widetilde{F}_{2}) &=  - i e^{-i \pi (a+b)} \eta \,  s_{0}(\widetilde{F}_{1})\, , &\qquad w \in (-\infty,0) \, .
    \end{aligned}
    \right.
\end{equation}

The structure of Stokes lines in the $w$-plane is displayed in figure~\ref{fig:simple-2F1-stokes}. The lines emanate from the ``singular'' points $w = 0,1$ and go to infinity along the real axis.
Out-going arrows from the singular points denote where $\widetilde{F}_{1}$ jumps while in-going arrows denote where $\widetilde{F}_{2}$ jumps.
In the main text, these ``fundamental'' Stokes lines are mapped into the Stokes lines in the $z$-plane of figures \ref{fig:4H-stokes-and-branch} and \ref{fig:2H2L-stokes} through the maps $w=r(z)$ and $w=v(z)$, respectively.

\subsection{Stokes jumps for large negative parameter}
\label{app:resurgence-eps-negative}

For $\epsilon < 0$ the relevant Stokes jumps for $\widetilde{F}_{1}$ or $\widetilde{F}_{2}$ happen when $t_k(w)$ or $t_k(1-w)$ (defined in~\eqref{eq:singularities-hatF1}), respectively, cross the \emph{negative} real axis.
It is simple to see that this happens precisely when one of the other is real and positive.
Thus, the Stokes lines of  $\widetilde{F}_{1,2}$ swap, $\widetilde{F}_1$ is not Borel summable for $w\in(-\infty,0)$ and $\widetilde{F}_2$ is not Borel summable for $w\in(1,\infty)$.

Since the asymptotic series associated with the singularity does not change with either the sign of $\epsilon$ or $w$, the Stokes jump itself is the same.
However, due to the analytic continuation in the branch cuts of the pre-factors, there is a small phase difference.
Summarily, we can write the Stokes jumps for $\epsilon<0$ as
\begin{equation}\label{eq:phi1-phi2-system-stokes-jumps-neg}
    \begin{aligned}
        (s_{\pi+}-s_{\pi-})(\widetilde{F}_{1})&= i e^{2 \pi i(a+b)} \eta \, s_{\pi}(\widetilde{F}_{2}) \, , &\qquad w \in (-\infty,0) \, , \\
        (s_{\pi+}-s_{\pi-})(\widetilde{F}_{2})&= - i e^{- \pi i(a+b)} \eta \,  s_{\pi}(\widetilde{F}_{1})\, , &\qquad w \in (1,+\infty) \, .
    \end{aligned}
\end{equation}
Note that the resummations of the asymptotic series differ substantially from the $\epsilon>0$ case.
This is to be expected since we cross multiple Stokes phenomena as we analytically continue from the positive to the negative real axis.
However, starting from the exact Borel transforms we can calculate the resummation with a change of variables analogous to~\eqref{eq:tildeF1-borel-resummation-deriv}, with the ray along the negative real axis instead.
We find
\begin{equation}\label{eq:resum-neg-eps}
    \begin{aligned}
        s_{\pi}(\widetilde{F}_1) &= 
        \, \hypgeo{2}{1}\left(a,b;\frac{1}{\epsilon };w\right) + \mathcal{D}(\epsilon)  w^{1-\frac{1}{\epsilon}}
        \, \hypgeo{2}{1}\left(1+a-\frac{1}{\epsilon},1+b-\frac{1}{\epsilon};2-\frac{1}{\epsilon };w\right), &\; w \in \mathbb{C} \setminus (-\infty,0]\\
        s_{\pi}(\widetilde{F}_2) &= 
        - \mathcal{D}'(\epsilon) \left(\frac{w}{w-1}\right)^{-\frac{1}{\epsilon}}\frac{w \, e^{-i \pi  (+a+b)}}{(w-1)^{a+b}}
        \, \hypgeo{2}{1}\left(1-a,1-b;2-\frac{1}{\epsilon };w\right)\,,   &\; w \in \mathbb{C} \setminus [1,+\infty)
    \end{aligned}
\end{equation}
where
\begin{equation}
    \begin{aligned}
    \mathcal{D}(\epsilon) 
    & = \frac{
        \Gamma \big(\tfrac{1}{\epsilon }-1\big) \Gamma \big(a-\tfrac{1}{\epsilon }+1\big) \Gamma \big(b-\tfrac{1}{\epsilon }+1\big) 
        }{
        \Gamma (a) \Gamma (b) \Gamma \big(1-\tfrac{1}{\epsilon }\big)
        } ,\, \\[0.5em]
    \mathcal{D}'(\epsilon) 
    & = \frac{\pi}{\sqrt{\sin (\pi  a) \sin (\pi  b)}} 
    \frac{
        \Gamma \big(a-\tfrac{1}{\epsilon }+1\big) \Gamma \big(b-\tfrac{1}{\epsilon }+1\big) 
        }{
        \Gamma (a) \Gamma (b) \Gamma \big(2-\tfrac{1}{\epsilon }\big) \Gamma \big(1-\tfrac{1}{\epsilon }\big)
        } \,.
    \end{aligned}
\end{equation}
For $\epsilon<0$ we see that $s_{\pi}(\widetilde F_1)\neq F_1$, but rather it equals a specific linear combination of the two independent solutions of the hypergeometric equation~\eqref{eq:2F1-ode}.
This combination is the only representative in the class of functions with asymptotic series $\widetilde F_1$ which has the correct analytic behaviour in $1/\epsilon$ in an open disc $D_R$ of radius $R$ with centre at $-\epsilon=R$.
Up to a numerical factor, the resummation of $\widetilde F_2$ in~\eqref{eq:resum-neg-eps} equals $w^{1-1/\epsilon}\hypgeo{2}{1}(1+a-1/\epsilon,1+b-1/\epsilon,2-1/\epsilon; w)$, the ``other'' solution around $w=0$.

\subsection{Large-order behaviour}
\label{subsec:lbo}

The large order behaviour of the coefficients $f_n$ in~\eqref{eq:2F1-expansion-large-c-no-shift} 
can be extracted from the Borel plane analysis,
as pointed out in~\eqref{eq:lboDer1} and~\eqref{eq:lboDer2}.
Since we have access to the exact Borel transform in~\eqref{eq:hatF1-derivation}, we can in principle extract the full large order behaviour.
To do so, we inspect the rational branch cuts, which are of the form\footnote{The relation~\eqref{eq:BorelLBO} is easy to generalize for integer $\beta$ for which it corresponds to pole singularities (if $\beta=0$) and logarithmic branch cuts ($\beta\geq 0$).}
\begin{equation}\label{eq:BorelLBO}
    \hatF(t) = \text{holomorphic part} + \frac{\pi}{\sin(\pi \beta)}(S-t)^{-\beta}\sum_{k\geq 0} \frac{c^{(S)}_k}{\Gamma(n+1-\beta)} (t-S)^k+\cdots\,.
\end{equation}
These singularities then show up in the coefficients $f_n$ of $\widetilde f$ as
\begin{equation}
    f_n \sim \sum_{k\geq 0} c_k^{(S)} \Gamma\big((n-k)-\beta\big) S^{-(n-k)+\beta}.
\end{equation}
For real power series $\widetilde{f}$, the leading contribution is given by either a single real singularity or a pair of complex conjugate singularities. The latter case can be simplified to
\begin{equation}\label{eq:fnLOB}
    f_n \sim \sum_{k\geq 0} 2\cos\!\big((n-k-\beta)\phi-\alpha_k\big)\Gamma\big((n-k)-\beta\big)\rho^{-(n-k)+\beta}\left|c_k^{(S)}\right|\,,
\end{equation}
where $S = \rho e^{i \phi}$, $c_k^{(S)} = \big|c_k^{(S)}\big| e^{i\alpha_k}$. 

The leading large-order behaviour is then determined by the largest term in 
the sum appearing in~\eqref{eq:fnLOB}, generally given by the $k=0$ term of the singularity with the smallest $\rho$ (closest to the origin).
Due to the many, $w$-dependent, singularities this leading behaviour depends on $w$. For simplicity, we focus on $w$ real.
We have three regimes for $w$ real:
\begin{equation}\label{eq:fn-LOB-cases}
    \resizebox{\textwidth}{!}{\(
    f_n(a,b,w) \approx
    \begin{dcases} 
         \frac{w \left((w-1)\log\frac{w}{w-1}\right)^{-a-b}}{\Gamma (a) \Gamma (b)} \Gamma (n+a+b-1) \left(\log\frac{w}{w-1}\right)^{1-n},
        & w<0 \vee w > \tfrac{1}{1-e^{-2 \pi }},  \\[0.5em]
        \begin{aligned}
            &\frac{w (1-w)^{-a-b}}{\Gamma (a) \Gamma (b)} \Gamma (n+a+b-1)
            \left| \log \frac{w}{w-1}\right| ^{1-n-a-b}\\
            &\qquad \times 2\cos \left((n+a+b-1) \arg \left(\log\frac{w}{w-1}\right)-\pi  (a+b)\right),
        \end{aligned}
        & 0<w<\tfrac{1}{1+e^{-\sqrt{3} \pi }}, \\[0.5em]
        2^{2-n} \pi ^{-n-1} \sin (\pi  a) \sin (\pi  b) \Gamma (n) \sin \left(\pi  (a+b)+\frac{\pi  n}{2}\right),
        & \tfrac{1}{1+e^{-\sqrt{3} \pi }} < w < \tfrac{1}{1-e^{-2 \pi }}
       .
    \end{dcases}
    \)}
\end{equation}
In the first and last cases, there is a single closest singularity at $t_0$.
In the middle case, there are two complex singularities which dominate, corresponding to $t_{0,1}=\log\!\big(\tfrac{w}{1-w}\big)\pm \pi i$.
Lastly, when $w$ is very close to $1$, $\big|\!\log\tfrac{w}{w-1}\big|$ becomes large and the closest singularities are those at $t=\pm 2 \pi i$.

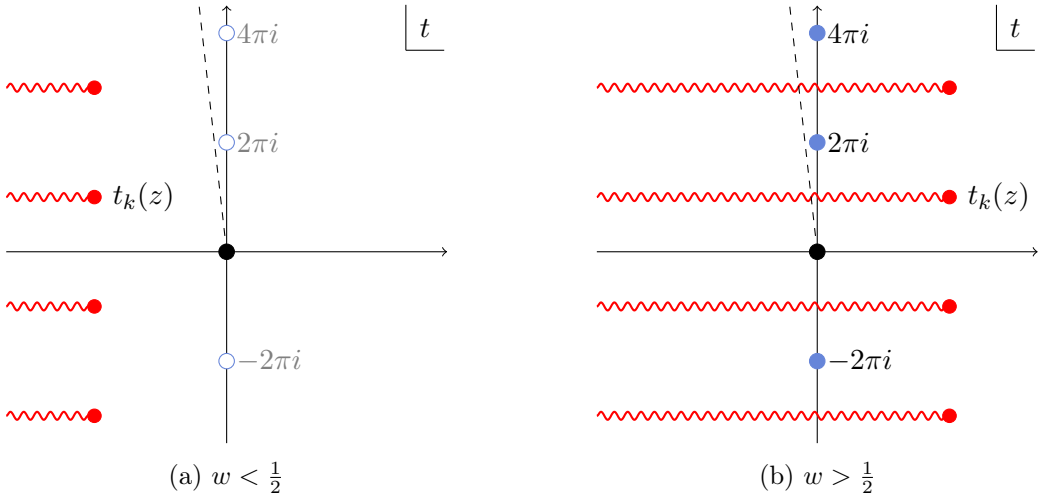
\begin{figure}[t!]
    \centering
    \hspace{1em}
    \begin{subfigure}[c]{0.45\textwidth}
        \centering
        \begin{tikzpicture}[scale=1.45]
            \draw[->] (-2,0) -- (2,0);
            \draw[->] (0,-1.75) -- (0,2.25);

            \node[draw=none, inner sep=5pt, append after command={
                \pgfextra{\draw (\tikzlastnode.south west) -- (\tikzlastnode.south east);}
                \pgfextra{\draw (\tikzlastnode.south west) -- (\tikzlastnode.north west);}
            }] at (1.8, 2.05) {$t$};

            \foreach \y in {-2,2,4} {
                \filldraw[stokes, fill=white] (0,{\y/2}) circle (2pt) node[anchor=west, gray] {\( \y \pi i \)};
            }
            \foreach \y in {-1.5,-0.5,0.5,1.5} {
                \draw[branch, thick, decorate, decoration={snake, segment length=5, amplitude=1.5}, red] (-2,\y) -- (-1.2,\y);
                \filldraw[red, fill=red] (-1.2,\y) circle (1.75pt);
            }
            \node[anchor=west] at (-1.2,0.5) {$\;t_k(z)$};

            \draw[black, dashed] (0,0) -- (-0.25, 2.25);
            \filldraw[black, fill=black] (0,0) circle (2pt);
        \end{tikzpicture}
        \caption{$w<\tfrac{1}{2}$}
    \end{subfigure}
    \hfill
    \begin{subfigure}[c]{0.45\textwidth}
        \centering
        \begin{tikzpicture}[scale=1.45]
            \draw[->] (-2,0) -- (2,0);
            \draw[->] (0,-1.75) -- (0,2.25);

            \node[draw=none, inner sep=5pt, append after command={
                \pgfextra{\draw (\tikzlastnode.south west) -- (\tikzlastnode.south east);}
                \pgfextra{\draw (\tikzlastnode.south west) -- (\tikzlastnode.north west);}
            }] at (1.8, 2.05) {$t$};

            \foreach \y in {-2,2,4} {
                \filldraw[stokes] (0,{\y/2}) circle (2pt) node[anchor=west, black] {\( \y \pi i \)};
            }
            \foreach \y in {-1.5,-0.5,0.5,1.5} {
                \draw[branch, thick, decorate, decoration={snake, segment length=5, amplitude=1.5}, red] (-2,\y) -- (1.2,\y);
                \filldraw[red, fill=red] (1.2,\y) circle (1.75pt);
            }
            \node[anchor=west] at (1.2,0.5) {$\;t_k(z)$};

            \draw[black, dashed] (0,0) -- (-0.25, 2.25);
            \filldraw[black, fill=black] (0,0) circle (2pt);
        \end{tikzpicture}
        \caption{$w>\tfrac{1}{2}$}
    \end{subfigure}
    \hspace{1em}
    \caption{Higher order stokes phenomenon in the Borel transform~\eqref{eq:hatF1-derivation}.
    The red dots are the singularities $t_k$ defined in~\eqref{eq:singularities-hatF1}, with their respective branch cuts drawn as wavy lines.
    The branch cuts are placed along the natural choice for $\hypgeo{2}{1}$ after the exponential map $u$.
    The singularities at non-zero multiples of $2\pi i$ are in the non-principal sheet of the hypergeometric function and are only reached by the ray of Borel summation if the ray crosses a branch.
    This happens for $w>\tfrac{1}{2}$ (right) but not for $w<\tfrac{1}{2}$ (left).
    We colour these singularities as grey when they are invisible and as blue otherwise.
    The dashed line represents the ray of~\eqref{eq:laplace-transform} for an angle close to $\theta=\pi/2$.\label{fig:Fig-higher-os}}
\end{figure}

One could be surprised that the singularities at $t=\pm 2 \pi i$ do not also dominate the large order behaviour for $w$ close to $0$ as $|t_0|\gg 1$.
In fact, they do not contribute at all due to a so-called higher-order Stokes phenomenon~\cite{HOStokes}.
Higher-order Stokes phenomena happen when two singularities in the Borel plane become collinear with the origin.
Then the branch cut emanating from the closer singularity affects whether rays emanating at the origin reach the further singularity.
The large-order behaviour is sensible to singularities which intersect the rays on which the Borel summation is performed for any angle, see~\eqref{eq:lboDer1}. In the case of aligned singularities, the closest one can eclipse all others, preventing them from showing up in the large-order behaviour.

In this case, the points $2\pi i k$ are not singularities of the hypergeometric function in~\eqref{eq:hatF1-derivation} in the principal sheet, but they are on other ones.
When $\re(w) <1/2$, the branch cuts starting at $t_k$ are such that the rays emanating from the origin remain in the principal sheet of the hypergeometric function along the imaginary axis.
At  $\re(w)=1/2$ the singularities at $t_k$ align with those $2\pi i k$, leading to a higher order Stokes phenomenon.
Thus, if $\re(w)>1/2$, a ray emanating from the origin crosses the branch cut and reaches $t=\pm 2\pi i$ in the other sheet, where the hypergeometric function is singular.
This is illustrated in figure~\ref{fig:Fig-higher-os}. Thus, the imaginary singularities only contribute to the large order behaviour when $\re(w)>1/2$.

\section{Summary of asymptotic expansions}
\label{app:summary2F1}

We define \emph{Stirling polynomials} of the first and second kind, respectively, as
\begin{align}
    \stirlingIpoly{n}{k}{x} &= \sum_{m=k}^{n} \stirlingI{n}{m}\binom{m}{k} x^{m-k}\, , & n \geq 0, \ 0 \leq k \leq n\, , \\[0.5em]
    \stirlingIIpoly{n}{k}{x} &= \sum_{m=k}^{n} \stirlingII{m}{k} \binom{n}{m} x^{n-m} \, , & n \geq 0, \ 0 \leq k \leq n \, ,
\end{align}
where $\stirlingI{n}{m}$ are \emph{unsigned} Stirling numbers of the first kind and $\stirlingII{m}{k}$ are Stirling numbers of the second kind.
When $x = 0$, the Stirling polynomials reduce to the corresponding Stirling numbers.

\paragraph{Large \texorpdfstring{$b$}{b} and \texorpdfstring{$c$}{c}} As $\lambda \to +\infty$, we have
\begin{gather}\label{eq:2F1-summary-large-b-and-c}
    \hypgeo{2}{1}(a, b_{0} + b_{1} \lambda , c + \lambda; z) 
    \sim (1-b_{1} z)^{-a} \left(1+\sum_{n=1}^{+\infty} g_{n}\left(a, b_{0}, b_{1}, c; \frac{z}{1-b_{1}z}\right)\lambda^{-n}\right) \, , \\
    \begin{aligned}        
        g_{n}(a, b_{0}, b_{1}, c; w) 
        &= \sum_{l=1}^{2n} d_{n,l}(a,b_{0},b_{1},c) \, w^{l} \, ,\\[0.5em]
        d_{n,l}(a,b_{0},b_{1},c)
        &= \frac{(a)_{l}}{l!}\sum_{r=1}^{l}\sum_{k=0}^{\min\{n,r\}}(-1)^{n+l+r-k}\binom{l}{r} \stirlingIIpoly{r+n-k-1}{r-1}{c} \stirlingIpoly{r}{r-k}{b_{0}}  b_{1}^{l-k}  \, .
    \end{aligned} \nn
\end{gather}
We have not investigated in detail the domain $z \in D \subseteq \mathbb{C}$ and parameters $a, b_{0}, b_{1}, c \in \mathbb{C}$ where this asymptotic approximation is valid.
This domain could be deduced by studying the web of Stokes and anti-Stokes lines associated with the differential equation. It would be interesting to fill this gap.

\paragraph{Large \texorpdfstring{$c$}{c}} As $\lambda \to +\infty$, we have
\begin{equation}\label{eq:2F1-summary-large-c}
    \begin{aligned}
        \hypgeo{2}{1}\left(a,b,c+\lambda; z\right) &\sim 1 + \sum_{n=1}^{+\infty} f_{n}(a,b,c;z) \lambda^{-n} \, , & z \in \mathbb{C}\setminus[1,+\infty), \\[0.5em]
        f_{n}(a,b,c;z) &= \sum_{k=1}^{n} (-1)^{n+k} \stirlingIIpoly{n-1}{k-1}{c} \frac{(a)_{k}(b)_{k}}{k!} z^{k} \, .
    \end{aligned}
\end{equation}
This can also be obtained as a special case of~\eqref{eq:2F1-summary-large-b-and-c} with $b_{1} = 0$.
As $\lambda \to -\infty$, the asymptotic form is the same but the domain is restricted to $\{z \in \mathbb{C} \setminus (-\infty,0] \ \colon \re{(z)} < 1/2\}$.

\paragraph{Special cases} Large parameter asymptotic expansion of $\hypgeo{2}{1}(-\lambda,1+\lambda,2+2\lambda;z)$ can be reduced to special cases of the asymptotic expansion~\eqref{eq:2F1-summary-large-c} after a change of variable.
As $\lambda \to +\infty$, we have
\begin{equation}
    \begin{aligned}
        \hypgeo{2}{1}(-\lambda,1+\lambda,2+2\lambda;z)
        &\sim e^{-\lambda S(z)} A(z)\left(1+ \sum_{n=1}^{+\infty} f_{n}(r(z)) \lambda^{-n}\right) \, , \\[0.5em]
        f_{n}(r) &= \sum_{k=1}^n (-1)^{n+k} \stirlingIIpoly{n-1}{k-1}{\frac{3}{2}} \frac{\big(\frac{1}{6}\big)_k \big(\frac{5}{6}\big)_k }{k!} r^k \, ,
    \end{aligned}    
\end{equation}
where
\begin{equation}
    S(z) = \frac{1}{2}\log\left(\frac{27}{16} \frac{z^{2}}{(1-z)^{2}}\frac{1-r(z)}{r(z)}\right) \, , \qquad \qquad 
    A(z) = \frac{e^{-\frac{1}{2}S(z)}}{(1-z+z^{2})^{\frac{1}{4}}} \, ,
\end{equation}
and 
\begin{equation}
    r(z) = \frac{1}{2} + \frac{(z+1)(2z-1) (2-z)}{4(1-z+z^2)^{\frac{3}{2}}} \, .
\end{equation}

Similarly, large parameter asymptotic expansion of $\hypgeo{2}{1}(2\gamma, 1+\lambda, 2+2\lambda)$ can be reduced to the asymptotic expansion~\eqref{eq:2F1-summary-large-c}. As $\lambda \to +\infty$, we have
\begin{equation}\label{eq:2F1-summary-large-c-2}
    \begin{aligned}
        \hypgeo{2}{1}(2\gamma, 1+\lambda, 2+2\lambda;z) 
        &\sim \left(1-\frac{z}{2}\right)^{-2\gamma}\left(1+ \sum_{n=1}^{+\infty} f_{n}\left(\gamma; \frac{z^2}{(2-z)^{2}}\right) \lambda^{-n}\right) \, , \\[0.5em]
        f_{n}(\gamma; w) &= \sum_{k=1}^{n} (-1)^{n+k} \stirlingIIpoly{n-1}{k-1}{\frac{3}{2}} \frac{(\gamma)_k (\gamma + 1/2)_k }{k!} w^k \, .
    \end{aligned}    
\end{equation}
See the discussion at the end of section~\ref{subsec:4H-eps-negative} and~\ref{subsec:2H2L-eps-negative} and the related figure~\ref{fig:4H-exact-vs-semiclassic} and~\ref{fig:2H2L-exact-vs-semiclassic}, respectively, for a discussion on the domain in $z$ of validity of~\eqref{eq:2F1-summary-large-c-2}, also in the case $\lambda \to -\infty$.

\bibliographystyle{JHEP}
\bibliography{Refs}

\end{document}